\documentclass[12pt]{article}
\usepackage{graphicx}
\usepackage{parskip} 
\usepackage{graphicx}
\usepackage{graphics}
\usepackage{amssymb}
\usepackage{amsmath}
\usepackage[pdftex]{hyperref}
\hypersetup{colorlinks,urlcolor=magenta,citecolor=red,linkcolor=blue,filecolor=black}
\usepackage{epsf,amsfonts,hyperref}

\usepackage{tikz}

\usepackage[
backend=biber,
style=alphabetic
]{biblatex}
\newcommand{\examplename}{Example}
\newcommand{\exampleref}[1]{\hyperref[#1]{\examplename~\ref*{#1}}}

\newcommand\encadremath[1]{\vbox{\hrule\hbox{\vrule\kern8pt 
\vbox{\kern8pt \hbox{$\displaystyle #1$}\kern8pt} 
\kern8pt\vrule}\hrule}}
\def\enca#1{\vbox{\hrule\hbox{
\vrule\kern8pt\vbox{\kern8pt \hbox{$\displaystyle #1$}
\kern8pt} \kern8pt\vrule}\hrule}}

\newcommand\figureframex[3]{
\begin{figure}[bth]
\hrule\hbox{\vrule\kern8pt 
\vbox{\kern8pt \vbox{
\begin{center}
{\mbox{\epsfxsize=#1.truecm\epsfbox{#2}}}
\end{center}
\caption{#3}
}\kern8pt} 
\kern8pt\vrule}\hrule
\end{figure}
}
\newcommand\figureframey[3]{
\begin{figure}[bth]
\hrule\hbox{\vrule\kern8pt 
\vbox{\kern8pt \vbox{
\begin{center}
{\mbox{\epsfysize=#1.truecm\epsfbox{#2}}}
\end{center}
\caption{#3}
}\kern8pt} 
\kern8pt\vrule}\hrule
\end{figure}
}

\makeatletter
\@addtoreset{equation}{section}
\makeatother
\newtheorem{theorem}{Theorem}[section]

\newtheorem{remark}{Remark}[section]
\newtheorem{proposition}{Proposition}[section]
\newtheorem{lemma}{Lemma}[section]
\newtheorem{corollary}{Corollary}[section]
\newtheorem{definition}{Definition}[section]
\def\br{\begin{remark}\rm\small}
\def\er{\end{remark}}
\def\bt{\begin{theorem}}
\def\et{\end{theorem}}
\def\bd{\begin{definition}}
\def\ed{\end{definition}}
\def\bp{\begin{proposition}}
\def\ep{\end{proposition}}
\def\bl{\begin{lemma}}
\def\el{\end{lemma}}
\def\bc{\begin{corollary}}
\def\ec{\end{corollary}}
\def\beaq{\begin{eqnarray}}
\def\eeaq{\end{eqnarray}}
\def\bex{\begin{example}}
\def\eex{\end{example}}
\newcommand{\proof}[1]{{\noindent \bf Proof:}\par
{#1} $\square$}
\newtheorem{example}{Example}[section]

\newcommand{\beq}{\begin{equation}}
\newcommand{\eeq}{\end{equation}}
\newcommand{\bea}{\begin{eqnarray}}
\newcommand{\eea}{\end{eqnarray}}

\renewcommand{\and}{{\qquad {\rm and} \qquad}}

 \newcommand{\Tr}{{\,\rm Tr}\:}

\newcommand{\Res}{\mathop{\,\rm Res\,}}

\newcommand{\td}[1]{{\tilde{#1}}}

\newcommand{\Pint}{{\int\kern -1.em -\kern-.25em}}

\usepackage{comment}
\def\w{w}
\def\wb{{\bar w}}
\begin{document}
\sloppy


\pagestyle{empty}
\addtolength{\baselineskip}{0.20\baselineskip}
\baselineskip 16pt 
\begin{center}
\begin{Large}\fontfamily{cmss}
\fontsize{20pt}{30pt}
\selectfont
\medskip
\textbf{Counting mobiles by integrable systems}
\end{Large}\\
\bigskip
\bigskip
{\sl M.\ Berg\`ere}${}^{1}$\hspace*{0.05cm},
{\sl B.\ Eynard}${}^{12}$\hspace*{0.05cm},
{\sl E.\ Guitter}${}^{1}$\hspace*{0.05cm},
{\sl S.\ Oukassi}${}^{1}$\hspace*{0.05cm}\\
${}^1$ Universit\'e Paris-Saclay, CEA, CNRS, Institut de physique th\'{e}orique\\91191 Gif-sur-Yvette, France.\\
${}^2$ CRM Centre de Recherches Math\'ematiques de Montr\'eal, QC, Canada.\\
\vspace{6pt}
\end{center}

\vspace{20pt}
\begin{center}
{\bf Abstract}
\end{center}
Mobiles are a particular class of decorated plane trees which serve as codings for planar maps. Here we address the question of enumerating mobiles in their most general flavor,  in correspondence with planar Eulerian (i.e., bicolored) maps. We show that the generating functions for such mobiles satisfy a number of recursive equations which lie in the field of integrable systems, leading us to explicit expressions for these generating functions as ratios of particular determinants. In particular we recover known results for mobiles associated with uncolored maps and prove some conjectured formulas for the generating functions of mobiles associated with $p$-constellations.

%





\vspace{26pt}
\pagestyle{plain}
\setcounter{page}{1}


\section{Introduction}
Mobiles denote a particular class of decorated plane trees carrying integer labels and subject to 
a number of local rules -- to be detailed below -- around their vertices. Mobiles were introduced 
in \cite{PMLM}, where it was shown that they provide a \emph{bijective coding} of general classes of \emph{planar maps}. 

Recall that a planar map is a connected graph drawn on the 
two-dimensional sphere (or equivalently on the plane) without edge crossings, and considered up to 
continuous deformations (see \cite{Schaeffer15} for a comprehensive introduction to maps). 
The problem of enumerating maps has attracted a lot of attention since the seminal work of Tutte in
the early 60's\cite{Tutte63}, and many enumeration techniques for maps with various topologies were
developed over the years, ranging from matrix integral techniques \cite{BIPZ} to the so-called topological recursion approach
\cite{Eynard2007,Eynard2016}. 

In this quest for enumerative formulas, the discovery of the coding by mobiles was an important step as it allowed one
to get a recursive construction of planar maps, transcribed into a set of recursive equations for their \emph{generating functions}.   
An interesting feature of the coding is that the decorations carried by the mobiles 
allow one to keep track of some of the geodesic distances between marked ``points'' (i.e., vertices, edges or faces) 
on the associated map. This property led to a number of results on the \emph{statistics of distances} between points within random maps  \cite{GDPG}.

Of particular interest is the \emph{generating function} $R_i$ for mobiles rooted at a vertex labelled by some positive integer $i$ which,
via the bijection with maps, enumerates planar maps with two marked points at a prescribed distance constrained to remain less than $i$ 
-- see the next section for a precise definition. Indeed, it was soon realized \cite{GDPG} that, for many map families of interest,
the recursive equations obeyed by the $R_i$'s for $i\geq 1$ turn out to be \emph{integrable} \cite{DFG05,GDPGIA}. 
In particular, fully explicit expressions for $R_i$ were obtained for a number of families of planar maps:
these include \emph{quadrangulations} -- in bijection with so-called well-labelled trees \cite{Schaeffer98} which are a particularly simple example of mobiles --,
\emph{triangulations}, and more generally \emph{maps with prescribed face degrees}. In this
latter case, an interesting link with continued fractions was established in \cite{PMCF}, which allows one to understand the precise form of the 
$R_i$'s as ratios of Hankel determinants.

The most general family of mobiles that were introduced in \cite{PMLM} are those mobiles which code for so-called \emph{planar Eulerian maps}.
Recall that a planar Eulerian map is a planar map whose all vertices have even degree, which, if the map is rooted, is equivalent to the condition that the map
be canonically face-bicolored, i.e., with faces colored in black and white with no two adjacent faces of the same color. Such maps are also called 
\emph{hypermaps} in the literature, to emphasize the fact that they encompass the case of usual uncolored general maps\footnote{Indeed, starting from a general uncolored map,
we get a bicolored one by first coloring all faces in white, and by then inflating all the
edges into black bivalent faces separating the original faces.}. A natural question is therefore that of finding an explicit expression for the $R_i$'s in this extended family of mobiles.
It is precisely the purpose of the present paper to obtain such a formula thanks to the machinery developed for solving integrable systems.

Before we proceed, let us recall that expressions for $R_i$ in the context of genuine Eulerian maps were conjectured in the
restricted case of $p$-constellations \cite{GDPG}, which are Eulerian maps with only $p$-valent black faces, and white faces 
whose degrees are multiples of $p$. So far, these conjectured formulas were proved only for the case of Eulerian triangulations, 
in connection with the theory of multicontinued fractions \cite{constfpsac}.
The approach developed here allows us: (i) to formulate the mobile enumeration problem in the framework of integrable systems 
(ii) to recover known results for general uncolored maps, (iii) to prove the conjectured formulas for $p$-constellations 
and (iv) to extend these formulas to the general family of mobiles coding  for planar Eulerian maps with bounded face degrees.

The paper is organized as follows: 

\autoref{sec:mobiles} is devoted to the definition of mobiles. We first discuss in \autoref{sec:EM2mob} the local rules on labels satisfied by mobiles in their most general flavor as obtained from their correspondence with Eulerian maps. 
We then define in \autoref{sec:enuMob} a number of generating functions for mobiles and half-mobiles and derive the set of recursive equations which determine them. These generating functions can be expressed as the entries of two semi infinite matrices $P$ and $Q$ which are shown to ``quasi-commute'' in \autoref{sec:comPQ}. We discuss in \autoref{sec:bounded} the case of mobiles with bounded vertex degrees in correspondence with maps with bounded face degrees and show that the $R_i$'s, as well as the other (half-mobile) generating functions, admit large $i$ limits which are formal power series in the weights attached to the mobile vertices. We end this section by discussing in \autoref{sec:Mwg} mobiles with a weight $g$ per labeled vertex, corresponding to a weight $g$ per vertex of the associated map: we show in particular that our recursive system of equations admits a unique \textit{combinatorial} solution for which $R_i$ and the other generating functions of interest are formal power series in $\sqrt{g}$. 

The goal of \autoref{sec:Intsys} is to place our equations within the framework of integrable systems and to use the associated machinery to obtain a number of properties that must be satisfied by the operators $P$ and $Q$. We first show that, thanks to their quasi-commutation property, we can build out of the operators $P$ and $Q$ an \textit{isospectral system}: more precisely we build in \autoref{sec:isosys} a \textit{discrete Lax equation}. We define and compute in \autoref{sec:spcurv} the associated \textit{spectral curve} $\mathcal{E}(x,y)$ as the characteristic polynomial of a suitable endomorphism expressing the action of $P$ in the eigenvector space of $Q$ associated with an arbitrary eigenvalue $x$. In \autoref{sec:Brptdblpt} we introduce the notion of branch points and double points of the spectral curve and prove some crucial identity involving these double points (\autoref{lemma:eqwbarw}). We then construct in \autoref{sec:asyeig} the matrix wave function associated to the discrete Lax equation whose entries are given by common right (resp left) eigenvectors $\psi$ (resp $\phi$) of $P$ and $Q$ at the branch points of generic $x$. In \autoref{sec:recformBAfct} we prove that, as limit of a well defined matrix determinant,  $\psi$ and $\phi$ take the form of Baker-Akhiezer functions via the so-called Reconstruction formula. 

\autoref{sec:mainth} presents the proof of our main theorem, namely \autoref{th:mainth}, giving in particular an explicit expression for $R_i$. 	Note that, appart from \autoref{lemma:eqwbarw}, this section does not make use of the results of \autoref{sec:Intsys} and is thus self-contained. Starting from the spectral curve and its double points we define in \autoref{sec:spcurv2} a set of specific Baker-Akhiezer functions $\psi$ and $\phi$ in the form of appropriate determinants.  Thanks to a suitable scalar product introduced in \autoref{sec:scprod} we reconstruct explicitly in \autoref{sec:constPQ}  the entries of the operators $P$ and $Q$ as scalar products of these Baker-Akhiezer functions. These entries are shown to satisfy half of the recursion relations wanted for the mobile generating functions. In order to show that the second half of the wanted equations is  also satisfied, we construct in the same way in \autoref{sec:constPQt} the entries of $Q$ and $P$ transposed. We conclude our proof in \autoref{sec:concproof}: in particular we obtain an explicit expression for the generating functions $R_i$ of mobiles as bi-ratios of determinants.

In \autoref{sec:apps} we present a number of applications of our main result: we first recover in \autoref{sec:genplamap} the expression for $R_i$ found in \cite{PMCF} for the case of mobiles associated to general (uncolored) planar maps with bounded face degrees. We then prove in \autoref{sec:constel} the formula for $R_i$ conjectured in \cite{GDPG} and \cite{GDPGIA} in the case of $p$-constellations. 

We gather in \autoref{sec:conc} a number of concluding remarks. Some technical points are discussed in various Appendices.

\section{Mobiles}  
\label{sec:mobiles}

\subsection{From Eulerian maps to mobiles}
\label{sec:EM2mob}
The rules which define mobiles are directly inherited from their correspondence with maps.
Several classes of mobiles may thus be defined, in connection with several classes of maps. 
In this paper, we discuss mobiles in their most general flavor, as obtained from their bijection 
with {\it Eulerian planar maps}.

\begin{figure}[h!]
\begin{center}
\includegraphics[width=6cm]{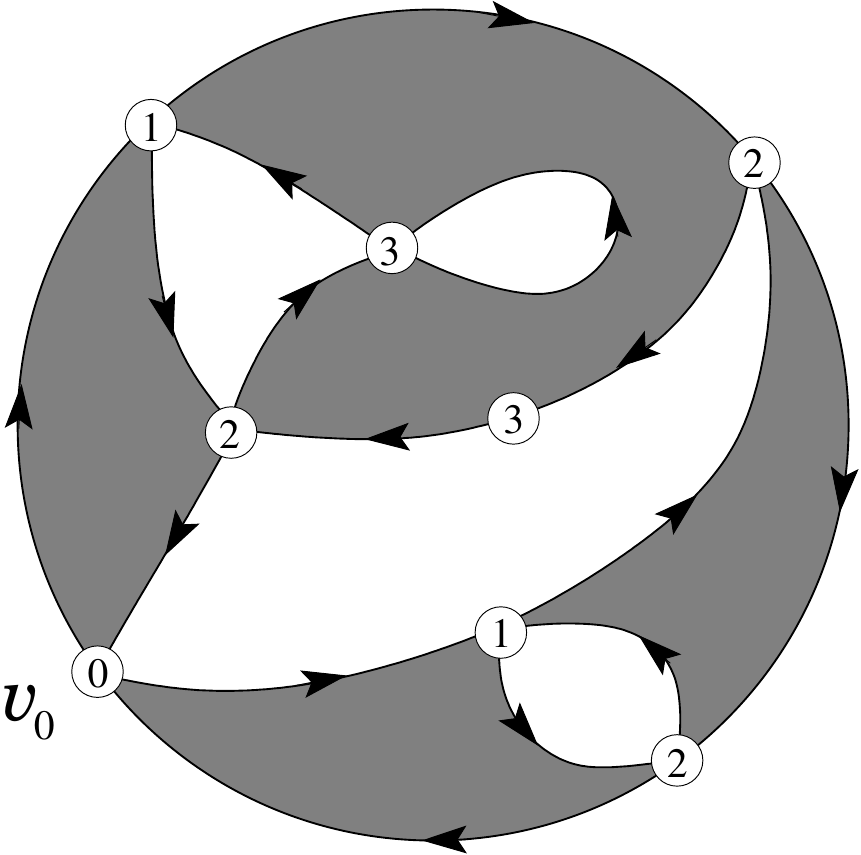}
\end{center}
\caption{\small{A pointed Eulerian (i.e. face bicolored) planar map with root vertex $v_0$. Edges are oriented clockwise around black faces and counterclockwise around white faces. 
Each vertex is labeled by its oriented geodesic distance from the root vertex.}} 
\label{fig:Pointedmap}
\end{figure}

Recall than a planar map is a cellular embedding of a graph in the sphere, considered up to continuous deformation. 
It is therefore made of vertices, edges and faces, the degree of a vertex (resp. a face) being the number of its incident half-edges
(resp. edge sides). A planar map is said Eulerian if {\it its vertices all have even degree}. Equivalently, 
this is a planar map which may be {\it face-bicolored}, i.e.\  whose faces may be colored, say in black and white 
so that no two adjacent faces are of the same color. Note that for a given Eulerian map, there are exactly two possible 
such colorings (related to each other by exchanging the colors). In the following, we shall always assume that our Eulerian 
maps are endowed with one of their two bi-colorings, i.e.\ a Eulerian map will in practice refer to a face-bicolored map. 
Let us now see how decorated trees emerge as a coding for Eulerian maps.

Since each edge in an Eulerian map separates a black and a white face, we may orient canonically all the edges of the 
map by demanding that their incident black face lies on their right when following the orientation. In other words, 
edges are oriented clockwise around the black faces and counterclockwise around the white faces (see \autoref{fig:Pointedmap}). 
If we now consider 
a {\it pointed} Eulerian map, i.e.\ with a distinguished {\it root vertex} $v_0$, we may label each vertex $v$ of the map by 
its {\it oriented geodesic distance} from the root vertex, defined as the length ($=$ number of edges) of any shortest path 
from $v_0$ to $v$ following edges of the map and respecting their orientation.
This label is a non-negative integer $i(v)$ satisfying $i(v_0)=0$ and
\begin{equation}
\hbox{for each edge $e$ oriented from $v$ to $v'$:} \qquad
i(v')\leq i(v)+1\ .
\end{equation}
\label{condlabel}

\begin{figure}[h!]
\begin{center}
\includegraphics[width=12cm]{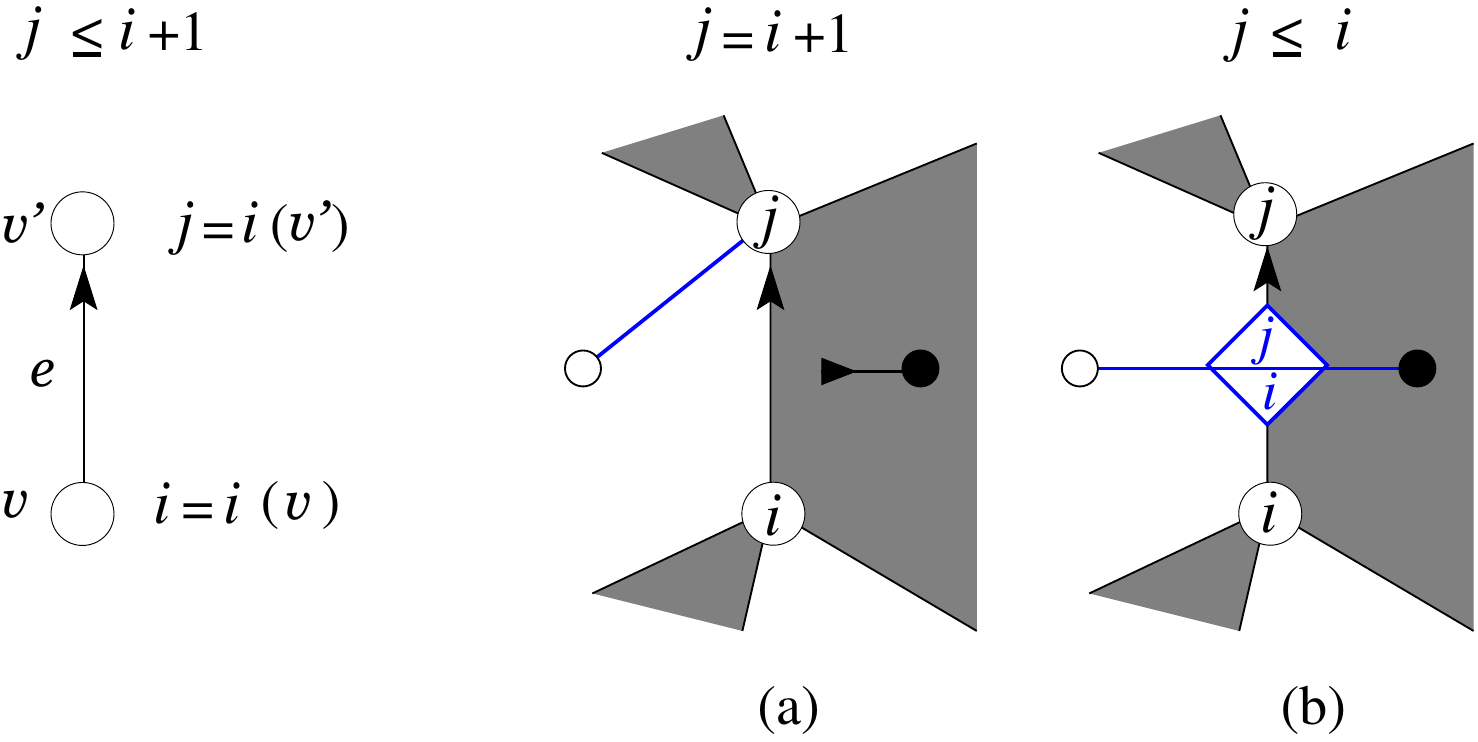}
\end{center}
\caption{\small{Local rule for the construction of a mobile, applied to each oriented edge $e$ of an Eulerian map, linking a vertex $v$ to a vertex $v'$ with respective labels $i$ and $j$.
Either $j=i+1$ and we do the construction (a) resulting in a labeled vertex with label $j$ connected by a regular edge to a white vertex and a bud ($\blacktriangleright\!\!\!-\!\!\!-$) connected to a black vertex,
or $j\leq i$ and we do the construction (b) resulting in a flagged edge with labels $i$, $j$ connecting a white to a black vertex.}} 
\label{fig:Constmobile}
\end{figure}

\begin{figure}[h!]
\begin{center}
\includegraphics[width=14cm]{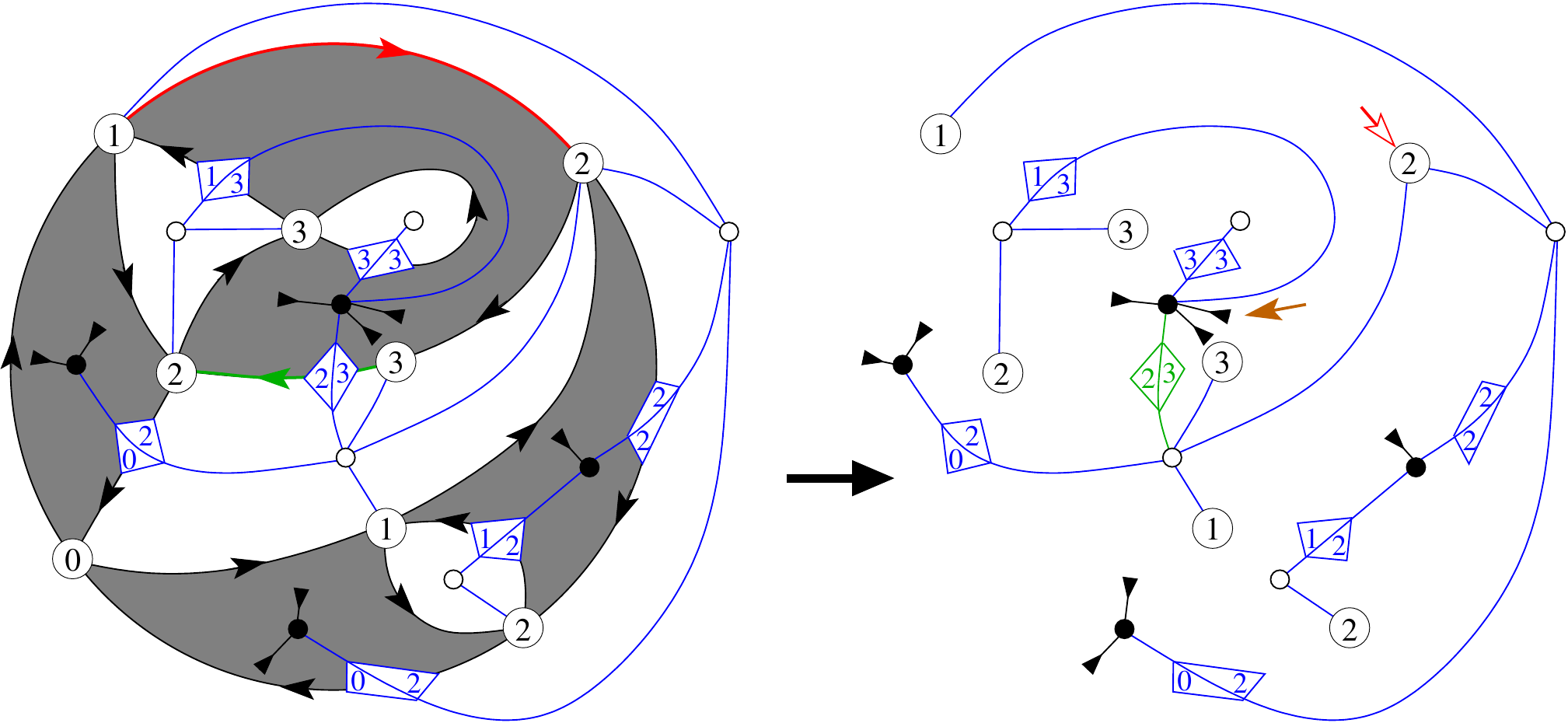}
\end{center}
\caption{\small{Applying the rules of \autoref{fig:Constmobile} to all the edges of a pointed Eulerian map results into a mobile (right). In red: an edge of type $i\to i+1$ is in correspondence with
a corner at a labeled vertex with label $i+1$. In green: an edge of type $i\to j$ with $j\leq i$ is in correspondence with a flagged edge with labels $i$, $j$. In brown: the bud associated 
to the marked red corner (see text).}} 
\label{fig:Maptomobile}
\end{figure}

A mobile is then constructed from the pointed Eulerian map as follows: we first add a black (resp. white)
vertex at the center of each black (resp. white) face. For each edge $e$ oriented from $v$ to $v'$, we do the following
construction: if $i(v')=i(v)+1$, we draw a new edge from $v'$ to the white vertex at the center of the white face
incident to $e$ and attach a {\it bud} (pointing towards $e$) to the black vertex at the center of the black face
incident to $e$ (see \autoref{fig:Constmobile}-(a)). 
If $i(v')\leq i(v)$, we draw instead a new edge between the black 
and white vertices at the center of
the black and white faces incident to $e$ and put labeled flags on both sides of this edge: a flag with label $i(v)$ on the side of
$v$ and a flag with label $i(v')$ on the side of $v'$ (see \autoref{fig:Constmobile}-(b)). It was shown in \cite{PMLM}
that the graph formed by this newly 
drawn edges forms a {\it connected plane tree} which spans all the original vertices of the map but the root vertex $v_0$ 
(see \autoref{fig:Maptomobile} for an example).
This tree, endowed with all its decorations (buds, flags) and labels forms the desired mobile.

We arrive at the following definition of mobiles, entirely dictated by their construction from the associated Eulerian maps.
\vskip .5cm
\bd\label{claim:defmob} A mobile is a plane tree formed of:
\begin{itemize}
\item three types of vertices: black vertices, white vertices and labeled vertices which carry {\it positive} integer labels;
\item two types of edges: regular edges and flagged edges which carry pairs of {\it non-negative} integer labels (one
on each side of the edge). Regular edges necessarily connect a white vertex to a labeled one. Flagged edges necessarily
connect a black vertex to a white one.
\end{itemize}
\noindent The labels obey the following local rules around black and white vertices:
\begin{itemize}
\item given a black vertex, the sequence of labels on its incident flagged edges, as read {\it clockwise} around this
vertex, is non-increasing at the crossing of each flagged edge and non-decreasing between two consecutive flagged edges.
\item given a white vertex, the sequence of labels on its incident flagged edges and adjacent labeled vertices,
 as read {\it clockwise} around this vertex, is non-decreasing at the crossing of each flagged edge, constant between a flag
 and the next flag or labeled vertex, and decreasing by $1$ between a labeled vertex and the next flag or labeled vertex.
 \item black vertices are decorated by buds by putting exactly $j-i$ buds between two consecutive flagged edges 
 with consecutive labels $i$ and $j$ (recall that $j\geq i$ by definition).
\end{itemize}
\vskip .15cm
\noindent The vertex maps reduced to a single (black, white or labeled) vertex are not considered as mobiles.
\ed
The label rules around black and white vertices directly follow from the mobile edge construction rules, as described above, applied to 
the sequence of edges around a given black or white face in an Eulerian map. Note that a mobile
necessarily contains a flagged edge (since the rule around a white vertex cannot be fulfilled if all its incident edges are regular),
hence it contains a black and a white vertex (since a flagged edge connects a black to a white vertex). Note also that,
from the label rules, the minimum label $i_{\rm min}$ on a mobile is necessarily carried by a flag and satisfies $i_{\rm min}\geq 0$. Note finally that buds clearly
constitute a redundant information but we introduce them to guaranty that the total degree of a black vertex (counting
buds as incident half edges) is identical to that of the associated black face.  
We have the following bijection \cite{PMLM}:
\bt\label{BDGbij} {\bf Bouttier Di Francesco Guitter (BDG) bijection}

The above mapping is a bijection between pointed Eulerian maps and mobiles with minimum label $i_{\rm min}=0$ (or equivalently mobiles having at least a flag labeled $0$).
\et
In particular, pointed Eulerian maps with a {\it marked edge} $e$ oriented from $v$ to $v'$ at respective
oriented distance $i(v)$ and $i(v')$ from the root vertex are in bijection with mobiles having at least a flag labeled $0$
with a {\it marked flagged edge} with flag labels $i(v)$ and $i(v')$ if $i(v')\leq i(v)$, or with a {\it marked corner}
at a labeled vertex with label $i(v')$ if $i(v')=i(v)+1$ (see \autoref{fig:Maptomobile}). 

Note that the constraint that the mobile has minimum label $i_{\min}=0$ follows from the fact that any edge oriented towards the root vertex 
will create a flagged edge with one of its labels equal to $0$.
For convenience, this constraint is {\it not} included in the above definition of mobiles, {\it whose minimal label
is only required to be non-negative}. We will see later how to simply re-instore the constraint when necessary.
Clearly the rules around black and white vertices are invariant by a global shift of the labels. We have thus 
the following corollary
\bc
Given some fixed integer $i_{\min}\geq 0$, mobiles whose minimum
label is $i_{\min}$ are in bijection with pointed Eulerian maps.
\ec
The correspondence between these mobiles and pointed Eulerian maps is moreover
obtained by exactly the same construction as above providing we now endow the Eulerian maps with a new labeling
of vertices equal to their oriented geodesic distance from the root vertex {\it plus} $i_{\rm min}$. 
 
 \vskip .5cm
 \begin{figure}[h!]
\begin{center}
\includegraphics[width=14cm]{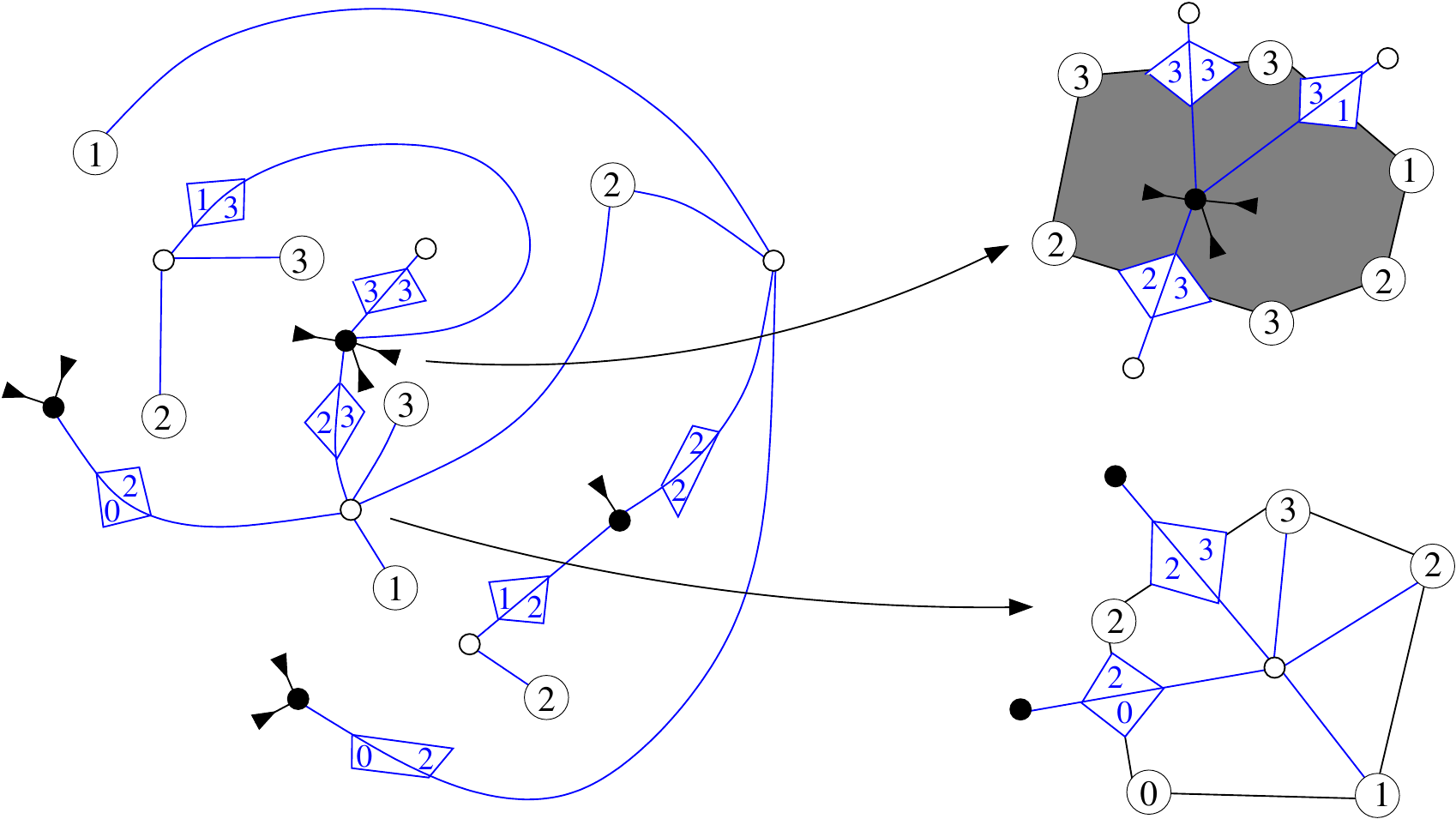}
\end{center}
\caption{\small{Top right: reconstruction of the vertex labels around a black face from the environment of the
associated black vertex in the mobile. Bottom right: reconstruction of the vertex labels around a white face from the environment of the
associated white vertex in the mobile.}} 
\label{fig:Mobile}
\end{figure}

To conclude this section, let us mention that the labels around black or white vertices in a mobile
are simply {\it two different but equivalent codings} for sequences of labels satisfying \eqref{condlabel} along the oriented edges 
around a black or a white face in a pointed Eulerian map. At a white vertex, the sequence is trivially recovered by going counterclockwise
around a vertex and ignoring the first encountered label of each flagged edge (which is identical to the
preceding label counterclockwise). In particular the degree of the associated white face is also, as it should, the degree of the
white vertex on the mobile (see \autoref{fig:Mobile} for an example). Around a black vertex, this sequence is 
recovered instead by considering the clockwise 
cyclic sequence of labels $i_1,i'_1,i_2,i'_2,\cdots i_p,i'_p$, where $i_k\to i'_k$ denotes the passage 
from a flagged edge to the next (hence  $i'_k\geq i_k$) and $i'_k\to i_{k+1}$ denotes the crossing of a given flagged edge
(hence $i_{k+1}\leq i'_k$ with the convention $i_{p+1}=i_1$), and replacing 
each step $i_k\to i'_k$ by a sequence of $i'_k-i_k$ steps, each with increment $+1$ (so as to obtain 
the sequence $i_1,i_1+1,i_1+2,\cdots i'_{1}$, $i_2,i_2+1,i_1+2,\cdots i'_2$, 
$\cdots i_p,i_p+1,i'_1+2,\cdots i'_p$ ). In particular, the degree of the associated black face in the Eulerian map 
is given by $\sum_{k=1}^p(i'_k-i_k+1)$, which is precisely, as it should, the degree of the black vertex (counting the buds)
(see \autoref{fig:Mobile} for an example).

\subsection{Enumeration of mobiles}
\label{sec:enuMob}
We now come to the enumeration of mobiles. More precisely we shall consider generating functions for mobiles
with a weight $g_k$ per white vertex of degree $k$ and a weight ${\tilde g}_k$ per black vertex of degree $k$. 
From the BDG bijection, this amounts to weigh the Eulerian maps with
a weight $g_k$ (resp. $\tilde{g}_k$) per white (resp. black) face of degree $k$.
\begin{figure}[h!]
\begin{center}
\includegraphics[width=8cm]{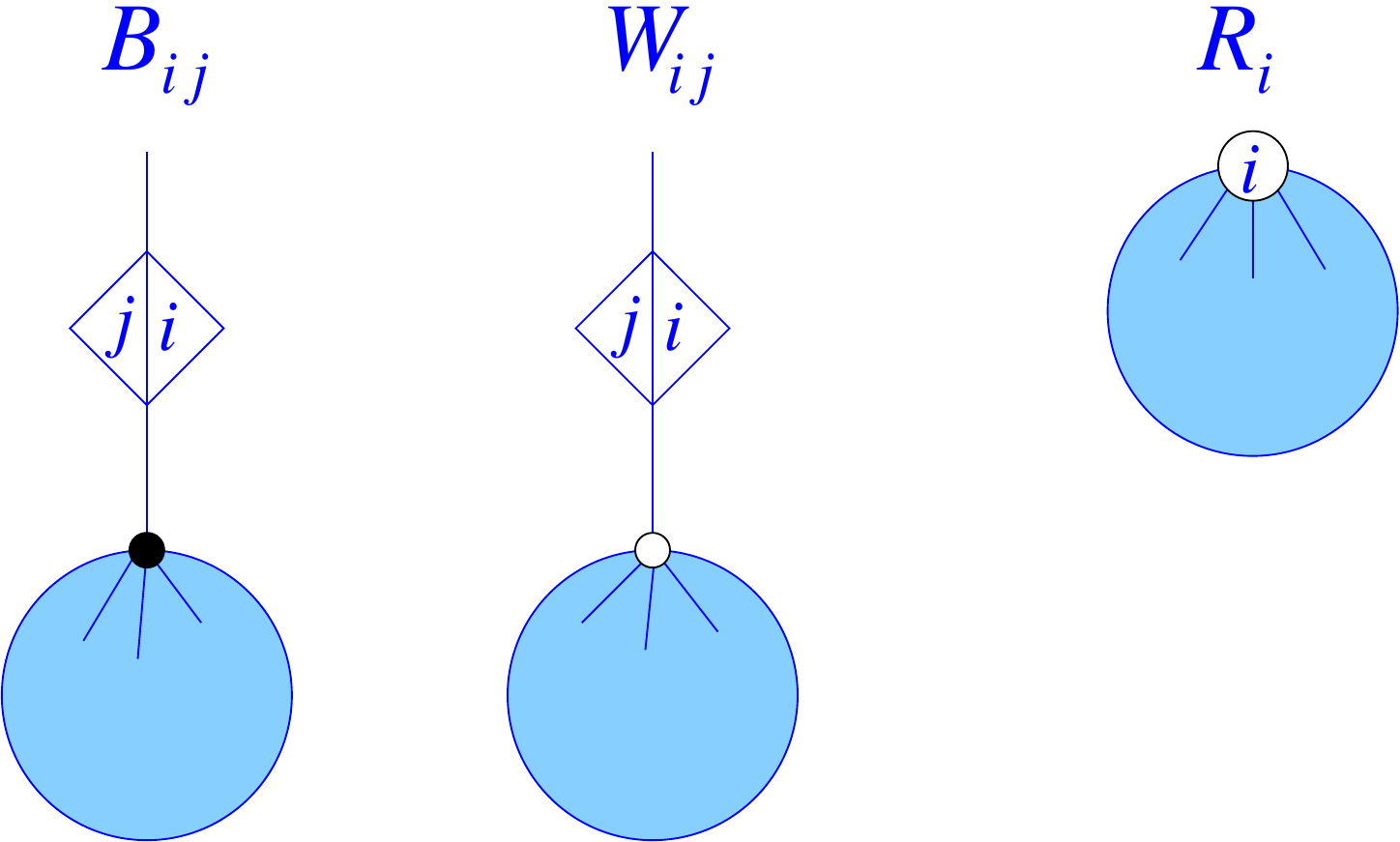}
\end{center}
\caption{\small{Schematic picture of a black half-mobile enumerated by $B_{ij}$, a white half-mobile enumerated by $W_{ij}$ 
and a mobile with a marked corner (or a single labeled vertex), as enumerated by $R_i$.}} 
\label{fig:Genfun}
\end{figure}
As was done in \cite{PMLM}, we shall consider three families of generating functions: 
\begin{itemize}
\item the generating functions $R_i$ ($i\geq 1$) for mobiles with a marked corner at a labeled vertex with label $i$.
For convenience, we incorporate also in $R_i$ a conventional additional term $1$, which 
we interprete as enumerating the vertex map made of 
a single isolated vertex with label $i$ (recall that this vertex map is not considered as a mobile -- the generating function
for genuine mobiles is thus $R_i-1$).  
\item the generating functions $W_{i,j}$ ($0\leq j\leq i$) for {\it white half-mobiles} as obtained by cutting a mobile at
a flagged edge with labels $i$ and $j$ and keeping only the part of the mobile lying on the side of its white incident vertex.
Note that we keep also in the half-mobile the flagged edge itself with its labels $i$ and $j$ so that  
$i$ denotes the flag label on the left when going {\it towards} the white vertex (see \autoref{fig:Genfun}).
\item the generating functions $B_{i,j}$ ($0\leq i\leq j$) for {\it black half-mobiles} as obtained by cutting a mobile at
a flagged edge with labels $i$ and $j$ and keeping only the part of the mobile lying on the side of its black incident vertex.
Again we keep in the half-mobile the flagged edge itself with its labels $i$ and $j$ so that
$i$ denotes the flag label on the left when going {\it towards} the black vertex (see \autoref{fig:Genfun}).
\end{itemize}

\begin{figure}[h!]
\begin{center}
\includegraphics[width=15cm]{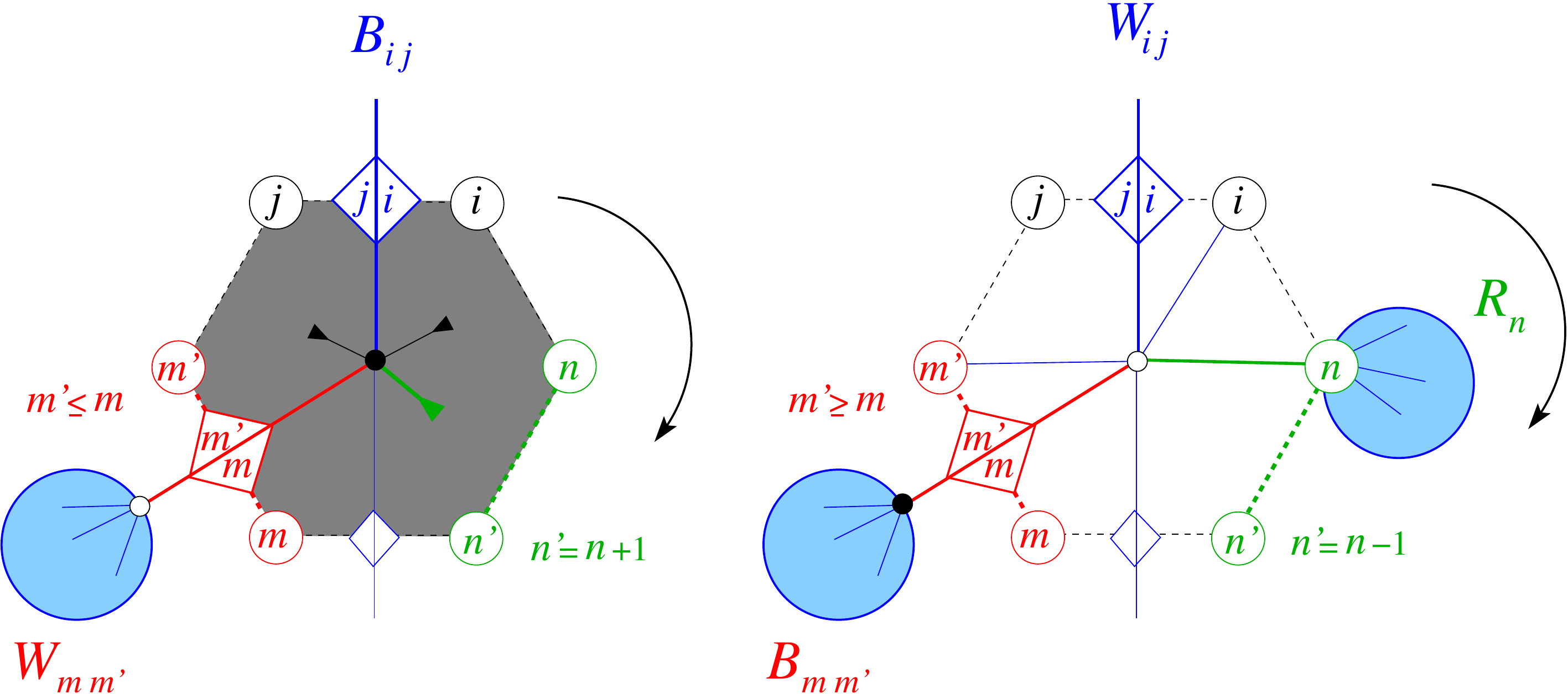}
\end{center}
\caption{\small{A schematic picture of the Equations~\eqref{eq:BWeq} giving the expressions of $B_{ij}$ and $W_{ij}$ respectively.}} 
\label{fig:EqsBW}
\end{figure}

\noindent The generating functions $R_i$, $B_{i,j}$ and $W_{i,j}$ satisfy a closed set of equations that determine them completely
as power series in the weights $g_k$ and $\tilde{g}_k$. These equations are best expressed upon introducing the
semi-infinite matrices $P$ and $Q$ with entries
\begin{equation}
P_{i,j}=\left\{
\begin{matrix}
B_{i,j} & {\rm if}\ j\geq i \\
R_i & {\rm if}\  j=i-1\\ 0 & {\rm if}\ j<i-1
\end{matrix} 
\right.
\qquad
Q_{i,j}=\left\{
\begin{matrix}
W_{i,j} & {\rm if}\ j\leq i \\
1& {\rm if}\  j=i+1\\ 0 & {\rm if}\ j>i+1
\end{matrix} 
\right.
\label{eq:BWexp}
\end{equation}
for $i,j\geq 0$.
We may then write the following two equations:
\begin{equation}
\begin{split}
& B_{i,j}=\sum_{k\geq 1} \tilde{g}_k (Q^{k-1})_{i,j} \quad {\rm for}\  i\leq j ,\\
& W_{i,j}=\sum_{k\geq 1} g_k (P^{k-1})_{i,j} \quad {\rm for}\  i\geq j .\\
\label{eq:BWeq}
\end{split}
\end{equation}
The first equation relies simply on a census of all possible environments of the first encountered black vertex in a black half mobile
enumerated by $B_{i,j}$ (see \autoref{fig:EqsBW}): 
if this vertex has degree $k$ in the decorated mobile (hence receives the weight $\tilde{g}_k$),
this environment is coded by a path of length $k-1$ from ``height" $i$ to height $j$ (the height being the label
around the associated black face). The height at each elementary step either increases by $1$, say from $n$ to $n'=n+1$,
giving rise to a bud with no weight ($=$ multiplicative factor $1$) or decreases weakly, say from $m$ to $m'\leq m$, giving
rise to a white half mobile enumerated by $W_{m,m'}$. The sum of factors over all path configurations is clearly counted
by $(Q^{k-1})_{i,j}$, hence the formula. As for the environments of the first encountered white vertex in a white half mobile
enumerated by $W_{i,j}$: if this vertex has degree $k$ (hence receives the weight $g_k$), it is again coded by 
path of length $k-1$ from height $i$ to height $j$, whose height at each elementary step may now either decrease by 
$1$, say from $n$ to $n'=n-1$, giving rise to a mobile with a distinguished corner at a vertex labeled $n$ or to an isolated 
vertex labeled $n$ (the two possibilities being enumerated by $R_n$) or it may increase weakly, say from $m$ to $m'\geq m$, giving
rise to a black half mobile enumerated by $B_{m,m'}$ (see \autoref{fig:EqsBW}). 
The sum of these factors over all path configurations is now counted
by $(P^{k-1})_{i,j}$. 

\begin{figure}[h!]
\begin{center}
\includegraphics[width=15cm]{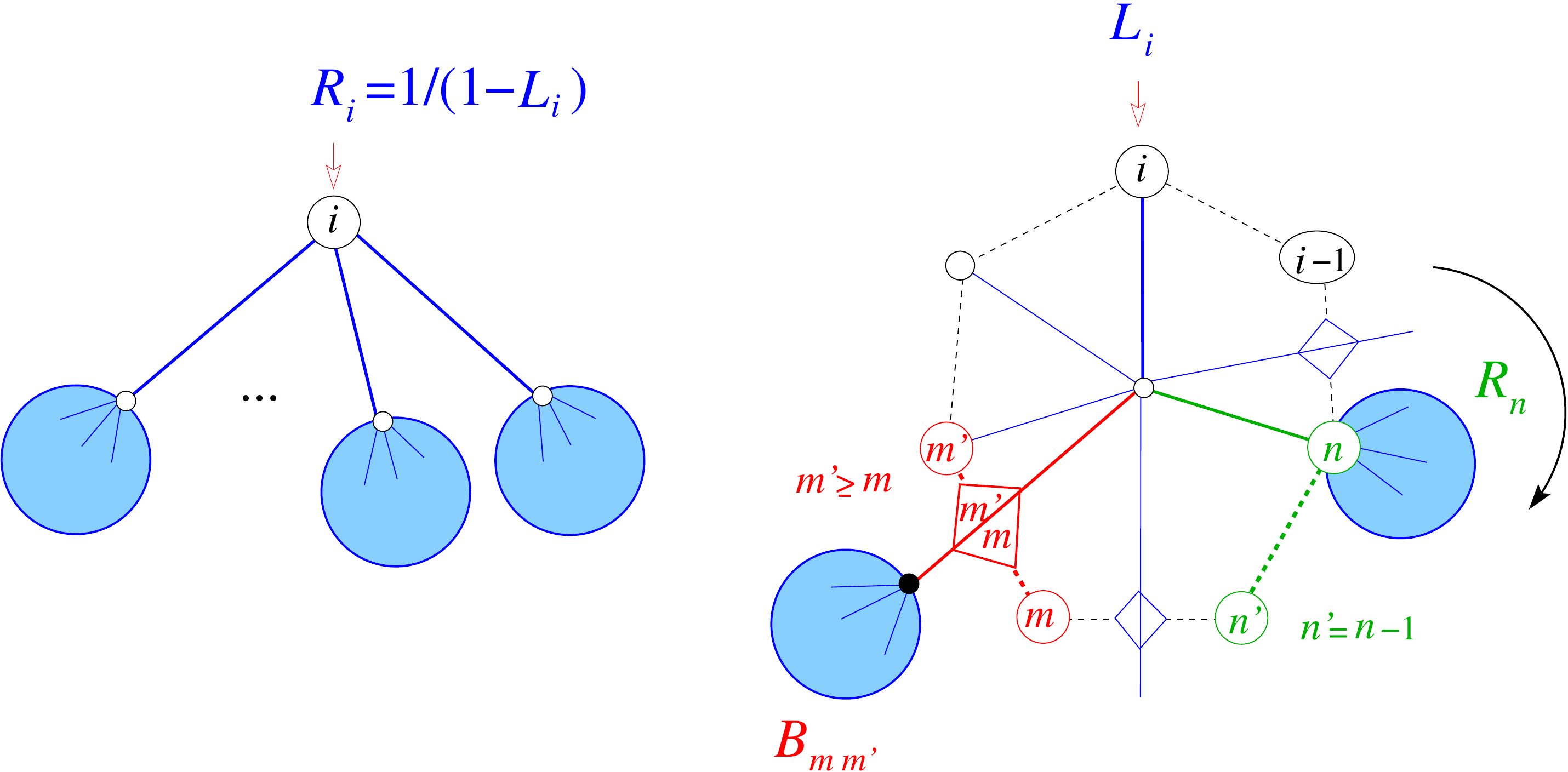}
\end{center}
\caption{\small{Left: the mobiles enumerated by $R_i$
form a sequence of mobiles planted at a univalent labeled vertex with label $i$, counted by $L_i$. This implies
the relation $R_i= 1/(1-L_i)$. Right: a schematic picture of the relation $L_i=\sum_{k\geq 1} g_k (P^{k-1})_{i-1,i}$.}} 
\label{fig:EqR}
\end{figure}

To close the system, we must also give the equation for $R_i$. It is obtained by noting that the mobiles counted by $R_i$
form a sequence of an arbitrary number of mobiles planted at a {\it univalent} labeled vertex with label $i$ (see \autoref{fig:EqR}-left). Denoting $L_i$ the generating function for these latter mobiles, we may therefore 
write $R_i=\sum_{k\geq 0} (L_i)^k = 1/(1-L_i)$ (this expression includes a first term $1$ accounting for the vertex map with label $i$,
as desired). As for $L_i$, it is determined along the same line as before upon inspecting the possible environments
of the white vertex incident to the univalent vertex $i$: if this vertex has degree $k$ (hence receives the weight $g_k$), the
environment is coded by path of length $k-1$ from height $i-1$ to height $i$ and the sum over all path configurations is 
therefore now counted by $(P^{k-1})_{i-1,i}$. We deduce:
\begin{equation}
 R_i=1/\left(1-\sum_{k\geq 1} g_k (P^{k-1})_{i-1,i}\right)
 \label{eq:Rexp}
\end{equation}
for $i\geq 1$.
It is interesting to note that $R_i$ may be alternatively obtained by a another equation, now depending
explicitly on the $\tilde{g}_k$'s instead, namely:
\begin{equation}
 R_i=1+\sum_{k\geq 1} \tilde{g}_k (Q^{k-1})_{i,i-1}
 \label{eq:Rexpbis}
\end{equation}
for $i\geq 1$.
This equation may be understood combinatorially as follows: $R_i-1$ enumerates mobiles with a marked corner at a vertex with label
$i$. If we call $i_{\rm min}$ the minimum label of any such mobile, this mobile is associated with a pointed Eulerian map
having a marked oriented edge $e$ pointing from a vertex $v$ with label $i-1$ to a vertex $v'$ with label 
$i$ under the labeling by oriented distances from the root vertex plus $i_{\min}$. This oriented edge selects in turn a bud in the mobile,
namely the bud incident to the black vertex at the center of the black face incident to $e$ and pointing towards $e$. 
We may therefore {\it re-root} the mobile at 
this bud (see \autoref{fig:Maptomobile} for an example of such re-rooting) and recover $R_i-1$ by enumerating these bud-rooted mobiles. The environment of the black vertex incident to the root bud, 
of arbitrary degree $k$ (hence weighted by $\tilde{g}_k$) is coded by a path of length $k-1$ from height $i$ to 
height $i-1$ (as obtained by reading the surrounding labels clockwise around the associated black face in the map, starting 
from the distinguished bud). Such environments are enumerated by $(Q^{k-1})_{i,i-1}$, leading directly to 
the above expression for $R_i$.

\subsection{Computation of $[P,Q]$}
\label{sec:comPQ}
The equations \eqref{eq:BWexp}--\eqref{eq:BWeq} may be rephrased as 
\begin{equation}
\left(P-\sum_{k\geq 1} \tilde{g}_{k} Q^{k-1}\right)_+=0\qquad
\left(Q-\sum_{k\geq 1} g_k P^{k-1}\right)_-=0
\end{equation}
where $(\cdot)_+$ (resp. $(\cdot)_-$) denotes the upper (resp. lower) part of the matrix at hand, namely
the upper (resp, lower triangular) matrix obtained by keeping only those elements $(\cdot)_{i,j}$ with
$j\geq i$ (resp. $i\geq j$).
This implies that $(P-\sum_k \tilde{g}_k Q^{k-1})$ is a {\it strictly} lower triangular matrix, whose 
commutator with $Q$ is therefore a lower triangular matrix. We deduce that $[P,Q]$ is a lower triangular matrix.
Similarly, $(Q-\sum_k g_k P^{k-1})$ is a {\it strictly} upper triangular matrix, whose 
commutator with $P$ is thus an upper triangular matrix. We deduce that $[P,Q]$ is also an upper triangular matrix,
hence it is diagonal.
We arrive at the following property:
\bp
\label{prop:PQdiag}
$[P,Q]$ is a diagonal matrix with diagonal elements
\begin{equation}
[P,Q]_{i,i}=-\delta_{i,0}=-(e_0e_0^t)_{i,i}
\label{eq:PQdiag}
\end{equation}
where $e_k$ is the vector whose coordinates are $(e_k)_j=\delta_{k,j}$
\ep
 \proof{
The diagonal elements may be computed via:
\begin{equation}
\begin{split}
[P,Q]_{i,i}& =[P-\sum_{k\geq 1} \tilde{g}_k Q^{k-1},Q]_{i,i}\\
& =\left\{
\begin{matrix}
(P-\sum\limits_{k\geq 1} \tilde{g}_k Q^{k-1})_{i,i-1}Q_{i-1,i}-Q_{i,i+1}(P-\sum\limits_{k\geq 1} \tilde{g}_k Q^{k-1})_{i+1,i} 
& {\rm for}\ i\geq 1 \\
& \\
-Q_{0,1}(P-\sum\limits_{k\geq 1} \tilde{g}_k Q^{k-1})_{1,0} 
& {\rm for}\ i=0 \\
\end{matrix} 
\right. \\
& =\left\{
\begin{matrix}
(R_i-\sum\limits_{k\geq 1} \tilde{g}_k (Q^{k-1})_{i,i-1})-(R_{i+1}-\sum\limits_{k\geq 1} \tilde{g}_k (Q^{k-1})_{i+1,i}) 
& {\rm for}\ i\geq 1 \\
& \\
-(R_1-\sum\limits_{k\geq 1} \tilde{g}_k (Q^{k-1})_{1,0}) 
& {\rm for}\ i=0 \\
\end{matrix} 
\right. \\
&= -\delta_{i,0}
\end{split}
\end{equation}
where we used \eqref{eq:Rexpbis}.
}

We could have obtained the same result via:
\begin{equation}
\begin{split}
[P,Q]_{i,i}& =[P,Q-\sum_{k\geq 1} g_k P^{k-1}]_{i,i}\\
& =\left\{
\begin{matrix}
P_{i,i-1}(Q-\sum\limits_{k\geq 1} g_k P^{k-1})_{i-1,i}-(Q-\sum\limits_{k\geq 1} g_k P^{k-1})_{i,i+1}P_{i+1,i} 
& {\rm for}\ i\geq 1 \\
& \\
-P_{0,1}(Q-\sum\limits_{k\geq 1}k g_k P^{k-1})_{1,0} 
& {\rm for}\ i=0 \\
\end{matrix} 
\right. \\
& =\left\{
\begin{matrix}
R_i(1-\sum\limits_{k\geq 1} g_k (P^{k-1})_{i-1,i})-R_{i+1}(1-\sum\limits_{k\geq 1} g_k (P^{k-1})_{i,i+1})
& {\rm for}\ i\geq 1 \\
& \\
-R_1(1-\sum\limits_{k\geq 1} g_k (P^{k-1})_{0,1}) 
& {\rm for}\ i=0 \\
\end{matrix} 
\right. \\
&= -\delta_{i,0}
\end{split}
\end{equation}
upon using now \eqref{eq:Rexp}.
\vskip .5cm
\noindent{\bf Combinatorial interpretation}

\begin{figure}[h!]
\begin{center}
\includegraphics[width=11cm]{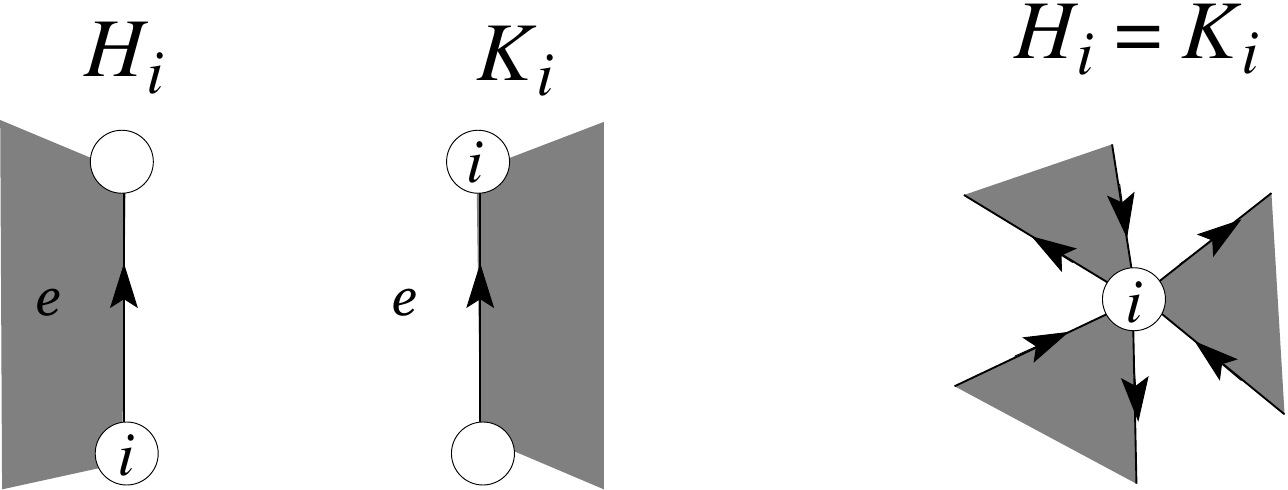}
\end{center}
\caption{\small{Schematic picture of $H_i$ and $K_i$ (left) and of the relation $H_i=K_i$ (right). }} 
\label{fig:HK}
\end{figure}

Interestingly enough, eq.~\eqref{eq:PQdiag} has a nice combinatorial explanation as follows:
consider the generating function $H_i$ of pointed Eulerian maps (with weights $g_k$ -- resp.\ $\tilde{g}_k$ --
per white face -- resp.\ black face of degree $k$) with a {\it marked oriented edge} $e$ {\it pointing away from a vertex at distance}
$i$ {\it from the root vertex} (see \autoref{fig:HK}). The marking of this edge is equivalent, in the associated mobile, to either the marking
of a flagged edge with labels $i$ and $j\leq i$ if the extremity of $e$ is at distance $j\leq i$ from the root vertex, or to
the marking of a corner at a labeled vertex with label $i+1$ if the extremity of $e$ is at distance $i+1$ from the root vertex.   
The first case corresponds to mobiles enumerated by $\sum_{j\leq i}W_{i,j}B_{j,i}$ if we ignore the constraint that the
mobiles in bijection with maps must have their minimum label equal to $0$, while the second case corresponds to mobiles enumerated by $R_{i+1}-1$ (recall that
$R_{i+1}$ incorporates a conventional term $1$ which does not account for any acceptable mobile). The constraint of having
minimum label $0$ is automatic if $i=0$ but, if $i\geq 1$, we must eliminate those mobiles having a minimum label larger than
$1$. Starting from these mobiles, we still get an acceptable mobile  by shifting all the labels by $-1$. In other
words, the mobiles to be subtracted are enumerated by  $\sum_{j\leq i-1}W_{i-1,j}B_{j,i-1}+R_i-1$. We end up with
\begin{equation}
\begin{split}
H_i & =\left\{
\begin{matrix}
\left(R_{i+1}-1+\sum\limits_{j\leq i} W_{i,j}B_{j,i}\right)-\left(R_i-1+\sum\limits_{j\leq i-1} W_{i-1,j}B_{j,i-1}\right)
& {\rm for}\ i\geq 1 \\
& \\
R_1-1+\sum\limits_{j\leq 0} W_{0,j}B_{j,0}
& {\rm for}\ i=0 \\
\end{matrix} 
\right. \\
& =\left\{
\begin{matrix}
(QP)_{i,i}-(QP)_{i-1,i-1}
& {\rm for}\ i\geq 1 \\
& \\
(QP)_{0,0}-1
& {\rm for}\ i=0 \\
\end{matrix} 
\right. \\
\end{split}
\end{equation}
Consider now the generating function $K_i$ of pointed Eulerian maps  with a {\it marked oriented edge} $e$ {\it pointing towards 
a vertex at distance} $i$ {\it from the root vertex} (see \autoref{fig:HK}).
The marking of this edge is now equivalent, in the associated mobile, to either the marking
of a flagged edge with labels $i$ and $j\geq i$ if the origin of $e$ is at distance $j\geq i$ from the root vertex, or to
the marking of a corner at a labeled vertex with label $i$ if the origin of $e$ is at distance $i-1$ from the root vertex.   
The first case corresponds to mobiles enumerated by $\sum_{j\geq i}B_{i,j}W_{j,i}$ if we ignore the constraint that the
desired mobiles must have their minimum label equal to $0$, while the second case corresponds to mobiles enumerated by $R_i-1$. Note that this second case is possible only 
if $i\geq 1$. The constraint of having minimum label $0$ is taken into account as above and we end up with
\begin{equation}
\begin{split}
K_i & =\left\{
\begin{matrix}
\left(R_i-1+\sum\limits_{j\geq i} B_{i,j}W_{j,i}\right)-\left(R_{i-1}-1+\sum\limits_{j\geq i-1} B_{i-1,j}W_{j,i-1}\right)
& {\rm for}\ i\geq 2 \\
& \\
\left(R_1-1+\sum\limits_{j\geq 1} B_{1,j}W_{j,1}\right)-\left(\sum\limits_{j\geq 0} B_{0,j}W_{j,0}\right)
& {\rm for}\ i=1 \\
& \\
\sum\limits_{j\geq 0} B_{0,j}W_{j,0}
& {\rm for}\ i=0 \\
\end{matrix} 
\right. \\
& =\left\{
\begin{matrix}
(PQ)_{i,i}-(PQ)_{i-1,i-1}
& {\rm for}\ i\geq 2 \\
& \\
(PQ)_{1,1}-(PQ)_{0,0}-1
& {\rm for}\ i=1 \\
& \\
(PQ)_{0,0}
& {\rm for}\ i=0 \\
\end{matrix} 
\right. \\
\end{split}
\end{equation}
Now, on a Eulerian map with its canonical edge orientation, and for each vertex on the map, there is an equal number of edges oriented away from this vertex and towards this vertex. Indeed, the orientations alternate when turning around the vertex (whose degree
is even by definition -- see \autoref{fig:HK}). We deduce in particular that 
\begin{equation}
H_i=K_i\ .
\end{equation} 
Equating the above expressions for $H_i$ and $K_i$, we obtain
 \begin{equation}
\begin{split}
&
[P,Q]_{i,i}-[P,Q]_{i-1,i-1}=0
\qquad {\rm for}\ i\geq 2 \\
& 
[P,Q]_{1,1}-[P,Q]_{0,0}=1 \\
&
[P,Q]_{0,0}=-1 \\
\end{split}
\end{equation}
from which \eqref{eq:PQdiag} follows immediately.

\subsection{Mobiles with bounded degrees}
\label{sec:bounded}
From now on, we will assume that our mobiles have {\it black and white vertices with bounded degrees}.
More precisely, we impose that the degree of a black vertex is bounded by $p$ and that of a 
white vertex bounded by $q$, where $p$ and $q$ are fixed integers such that $(p-1)(q-1)>1$.
Clearly those mobiles code for Eulerian maps with bounded {\it face degrees}, with black faces of
degree at most $p$, and white faces of degree at most $q$. Note that the degrees of the labeled vertices in 
the mobiles remain unbounded.

In practice, the degree restriction is achieved in the generating functions $R_i$, $B_{ij}$ and $W_{ij}$
by keeping non vanishing $g_k$'s only for $1\leq k\leq q$ and non vanishing $\tilde{g}_k$'s only for $1\leq k\leq p$.
From now on, it is thus implicitly assumed that
\begin{equation}  
R_i:=R_i(g_1,\ldots,g_q,\tilde{g}_1,\ldots,\tilde{g}_p)\quad \hbox{for $i\geq 1$,} 
\end{equation}
and similarly for $B_{ij}$ ($j\geq i\geq 0$) and $W_{ij}$ ($i\geq j\geq 0$).

From \eqref{eq:BWexp}--\eqref{eq:BWeq}, we immediately see that $B_{i,j}=0$ for $j>i+p-1$, while
$W_{ij}=0$ for $j<i-q+1$: the matrix $P$ therefore has one band below the diagonal (with entries $R_i$) and $p$ bands on or above the diagonal
(with entries $B_{i,i+k}$,  $0\leq k\leq p-1$), and the matrix $Q$ has one band above the diagonal (with entries $1$) and $q$ bands on or below the diagonal 
(with entries $W_{i,i-k}$, $0\leq k\leq q-1$). From their interpretation as mobile generating functions,
all these non-vanishing entries are formal power series of the coupling constants $g_k$ and $\tilde{g}_k$ with non-negative integer coefficients.

Recall that, by definition, the minimum label in a mobile or a half-mobile in non-negative: let us denote by 
$R_i^{(0)}$ (respectively $W_{ij}^{(0)}$ and $B_{ij}^{(0)}$) the generating function of those mobiles (respectively white
and black half-mobiles) in the set enumerated by $R_i$ (respectively by $W_{ij}$ and by $B_{ij}$)
and {\it whose minimum label is $0$}. We note that, in a mobile in the set enumerated by $R_i$ ($i\geq 2$) and whose minimum label is {\it positive},
we may shift all labels by $-1$ and get a mobile in the set enumerated by $R_{i-1}$: we deduce the relations
\begin{equation}
\begin{split}
R_1 &= R_1^{(0)}+1\ , \\
R_i &= R_{i}^{(0)}+R_{i-1}\quad \hbox{for $i\geq 2$}\ .\\
\end{split}
\end{equation}
Similarly we have
\begin{equation}
\begin{split}
W_{i,i-k}&=W^{(0)}_{i,i-k}+W_{i-1,i-1-k}\quad \hbox{for $i\geq k+1$ and $0\leq k\leq q-1$}\ ,\\
B_{i,i+k}&=B^{(0)}_{i,i+k}+B_{i-1,i-1+k}\quad \hbox{for $i\geq 1$ and $0\leq k\leq p-1$}\ .\\
\end{split}
\end{equation}
Now it is easily seen that, if we fix the numbers $m_\ell$ ($1\leq \ell \leq q$) of white vertices 
of degree $\ell$ of and the numbers $n_\ell$ ($1\leq \ell\leq p$) of black vertices of degree
$\ell$ in a mobile, the maximal difference of labels in this mobile is {\it bounded}. This
implies that, {\it for $i$ large enough}, 
\begin{equation}
\begin{split}
\left[g_1^{m_1}\cdots g_q^{m_q} \tilde{g}_1^{n_1}\cdots \tilde{g}_p^{n_p}\right]R_i^{(0)} & =0 \\
& \hskip -1cm \hbox{i.e.\ } \left[g_1^{m_1}\cdots g_q^{m_q} \tilde{g}_1^{n_1}\cdots \tilde{g}_p^{n_p}\right]R_i =
\left[g_1^{m_1}\cdots g_q^{m_q} \tilde{g}_1^{n_1}\cdots \tilde{g}_p^{n_p}\right]R_{i-1}\ .\\
\end{split}
\end{equation}
More precisely, we have the property:
\bp
For fixed non-negative integers $m_\ell$ ($1\leq \ell\leq q$) and $n_\ell$ ($1\leq \ell\leq p$),
there exist a positive integer $i_0:=i_0(m_1,\ldots,m_q,n_1,\ldots,n_p)$ and a non-negative integer
$r(m_1,\ldots,m_q,n_1,\ldots,n_p)$ such that
\begin{equation}
\forall i\geq i_0, \  \left[g_1^{m_1}\cdots g_q^{m_q} \tilde{g}_1^{n_1}\cdots \tilde{g}_p^{n_p}\right]R_i
=r(m_1,\ldots,m_q,n_1,\ldots,n_p)\ .
\end{equation}
\ep
Introducing the formal power series
\begin{equation}
R:=R(g_1,\ldots,g_q,\tilde{g}_1,\ldots,\tilde{g}_p)=\!\!\!\!\sum_{m_1,\ldots,m_q \geq 0\atop n_1,\ldots n_p \geq 0}
r(m_1,\ldots,m_q,n_1,\ldots,n_p)g_1^{m_1}\cdots g_q^{m_q} \tilde{g}_1^{n_1}\cdots \tilde{g}_p^{n_p}\ ,
\end{equation}
we thus have, as formal power series,
\begin{equation}
\lim_{i\to\infty} R_i= R\ .
\label{eq:limitR}
\end{equation}
The same argument holds for black or white half-mobiles.
\bp
For fixed $k$, $0\leq k\leq q-1$, for fixed non-negative integers $m_\ell$ ($1\leq \ell \leq q$) and $n_\ell$ ($1\leq \ell\leq p$),
there exist a positive integer $j_k:=j_k(m_1,\ldots,m_q,n_1,\ldots,n_p)$ and a non-negative integer
$a_k(m_1,\ldots,m_q,n_1,\ldots,n_p)$ such that
\begin{equation}
\forall i\geq j_k, \  \left[g_1^{m_1}\cdots g_q^{m_q} \tilde{g}_1^{n_1}\cdots \tilde{g}_p^{n_p}\right]W_{i,i-k}
=a_k(m_1,\ldots,m_q,n_1,\ldots,n_p)\ .
\end{equation}
 \ep
Introducing the formal power series
\begin{equation}
\alpha_k:=\!\!\!\!\sum_{m_1,\ldots,m_q \geq 0\atop n_1,\ldots n_p \geq 0}
a_k(m_1,\ldots,m_q,n_1,\ldots,n_p)g_1^{m_1}\cdots g_q^{m_q} \tilde{g}_1^{n_1}\cdots \tilde{g}_p^{n_p}\ ,
\end{equation}
we thus have, as formal power series,
\begin{equation}
\lim_{i\to\infty} W_{i,i-k} = \alpha_k\ .
\label{eq:limitalpha}
\end{equation}

\bp
For fixed $k$, $0\leq k\leq p-1$, for fixed non-negative integers $m_\ell$ ($1\leq \ell \leq q$) and $n_\ell$ ($1\leq \ell\leq p$),
there exist a positive integer $\ell_k:=\ell_k(m_1,\ldots,m_q,n_1,\ldots,n_p)$ and a non-negative integer
$b_k(m_1,\ldots,m_q,n_1,\ldots,n_p)$ such that
\begin{equation}
\forall i\geq \ell_k, \  \left[g_1^{m_1}\cdots g_q^{m_q} \tilde{g}_1^{n_1}\cdots \tilde{g}_p^{n_p}\right]B_{i+k,i}
=b_k(m_1,\ldots,m_q,n_1,\ldots,n_p)\ .
\end{equation}
 \ep
Introducing the formal power series
\begin{equation}
\beta_k:=\!\!\!\!\sum_{m_1,\ldots,m_q \geq 0\atop n_1,\ldots n_p \geq 0}
b_k(m_1,\ldots,m_q,n_1,\ldots,n_p)g_1^{m_1}\cdots g_q^{m_q} \tilde{g}_1^{n_1}\cdots \tilde{g}_p^{n_p}\ ,
\end{equation}
we have, as formal power series,
\begin{equation}
\lim_{i\to\infty} B_{i,i+k} = \beta_k\ .
\label{eq:limitbeta}
\end{equation}

The formal power series $R$, $\alpha_k$ and $\beta_k$ are linked via the following identity: 
\bp
\label{prop:sumabR0}
We have
\beq\label{eq:sumabR0}
R-1=\sum_{k=1}^{\min(q-1,p-1)} k\alpha_k \beta_k .
\eeq
\ep
\proof{
From the relation $R_i=R_i^{(0)}+R_{i-1}$, we deduce that, for $i\geq 0$, $R_{i+1}-1=\sum_{j=0}^{i}R_{j+1}^{(0)}$ and therefore, from the BDG bijection, $R_{i+1}-1$ is the generating function for pointed Eulerian
planar maps with a marked oriented edge of type $j\to j+1$ (after labeling vertices by their oriented distance from the root vertex) for some $j$ satisfying $0\leq j\leq i$. Taking the $i\to \infty$ limit, we deduce that $R-1$
is the generating function for pointed Eulerian planar maps with a marked oriented edge of type $j\to j+1$ for arbitrary $j\geq 0$. By a similar argument, $W_{i,i-k} B_{i-k,i}$ enumerates pointed Eulerian planar maps 
with a marked oriented edge of type $j\rightarrow j-k$ for some $j$ satisfying $k\leq j\leq i$.  Taking the $i\to \infty$ limit, we deduce that $\alpha_k\beta_k$
is the generating function for pointed Eulerian planar maps with a marked oriented edge of type $j\to j-k$ for arbitrary $j\geq k$. Consider now a white face $f$ in a pointed Eulerian map and denote by 
$t(f)$ the number of oriented edges incident to $f$ that are of type $j \rightarrow j+1$ for some arbitrary unfixed $j$ (i.e., those oriented edges along which the label increases by $1$). Similarly, denote by $s_k(f)$ the number of
oriented edges incident to $f$ that are of type $j \rightarrow j-k$ for some arbitrary unfixed $j$ (i.e., those oriented edges along which the label decreases by $k$).  In order to recover the same label after one turn around $f$, we
have the consistency relation $t(f)=\sum_{k=1}^{\min(p-1,q-1)}k\, s_k(f)$. Summing over all white faces $f$ and all pointed maps, the left hand side adds up to $R-1$ 
(enumerating pointed maps with a marked edge along which the label
increases) while the right hand side adds up to $\sum_{k=1}^{\min(q-1,p-1)} k \alpha_k\beta_k$ (pointed maps with a marked edge along which the label decreases by $k$, weighted by $k$ and summed over $k$). 
The desired identity \eqref{eq:sumabR0} follows immediately.
}

\subsection{Mobiles with a weight $\boldsymbol{g}$ per labeled vertex}

\label{sec:Mwg}
We will specialize to the case where $g_1=\td{g}_1=0$ and 
\beq
\label{eq:reng,tdg}
\begin{split}
g_k&=g^{\frac{k-2}{2}} \lambda_k\ , \qquad  k=2,\ldots, q\ ,\\
\tilde{g}_k&=g^{\frac{k-2}{2}}  \td{\lambda}_k\ , \qquad  k=2,\ldots, p\ .\\
\end{split}
\eeq
In the Eulerian map language this corresponds to forbidding faces of degree $1$ and giving a weight $\lambda_k$ (resp $\td{\lambda}_k$) per white (resp black) face of degree $k$. Assuming that there are $n_k$ (resp. $m_k$) such faces we also have an overall factor $g^{\sum_{k\geq 2} (\frac{k-2}{2})(n_k+m_k)}=g^{E-F}=g^{V-2}$ if $E$,$F$,$V$ are the total numbers of edges, faces and vertices. Here we used 
\beq
\sum_{k\geq 2} \frac{k-2}{2}(n_k+m_k)=\frac{1}{2}\sum_{k\geq 2}(k\,n_k+k\,m_k)-\sum_{k\geq 2}(n_k+m_k)=\frac{1}{2}(2E)-F
\eeq 
 and the Euler relation $V-E+F=2$ for planar maps. In the mobile language this correspond to assigning the weight $\lambda_k$ (resp. $\td{\lambda}_k$) per white (resp. black) vertex of degree $k$ (where the degree incorporates the number of buds for black vertices), together with an overall factor $g^{\mathcal{V}-1}$ where $\mathcal{V}$ is total number of labeled vertices in the mobile. Recall that we have indeed $\mathcal{V}=V-1$ since the root vertex of the (pointed) map is absent from the mobile. 

Lets us now show that:
\bp
\label{prop:psRBW}
Using the scaling \eqref{eq:reng,tdg}, the equations \eqref{eq:BWeq} and \eqref{eq:Rexpbis} have a unique solution for which $R_i$ ($i \geq 1$), $W_{i,j}$ ($i\geq j \geq 0$) and $B_{i,j}$ ($j\geq i \geq 0$)  are formal power series in $\sqrt{g}$.

This solution is precisely the one that we are looking for when we are interested in the enumeration of mobiles. 
\ep

\proof{
 The proof is given in \autoref{app:prop2.5}}.
 
From its definition as a mobile generating functions we can furthermore infer that $R_i$ has an expansion in integer powers of $g$ whose general term is polynomial in $\lambda_2,...,\lambda_q,\td{\lambda}_2,...,\td{\lambda}_q$ and $\frac{1}{1-\td{\lambda}_2\lambda_2}$ with nonnegative integer coefficients. In the case of half-mobile generating functions $B_{i,j}$ or $W_{i,j}$, it is simple exercise to show that their overall power in $\sqrt{g}$ is
\beq
\sqrt{g}^{j-i-1+2\mathcal{V}}
\eeq
 where $\mathcal{V}$ is now the total number of labeled vertices in the half-mobile. In particular if $j-i$ is even then the expansion of $B_{i,j}$ and $W_{i,j}$ has only half-integer powers of $g$ while if $j-i$ is odd it has only integer powers. Again the general terms in these expansions are polynomials in $\lambda_2,...,\lambda_q,\td{\lambda}_2,...,\td{\lambda}_q$ and $\frac{1}{1-\td{\lambda}_2\lambda_2}$ with nonnegative integer coefficients.

\begin{remark}
To summarize, for $g_1=\tilde{g}_1=0$, it will be enough to show that the expressions that we obtain for the generating functions $R_i$, $B_{i,j}$ and $W_{i,j}$ have power series expansions in $\sqrt{g}$ for the scaling \eqref{eq:reng,tdg} to guarantee that they also have power series expansions in all the $g_k$'s and $\tilde{g}_k$'s for $k\geq 2$ with positive integer coefficients.
\end{remark}

\section{Integrable systems}
\label{sec:Intsys}
From the preceeding section, the enumeration of mobiles in the restricted context of bounded degrees boils down to finding the solution of the following problem dictated by eqs \eqref{eq:BWexp} to \eqref{eq:Rexpbis}:
find two semi--infinite band matrices, $(Q_{n,m})_{n,m \geq 0}$  having one band above and $q-1$ bands below diagonal, and $(P_{n,m})_{n,m \geq 0}$  having one band below and $p-1$ bands above diagonal, such that
\beq
\begin{split}
\label{eq:PQ}
&\quad P_{n,n-1} = R_n,\qquad\qquad\qquad\qquad\qquad 
 Q_{n,n+1}=1,\\
&\left( P-\sum_{k=0}^{p-1} \tilde g_{k+1} Q^k\right)_+ = 0,\qquad\qquad \left( Q-\sum_{k=0}^{q-1} g_{k+1} P^k\right)_- = 0,\\
& \left( P-\sum_{k=0}^{p-1} \tilde g_{k+1} Q^k\right)_{n,n-1} = 1
,
 \qquad \, \left( Q-\sum_{k=0}^{q-1} g_{k+1} P^k\right)_{n,n+1} = \frac{1}{R_{n+1}}.
\end{split}
\eeq
Note that this algebraic problem could possibly have many solutions, but combinatorics of mobiles guarantees that there is a unique one such that $P_{n,m}$ and $Q_{n,m}$ are formal power series of the coupling constants $g_k$ and $\tilde g_k$.
Moreover this solution is such that $R_n$ has a limit as $n\to\infty$.

As we have seen in \autoref{prop:PQdiag}, these relations imply that $Q$ and $P$ quasi commute, namely
\beq
[P,Q]=-e_0e_0^t,
\eeq
which is a Lax equation, much studied in the literature of integrable systems, and whose general solution \cite{Krichever,Krichever_1977,Krichever_1978, BBT} is expressed through algebraic geometry. However, we cannot directly use the general solution here, because the mobiles correspond to highly degenerate initial conditions, and although the method is very similar to the general solution of \cite{Krichever,Krichever_1977,Krichever_1978} a new proof is required here in the context of mobiles.

Our approach uses the general framework of integrable systems:
in the present section, we shall explain in a sketchy way how necessary conditions lead us to some isospectral integrable system, in Lax form, and to Baker-Akhiezer functions.
Then, we will proceed backwards In \autoref{sec:mainth}, and show that specific Baker-Akhiezer functions indeed produce a solution to our combinatorial problem of counting mobiles.

The reader who is not interested in knowing how the solution emerges from the general framework of integrable systems may skip this section and go directely to \autoref{sec:mainth}.

\subsection{Isospectral system}
\label{sec:isosys}

Let us describe a number of general facts concerning band matrices.

\bd
Let $A=(A_{i,j})_{i,j\geq 0}$ be a semi--infinite band matrix with $a_+$ lines above diagonal and $a_-$ lines below.
A  semi--infinite vector $\vec\psi$ is called a right (resp. left) eigenvector of $A$ with eigenvalue $x$, iff
\beq
A\vec\psi = x\vec\psi
\qquad {\rm (resp.}\quad
\vec\psi^t A = x\vec\psi^t 
\, ).
\eeq
We shall denote by $\mathcal V_x(A)$ (resp.$\td{\mathcal V}_x(A)$)  the vector space of right (resp. left) eigenvectors of $A$ for the eigenvalue $x$. 
\ed

We have the following theorem:
\bt
If the upper-most and lower-most diagonals of $A$ have non--vanishing entries, then for any $x\in \mathbb{C}$, the space $\mathcal V_x(A)$ (resp.$\td{\mathcal V}_x(A)$)  of right (resp. left) eigenvectors of $A$ for the eigenvalue $x$ is a vector space of dimension $a_++a_-$, namely
\beq
\dim \mathcal V_x(A) = \dim \td{\mathcal V}_x(A) = a_++a_-.
\eeq
More precisely, let us call the set of $a_++a_-$ consecutive integers ${\mathcal W_n}=\{n-a_-,\dots,n+a_+-1\}$ the $n^{\rm th}$  ``window".
Take $n\geq a_{-}$. Let $\vec\psi$ be a right eigenvector, the map
\bea
\mathcal L_{\mathcal W_n}(x):\qquad \mathcal V_x(A) & \rightarrow & \mathbb C^{\mathcal W_n}  \cr
\vec \psi &\mapsto & \{\psi_i\}_{i\in \mathcal W_n}
\eea
is an isomorphism.

The map $\Lambda_n(x):=\mathcal L_{\mathcal W_{n+1}}(x) \mathcal L_{\mathcal W_{n}}^{-1}(x)$ from $\mathbb C^{\mathcal W_n}$  to $\mathbb C^{\mathcal W_{n+1}}$ is a companion matrix of size $a_++a_-$, linear in $x$. More precisely we have, for $n \geq a_-$:
\beq
\label{eq:Lambdan}
\Lambda_n(x)=\mathcal L_{\mathcal W_{n+1}}(x) \mathcal L_{\mathcal W_{n}}^{-1}(x) =
\begin{pmatrix}
0 & 1 & 0 \dots &  \cr
\vdots &  & \ddots & 0\cr
0 &  \dots & 0 & 1 \cr
\frac{-A_{n,n-a_-}}{A_{n,n+a_+}} &  \dots &  \frac{x -A_{n,n}}{A_{n,n+a_+}}  \,\,\, \dots & \frac{-A_{n,n+a_+-1}}{A_{n,n+a_+}}
\end{pmatrix}.
\eeq 

\et

\proof{ 
If $\vec \psi$ is a right eigenvector of $A$ for the eigenvalue $x$, i.e. $A\vec \psi = x\vec\psi$, then for every $m>0$, we have:
\beq
\psi_{m-1} = \frac{1}{A_{m-1+a_-,m-1}} \left( x\psi_{m-1+a_-} - \sum_{j=0}^{a_++a_--1} A_{m-1+a_-,m+j} \psi_{m+j} \right),
\eeq
and, for every $m\geq a_{+}+a_ {-}-1$:
\beq
\psi_{m+1} = \frac{1}{A_{m+1-a_+,m+1}} \left( x\psi_{m+1-a_+} - \sum_{j=0}^{a_++a_--1} A_{m+1-a_+,m-j} \psi_{m-j} \right).
\eeq
Using these equations, we see that every $\psi_m$ can be expressed as a linear combination of $\psi_i$ with $i$ restricted to the $n^{\rm th}$   window $\mathcal{W}_n$.
This implies that the vector space $\mathcal V_x(A)$ of eigenvectors of $A$ for the eigenvalue $x$ is isomorphic to $\mathbb C^{\mathcal W_n}$, hence has dimension $a_++a_-$. The same is true for left eigenspaces.

The fact that $\Lambda_n(x)$ is a companion matrix of the form \eqref{eq:Lambdan} is obtained by a simple computation.
}

As an application, since $Q$ is a band matrix with 1-band above and $q-1$ bands below diagonal, we have
\beq
\dim \mathcal V_x(Q)=\dim \tilde{\mathcal V}_x(Q) = q.
\eeq
Similarly, $P$ is a band matrix with $p-1$ bands above and $1$ band below diagonal, hence we have
\beq
\dim \mathcal V_y(P)=\dim \tilde{\mathcal V}_y(P) = p.
\eeq

Let us now discuss the action of $P$ on $\mathcal{V}_x(Q)$. We have:
\bt
If $\vec\psi\in \mathcal V_x(Q)$, then $\vec\phi=P\vec\psi$ satisfies
\beq
Q\vec\phi = x\vec\phi  + \psi_0 e_0
\eeq
i.e.
\beq
\forall n\geq 0 \ , \quad \sum_{m\geq 0} Q_{n,m} \phi_m = x\phi_n + \psi_0 \delta_{n,0}.
\eeq
The map
\bea
\pi: \quad P\mathcal V_x(Q) & \to & \mathcal V_x(Q) \cr
\vec\phi & \mapsto & {\mathcal L}_{\mathcal W_n}(x)^{-1}(\phi_{n+1-q},\dots,\phi_{n-1},\phi_n)
\eea
is well defined, and independent of $n$ for $n\geq q$.
 The map $\pi\circ P$ is an endomorphism of $\mathcal V_x(Q)$.
\et

\proof{
Let $\vec\psi\in \mathcal V_x(Q)$, we have:
\beq
Q P\vec\psi = P Q\vec\psi + [Q,P]\vec\psi = P(x\vec \psi) + \psi_0 e_0 = x P\vec \psi  + \psi_0 e_0 .
\eeq
If $\vec\phi=P\vec\psi$ we thus have:
\beq
\forall n\geq 0 \ , \quad \sum_{m\geq 0} Q_{n,m} \phi_m = x\phi_n + \psi_0 \delta_{n,0}.
\eeq
Let $n\geq q$, and let $\vec{\hat\phi} = {\mathcal L}_{\mathcal W_n}(x)^{-1}(\phi_{n+1-q},\dots,\phi_{n-1},\phi_n)$.
By definition of ${\mathcal L}_{\mathcal W_n}(x)$, we have $\vec{\hat\phi}\in \mathcal V_x(Q)$.
Since $n\geq q$ then $0\notin \mathcal W_n$, and thus ${\hat\phi}_m=\phi_m$ for all $m\in \mathcal W_n$. Moreover, $ \hat\phi_m$ and $\phi_m$ satisfy the same recursion relation of order $q$ for all $m>0$. In particular they coincide in any other window that doesn't contain $0$.
This shows that $\pi$ is independent of $n$ for all $n\geq q$.
}

\bd
For $n\geq q$, we may write $\pi\circ P$ in the canonical basis of $\mathbb C^{\mathcal W_n}$ as a $q\times q$ matrix, polynomial in $x$:
\beq
\mathcal D_n(x) := \mathcal L_{\mathcal W_{n}}(x) .\,\pi\circ P\,. \mathcal L_{\mathcal W_{n}}(x)^{-1}.
\eeq
More explicitly, we have:
\beq
\label{eq:DnexplicitL}
\begin{split}
\mathcal D_n(x) &= {\rm diag}(R_{n-q+1},\dots,R_n) \Lambda_{n-1}(x)^{-1} + {\rm diag}(P_{n-q+1,n-q+1},\dots,P_{n,n})  \cr
&+ \sum_{j=1}^{p-1} {\rm diag}(P_{n-q+1,n-q+1+j},\dots,P_{n,n+j})  \left(\Lambda_{n+j-1}(x)\dots \Lambda_{n+1}(x)\Lambda_n(x)\right) \cr
\end{split}
\eeq
with $\Lambda_n$ as in eq \eqref{eq:Lambdan} for $A=Q$, namely
\beq
\label{eq:QLambdan}
\Lambda_n(x)=
\begin{pmatrix}
0 & 1 & 0 \dots &  \cr
\vdots &  & \ddots & 0\cr
0 &  \dots & 0 & 1 \cr
-Q_{n,n-q+1} &  -Q_{n,n-q+2} & \dots & x-Q_{n,n}
\end{pmatrix}.
\eeq 
\ed
\bex
[Quadrangulations ($q=2,p=4$ and $g_1=\tilde{g}_1=\tilde{g}_3=0$)]
\begin{multline}\label{eq:DnQuadrang}
\mathcal D_n(x) = 
\tiny{
\begin{pmatrix}
x(\frac{R_{n-1}}{Q_{n-1,n-2}}-Q_{n,n-1}P_{n-1,n+2} ) & P_{n-1,n+2}(x^2-Q_{n+1,n}) + P_{n-1,n} - \frac{R_{n-1}}{Q_{n-1,n-2}} \cr 
-P_{n,n+3}Q_{n,n-1}(x^2-Q_{n+2,n+1})-P_{n,n+1}Q_{n,n-1}+R_n
& x P_{n,n+3}(x^2-Q_{n+2,n+1}-Q_{n+1,n})+P_{n,n+1} x
\end{pmatrix}.
}
\end{multline}

\eex

\bl
Changing the window $n\to n+1$ amounts to performing the conjugation
\beq\label{eq:discreteLax}
\mathcal D_{n+1}(x) = \Lambda_n(x) \mathcal D_{n}(x)\,\Lambda_n(x)^{-1}.
\eeq

\el
\bt[Isospectral system]

The eigenvalues of $\mathcal D_{n}(x)$ are independent of $n$ for $n\geq q$.

\et
\proof{This is an immediate consequence of the previous lemma.}

Eq~\eqref{eq:discreteLax} is a discrete {\bf Lax equation}\footnote{In a continuous time $t$, the Lax equation would be $\partial_t {\cal D} = [{\cal L}',{\cal D}]$, which implies that the eigenvalues are conserved $\partial_t \det(y-{\cal D}(x,t))=0$. Here we have its discrete time analog. Indeed the eigenvalues of ${\cal D}_n(x)$ are independent of $n$.}, and the matrices $({\cal D}_n(x), \Lambda_n(x))$ form a {\bf Lax pair} with the discrete time $n$, and spectral parameter $x$.
${\cal D}_n(x)$ is an {\bf isospectral} Lax matrix (meaning that its spectrum is independent of $n$).
We start entering the realm of integrable systems.

\subsection{Spectral curve}
\label{sec:spcurv}

\bd[Spectral curve] Let $n\geq q$. We define the spectral curve as the characteristic polynomial of the endomorphism $\pi\circ P$, namely:
\beq
\label{eq:Ebivpol}
\mathcal E(x,y) := \det\left( y\,{\rm Id}_{\mathcal V_x(Q)}-{\pi\circ P}_{\mathcal V_x(Q)} \right)=  \det(y\,{\rm Id}_{q\times q}-\mathcal D_n(x)).
\eeq
The spectral curve is the locus of the eigenvalues of $Q$ and $\pi \circ P$ for their common eigenvectors.
\beq
\mathcal V_x(Q)\cap \mathcal V_y(\pi\circ P) \neq \{0\}
\qquad \Leftrightarrow \qquad
\mathcal E(x,y)=0.
\eeq
Notice that $\mathcal E(x,y)$ is a polynomial of $x$ and $y$, and is independent of $n$ since $n\geq q$.
\ed

\bd
From eqs~\eqref{eq:BWeq} and \eqref{eq:limitR} to \eqref{eq:limitbeta} the following limits exist for the combinatorial solutions that we are looking for:
\beq
\alpha_k :=  \lim_{n\to\infty} Q_{n,n-k} 
\eeq
and
\beq
\beta_k := \lim_{n\to\infty} P_{n,n+k} 
, \ \ R := \lim_{n\to\infty} R_n.
\eeq
We define the following Laurent polynomials $\in \mathbb C[z,1/z]$:
\beq
X(z) := z+\sum_{i=0}^{q-1} \alpha_i z^{-i}  
 , \quad
Y(z) := \frac{R}{z}+\sum_{i=0}^{p-1} \beta_i z^{i}. 
\eeq
\ed

\bd[Potentials]
We also define the following polynomials, called the ``potentials":
\beq
\tilde V(x) := \sum_{k=1}^p \frac{\tilde g_k}{k} x^k
  , \qquad 
V(y) := \sum_{k=1}^q \frac{g_k}{k} y^k.
\label{eq:potentials}
\eeq

\ed

\bt\label{th:alphabeta}
The $\alpha_i$, $\beta_i$ can be found as follows:
 the equations \eqref{eq:PQ} imply that the generating functions $X(z)$ and $Y(z)$ satisfy the system of algebraic equations
\beq
\begin{split}
\tilde V'(X(z))_+ = Y(z)_+ 
&\quad , \quad
V'(Y(z))_- = X(z)_- 
\quad , \quad \cr
 Y(z)-\tilde{V'}(X(z)) &\mathop{\sim}_{z\to\infty} \frac{1}{z} +O(1/z^2) ,\cr
 X(z)-V'(Y(z))  &\mathop{\sim}_{z\to 0} \frac{z}{R} +O(z^2).
 \end{split} 
\eeq
The last two conditions are not independent, they can be obtained by computing $-\Res_{z\to\infty} YdX = \Res_{z\to 0} YdX = 1$, and can be reformulated as 
\beq\label{sumabR}
\sum_{k=1}^{\min(q-1,p-1)} k\alpha_k \beta_k = R-1.
\eeq
\et
This identity is nothing but Equation \eqref{eq:sumabR0}, whose combinatorial interpretation was given in \autoref{sec:bounded}.
\bex[Quadrangulations ($q=2,p=4$ and $g_1=\tilde{g}_1=\tilde{g}_3=0$)]
Take $\tilde V(x) = \frac{\tilde g_4}{4}x^4 + \frac{\tilde g_2}{2}x^2$ and $V(y) = \frac{g_2}{2}y^2$.
We have
\beq
X(z) = z + \alpha_1 z^{-1}\qquad,\qquad
Y(z) = \frac{R}{z} + \beta_1 z+\beta_3 z^3.
\eeq
They have to satisfy:
\beq
\beta_3 = \tilde g_4,\quad \beta_1 = 3 \tilde g_4 \alpha_1 + \tilde g_2,\quad \alpha_1 =  Rg_2,\quad
\alpha_1 \beta_1 = R-1.
\eeq
This gives
\beq
Rg_2 (\tilde g_2 + 3 \tilde g_4 Rg_2) = R-1
\eeq
and thus
\beq
R = \frac{1}{6 \tilde g_4 g_2^2}\left(1-\tilde g_2g_2 - \sqrt{\left(1-\tilde g_2g_2\right)^2-12g_2^2\tilde g_4 } \right).
\eeq
Here we choose the unique branch which is a power series of the coupling constants (i.e. the minus sign in front of the square-root), namely:
\beq
R = 1+\tilde g_2g_2 +\tilde g_2^2g_2^2+ 3 \tilde g_4g_2^2 + \dots
\eeq
\eex

Substituting $Q_{n,m}\to \alpha_{n-m}$ and $P_{n,m}\to\beta_{m-n}$, in \eqref{eq:DnexplicitL}  and \eqref{eq:QLambdan} we see that
$\mathcal D_n(x)$ and $\Lambda_n(x)$ have large $n$ limits:
\beq\label{eq:Lambdainfty}
\Lambda_\infty(x) = \begin{pmatrix}
0 & 1 & 0 \dots &  \cr
\vdots &  & \ddots & 0\cr
0 &  \dots & 0 & 1 \cr
-\alpha_{q-1} &  -\alpha_{q-2} & \dots & x-\alpha_0
\end{pmatrix}
\eeq
\beq\label{eq:Dinfty}
\mathcal D_\infty(x) = R\,\Lambda_\infty(x)^{-1} + \sum_{k=0}^{p-1} \beta_k \Lambda_\infty(x)^k.
\eeq
\bt[Spectral curve as a resultant]
$(x,y)\in \mathbb C\times \mathbb C $ belongs to the spectral curve iff there exists $z\in\mathbb C^*$ such that $x=X(z)$ and $y=Y(z)$:
\beq
\{(x,y)\,|\,\mathcal E(x,y)=0\}
\quad=\quad
\{(X(z),Y(z))\,|\,z\in \mathbb C^*\}. 
\eeq
Since it can be parametrized by a complex variable, this implies that the spectral curve is a genus zero Riemann surface.
As a consequence we have
\beq
\mathcal E(x,y) = \operatorname{Resultant}( z^{q-1}(X(z)-x),z(Y(z)-y))
\eeq
namely
\beq
\label{eq:Matres}
\mathcal E(x,y) = \frac{1}{\alpha_{q-1}}
\det \begin{pmatrix}
1 & \alpha_0-x & \alpha_1 & \dots & \alpha_{q-1} &  &   \cr
& \ddots & \ddots & & & \ddots &  & &   \cr
 & & 1 & \alpha_0-x & \alpha_1 & \dots & \alpha_{q-1}      \cr
\beta_{p-1} & \dots & \beta_0-y & R & &  &   \cr
& \ddots &  & \ddots & \ddots & &     \cr
& & \beta_{p-1} & \dots & \beta_0-y & R &    \cr
& & & \beta_{p-1} & \dots & \beta_0-y & R     \cr
\end{pmatrix}
\eeq
with $p$ lines in the upper part and $q$ lines in the lower part.
\et
\proof{
By definition, the resultant of two polynomials vanishes if and only if they have a common zero.
The determinant form \eqref{eq:Matres} is the standard formula for the resultant. Notice that if $z$ is a common zero of $X(z)-x$ and $Y(z)-y$, then the vector $(z^{p+q-1},\dots,z,1)$ is in the kernel of the resultant matrix, and vice-versa, any vector in the kernel must be of that form.
}

The two poles at $z=\infty$ and $z=0$ with respective asymptotics $Y\sim \tilde V'(X)$ and $X\sim V'(Y)$ imply that the spectral curve must be of the form
\beq
\mathcal E(x,y) = \frac{-1}{g_q} \left( (\tilde V'(x)-y)(V'(y)-x) - \sum_{i=0}^{p-2}
\sum_{j=0}^{q-2} C_{i,j} x^i y^j \right).
\eeq
Moreover, the condition \eqref{sumabR} implies that $C_{p-2,q-2}=g_q \tilde g_p$.

\bd[Newton polytope]
Let us write
\beq
\mathcal E(x,y) = \sum_{(i,j) \in \mathcal N} \mathcal E_{i,j} x^i y^j.
\eeq
where $\mathcal N\subset \{0,\ldots,p\}\times \{0,\ldots,q\}$ is a finite set of integer points in  $\mathbb Z^2$, called the Newton polytope of $\mathcal E$.

Remark that $\mathcal E_{0,q}=1$, $\mathcal E_{p,0}=\frac{\tilde g_p}{g_q}$, $\mathcal E_{i,q-1}=\frac{g_{q-1}\delta_{i,0}-\tilde g_{i+1}}{g_q} $, $\mathcal E_{p-1,j}=\frac{\tilde g_{p-1}\delta_{j,0}-g_{j+1}}{g_q} $, and
$\mathcal E_{p-2,q-2} = \tilde g_p$ are trivial functions of the coupling constants $g_k$ and $\tilde g_k$.

The remaining $N=(p-1)(q-1)-1$ coefficients $\mathcal E_{i,j}$ with $i\leq p-2$ and $j\leq q-2$ and $(i,j)\neq (p-2,q-2)$, are non-trivial combinations.

In general the points $(i,j)\in \mathcal N$ are called as follows:
\begin{itemize}
\item if $(i+1,j+1)$ is on the boundary or outside of the convex envelope of $\mathcal N$, then $\mathcal E_{i,j}$ is called a Casimir. It is a rational fraction of the coefficients $g_i, \tilde g_i$.
\item if $(i+1,j+1)$ is strictly inside the convex envelope of $\mathcal N$, then $\mathcal E_{i,j}$ is called a conserved Hamiltonian.

\end{itemize}

\ed

\bp[Conserved quantities] The coefficients of $\mathcal  E_{i,j}$ with $(i+1,j+1)$ strictly inside the convex envelope of $\mathcal N$, written as polynomials of the $Q_{i,j}$'s and $P_{i,j}$'s using eq \eqref{eq:Ebivpol}, i.e. --after elimination--  as polynomials of the $R_i$'s, are thus conserved quantities. In fact these generate {\bf ALL the conserved quantities}.
\ep
\proof{
Since $\mathcal E(x,y)$ as given by eq \eqref{eq:Ebivpol} is actually independent of $n$ for $n\geq q$, it is clear that all $\mathcal  E_{i,j}$ are conserved quantities. The restriction of being inside the convex envelope, is just because those that are outside (the so called Casimirs) are independent of the $Q_{i,j}$ or $P_{i,j}$: they are obvious constants.
The fact that these generate all conserved quantities is a classical result in integrable systems, and outside the scope of this article, we admit it here.
}
\bex
 [Quadrangulations ($q=2,p=4$ and $g_1=\tilde{g}_1=\tilde{g}_3=0$)] 
 From eq~\eqref{eq:DnQuadrang}, we have
\bea
\Tr \mathcal D_n(x)
&=& x\left( \frac{R_{n-1}}{Q_{n-1,n-2}}-Q_{n,n-1}P_{n-1,n+2} 
-P_{n,n+3}(Q_{n+2,n+1}+Q_{n+1,n})+P_{n,n+1}
\right) \cr
&& + x^3 \, P_{n,n+3}
\eea
\bea
\det \mathcal D_n(x) 
&=& 
\Big(R_n P_{n-1,n+2} Q_{n+1,
n}-\frac{R_{n-1} P_{n,n+1} Q_{n,n-1}}{Q_{n-1,n-2}}
   +\frac{R_{n-1} P_{n,n+3} Q_{n,n-1} Q_{n+2,n+1}}{Q_{n-1,n-2}} \cr
	&&   +P_{n-1,n} P_{n,n+1} Q_{n,n-1}
   -P_{n-1,n+2} P_{n,n+1} Q_{n,n-1} Q_{n+1,n}\cr
	&&   -P_{n-1,n} P_{n,n+3} Q_{n,n-1} Q_{n+2,n+1}
   +P_{n-1,n+2} P_{n,n+3} Q_{n,n-1} Q_{n+1,n} Q_{n+2,n+1} \cr
   &&   - R_n P_{n-1,n}+\frac{R_{n-1}
   R_n}{Q_{n-1,n-2}}\Big) \cr
&&   +\frac{x^2}{Q_{n-1,n-2}}\,\Big( 
  -R_n P_{n-1,n+2} Q_{n-1,n-2}
  -R_{n-1} P_{n,n+3} Q_{n,n-1} \cr 
   &&   -R_{n-1} P_{n,n+3} Q_{n+1,n}
   -R_{n-1} P_{n,n+3} Q_{n+2,n+1} \cr
   && +P_{n-1,n} P_{n,n+3} Q_{n-1,n-2} Q_{n,n-1}
   +R_{n-1} P_{n,n+1} \Big) \cr
&& +x^4\,\frac{R_{n-1} P_{n,n+3}}{Q_{n-1,n-2}}
\eea
There are thus five invariants (the coefficients of $x$ and $x^3$ in $\Tr \mathcal D_n(x)$ and the coefficients of $x^0,x^2,x^4$ in $\det \mathcal D_n(x)$), but some of them are trivialy independent of $n$, for instance $P_{n,n+3}=\tilde g_4$ and $Q_{n,n-1}/R_n= g_2$ .
\eex

\bp[Left eigenvectors]

We could have proceeded in the same way with left eigenvectors in $\td{\mathcal V}_x(Q)$, and in the windows $\td{\mathcal W}_n=\{n-1,n,n+1,\dots,n+q-2\}=\mathcal W_{n+q-2}$ for $n\geq 1$, and define
\beq
\td{\mathcal D}_n(x) = \mathcal L_{\td{\mathcal W}_{n}}(x) \td\pi\circ P^{t}  \mathcal L_{\td{\mathcal W}_{n}}(x)^{-1}
\eeq
where $\td\pi$ is defined as:
\bea
\td\pi: \quad P^{t}\td{\mathcal V}_x(Q) & \to & \td{\mathcal V}_x(Q) \cr
\vec\phi & \mapsto & {\mathcal L}_{\td{\mathcal W}_n(x)^{-1}}(\phi_{n-1},\dots,\phi_{n+q-2})
\eea
such that the map $\td\pi\circ P^{t}$ is an endomorphism of $\td{\mathcal V}_x(Q)$.
This yields the same spectral curve
\beq
\det(y\,{\rm Id}_{q\times q}-\td{\mathcal D}_n(x)) = \det(y\,{\rm Id}_{q\times q}-\mathcal D_n(x)) = \mathcal E(x,y).
\eeq
\ep
\proof{
Let $\tilde{\mathcal E}(x,y)= \det(y-\td{\mathcal D}_n(x))$.
For the same reason as for $\mathcal E(x,y)$, it is also independent of $n$ for $n$ large enough, and can thus be obtained from the large $n$ limit: 
in this limit it is the same as $\mathcal E(x,y)$. 
}
\begin{remark}
Changing the window $n\to n+1$ amounts to a conjugation with transition matrix
$\td\Lambda_n(x)=\mathcal L_{\td{\mathcal W}_{n+1}}(x) \mathcal L_{\td{\mathcal W}_{n}}^{-1}(x)$ as in eq \eqref{eq:Lambdan} with $a_+=q-1$, $a_-=1$, namely
\beq
\label{eq:QtdLambdan}
\td\Lambda_n(x)=
\begin{pmatrix}
0 & 1 & 0 \dots &  \cr
\vdots &  & \ddots & 0\cr
0 &  \dots & 0 & 1 \cr
-\frac{1}{Q_{n+q-1,n}} &  \frac{x-Q_{n,n}}{Q_{n+q-1,n}} & \dots & -\frac{Q_{n+q-2,n}}{Q_{n+q-1,n}}
\end{pmatrix}.
\eeq
\end{remark}

\subsection{Branchpoints and double points}
\label{sec:Brptdblpt}

For generic $(x,y)$ on the spectral curve we have that
\beq
\dim \mathcal V_x(Q)\cap \mathcal V_y(\pi \circ P) = 1.
\eeq

Given $x$, there are generically $q$ distinct values of $z$, denoted $z^i(x), i=0,1,\dots,q-1$ (the ordering doesn't matter, an ordering can be defined locally in each simply connected open domain of the spectral curve not containing singularities of $X$ or $X^{-1}$), such that 
\beq
X(z^i(x))=x
\quad , \quad
\{z^0(x),z^1(x),\dots,z^{q-1}(x)\} = X^{-1}(x).
\eeq

{\bf Branch points}

There exist non--generic points at which the $z^i(x)$ are not distinct: these are called \textit{branchpoints}. There are $q$ branchpoints, denoted $a_0,\dots,a_{q-1}$ on the spectral curve, which are the $q$ solutions of:
\beq
X'(a_i)=0.
\eeq
Generically branchpoints are simple, i.e. $X'(z)$ has a simple zero at $z=a_i$.
From now on, we assume that the coupling constants $g_k$ and $\tilde g_k$ are generic and all branchpoints are simple. The case of non-simple branchpoints is similar with more technical details, and anyway it can be obtained by analytic continuation.

Simple branchpoints are also solutions of
\beq
\mathcal E_y(X(a_i),Y(a_i)):=\partial_y \mathcal E(X(a_i),Y(a_i)) = 0.
\eeq

{\bf Double points}

There also exist non--generic points at which the $z^i(x)$ are distinct, but the $Y(z^i(x))$ are not distinct, these are called double points.
Double points are pairs $(w_a,\bar w_a)$ such that:
\beq
   X(w_a)=X(\bar w_a)\,\,,\,\,Y(w_a)=Y(\bar w_a) \,\,\,{\rm and}\,\, w_a\neq \bar w_a .
\eeq
By convention, we shall call $w_a$ the one with the lowest modulus
\beq
|w_a|\leq |\bar w_a|,
\eeq
and for generic $g_k$'s and $\tilde g_k$'s we have $|w_a|<|\bar w_a|$, that we shall assume in the following.

Double points are also solutions of
\beq
 \mathcal E_y(X(w_a),Y(w_a)) = 0 = \mathcal E_y(X(\bar w_a),Y(\bar w_a)) ,
\eeq
as well as
\beq
\mathcal E_x(X(w_a),Y(w_a)) = 0 = \mathcal E_x(X(\bar w_a),Y(\bar w_a)) .
\eeq
Let $N$ be the number of double points and define
\beq
\Delta(z) := \prod_{a=1}^N (z-w_a)
\quad , \quad
\bar\Delta(z) := \prod_{a=1}^N (z-\bar w_a).
\eeq
They are such that
\bl The number of double points is 
\beq
N=(p-1)(q-1)-1,
\eeq
and
\beq
\label{eq:Eyformz}
z^{N-1} \mathcal E_y(X(z),Y(z)) = (\tilde g_p)^{q-1} \Delta(z)\,\bar\Delta(z)\,X'(z)  \,\, .
\eeq
\el
\proof{
The left hand side of eq \eqref{eq:Eyformz} is a Laurent polynomials of $z$, it could have negative powers of $z$.
As $z\to 0$ we have $X(z)\sim \alpha_{q-1} z^{1-q}$ and $Y(z)\sim R/z$, and 
$\mathcal E_y(x,y)= (q-1) \mathcal E_{p-1,q-1} x^{p-1}y^{q-2} (1+ o(1)) $ 
which behaves as
$\mathcal E_y(X(z),Y(z))\sim (q-1) \mathcal E_{p-1,q-1} \alpha_{q-1}^{p-1}R^{q-2} z^{-(p-1)(q-1)} z^{-(q-2)} $.
If we define $N=(p-1)(q-1)-1$, we see that
\beq
z^{N+q-1} \mathcal E_y(X(z),Y(z))
\eeq
is a polynomial of $z$, it has no negative powers, and is not vanishing at $z=0$.
Moreover, at $z\to\infty$, we have $X(z)=O(z^{1}) $ and $Y(z)=O(z^{p-1})$, so  it behaves like
\beq
O(z^{N+q-1} X(z)^{p-1}Y(z)^{q-2}) = O(z^{N+q-1} z^{p-1} z^{(p-1)(q-2)}) = O(z^{2N+q}).
\eeq
In other words, $z^{N+q-1} \mathcal E_y(X(z),Y(z))$ is a polynomial of $z$ of degree $2N+q$.

Its zeros must be either the branch-points, i.e. the $q$ zeros of the polynomial $z^qX'(z)$, or the double points.
This implies that there are $N$ pairs of double points.
We have that
\beq
z^{N+q-1} \mathcal E_y(X(z),Y(z)) \propto z^q X'(z)  \Delta(z)\bar \Delta(z)
\eeq
and
\beq
\deg \Delta = \deg\bar\Delta = N.
\eeq

}

\textbf{Remark:} The number $N$ of double points is the number of interior points in the Newton polytope of $\mathcal E$.
In fact this is a general result in algebraic geometry: when the curve has genus zero (it has a rational parametrization), the number of double points is always equal to the number of interior points of the Newton polytope.

\bl\label{lemma:eqwbarw}
For $a=1,\dots,N$ we have:
\beq
\label{eq:eqwbarw1}
\frac{\bar w^{N-1}_a}{\Delta(\bar w_a) \bar\Delta'(\bar w_a)} = -\frac{w^{N-1}_a}{\Delta'(w_a) \bar\Delta(w_a)}\ .
\eeq
\el

\proof{
Let $(x,y)$ be a point on the spectral curve. For a small deviation around this point along the spectral curve we have: 

\begin{equation}
\begin{split}
    0=\mathcal E(x+\delta x,y+\delta y)=&\delta x \mathcal{E}_x(x,y)+\delta y \mathcal{E}_y(x,y)\\
    &+\frac{1}{2}\left((\delta y)^2\mathcal E_{yy}(x,y)+2\delta y \delta x \mathcal E_{xy}(x,y)+(\delta x)^2\mathcal E_{xx}(x,y)\right)+...
    \end{split}
\end{equation}
For double points we have:
\begin{equation}
    \mathcal{E}_x=\mathcal{E}_y=0
\end{equation}
and the second order expansion vanishes, namely
\begin{equation}
    (\delta y)^2\mathcal E_{yy}(X(w_a),Y(w_a))+2\delta y \delta x \mathcal E_{xy}(X(w_a),Y(w_a))+(\delta x)^2\mathcal E_{xx}(X(w_a),Y(w_a))=0.
\end{equation}
This is a second order equation for the variable $\frac{\delta y}{\delta x}$, therefore its two solutions satisify: 
\begin{equation}
\label{eq:sumzeros}
    \left(\frac{\delta y}{\delta x}\right)_1 + \left(\frac{\delta y}{\delta x}\right)_2 =-2 \frac{\mathcal E_{xy}}{\mathcal E_{yy}}
\end{equation}
with $\mathcal E_{xy}$ and $\mathcal E_{yy}$ evaluated at $(X(w_a),Y( w_a))=(X(\bar w_a),Y(\bar w_a))$.

At the double points ($w_a$,$\bar w_a$) the equation \eqref{eq:sumzeros} becomes 
\begin{equation}
\label{eq:wabarwaid}
    \frac{Y'(w_a)}{X'( w_a)}+\frac{Y'(\bar w_a)}{X'( \bar w_a)}=-2 \frac{\mathcal E_{xy}}{\mathcal E_{yy}}.
\end{equation}
Let's calculate the residue of the one form $\frac{dx}{\mathcal E_y}$:
\begin{equation}
    \Res_{z\to w_a} \frac{dx}{\mathcal E_y}=\frac{X'(w_a)}{X'(w_a)\mathcal{E}_{xy}+Y'(w_a)\mathcal{E}_{yy}}.
\end{equation}
Using eq \eqref{eq:wabarwaid} we obtain the identity:

\begin{equation}
   \frac{X'(w_a)}{X'(w_a)\mathcal{E}_{xy}+Y'(w_a)\mathcal{E}_{yy}}=-\frac{X'(\bar w_a)}{X'(\bar w_a)\mathcal{E}_{xy}+Y'(\bar w_a)\mathcal{E}_{yy}}
\end{equation}
and therefore
\begin{equation}
    \Res_{z\to w_a} \frac{dx}{\mathcal E_y}=-\Res_{z\to\bar w_a} \frac{dx}{\mathcal E_y}.
\end{equation}
 Using \bea
  \Res_{z\to w_a} \frac{dx}{\mathcal E_y}=\Res_{z\to w_a} \frac{X'(z)dz}{\mathcal E_y(X(z),Y(z))}
= \Res_{z\to w_a} \frac{z^{N-1}dz}{\Delta(z)\bar\Delta(z)}
=  \frac{w^{N-1}_a}{\Delta'(w_a) \bar\Delta(w_a)}
\eea
 and a similar equation for the residue at $\bar w_a$, we obtain the desired equation \eqref{eq:eqwbarw1}.

}

\subsection{Asymptotic behavior of eigenvectors}
\label{sec:asyeig}

At this stage, we know the spectral curve, i.e. the eigenvalues of $Q$ and $\pi \circ P$.

The so-called ``reconstruction method" in integrable systems consists in recovering the eigenvectors from the spectral curve.
Knowing both the eigenvalues and the eigenvectors, we can recover the full operators $Q$ and $\pi\circ P$ (or $\mathcal D_n(x)$).

The reconstruction method relies on the fact that eigenvectors must be (by Cramer's formula), rational functions of $z$ and, if we know their poles and zeros, we can find them explicitely.

\paragraph{Reconstruction by necessary conditions.}

Let $y$ be an eigenvalue of $\mathcal D_n(x)$, i.e. $(x,y)$ is on the spectral curve $\mathcal E(x,y)=0$, and therefore  there exists $z$ such that $x=X(z)$ and $y=Y(z)$.
Let us denote $\vec \psi_n(z)$ be the corresponding eigenvector of $\mathcal D_n(x)$, i.e.
\beq
\mathcal D_n(x) . 
\begin{pmatrix}
\psi_{n-q+1}(z) \cr
\vdots \cr
\psi_n(z)
\end{pmatrix}
= y\,\, 
\begin{pmatrix}
\psi_{n-q+1}(z) \cr
\vdots \cr
\psi_n(z)
\end{pmatrix}.
\eeq
By definition the $q\times q$ matrix $y\,{\rm Id}_{q\times q}-\mathcal D_n(x)$ is not invertible, but, for generic $z$, one can invert its $q-1\times q-1$ minors, and the formula of eigenvectors is a Cramer's like formula, a ratio of minors:
\bp[Cramer's formula for eigenvectors]
\beq\label{ratiopsimpsin}
\forall\,m\in \mathcal W_n \, , \qquad
\psi_m(z) = (-1)^{n-m}\,\psi_n(z) \,\frac{\det_{i\neq n; j\neq m}(y\,{\rm Id}_{q\times q}-\mathcal D_n(x))}{\det_{i\neq n; j\neq n} (y\,{\rm Id}_{q\times q}-\mathcal D_n(x))}.
\eeq

It is possible to normalize the eigenvectors so that for every $n$, $\psi_n(z)$ is a Laurent polynomial of $z$.

\ep
\proof{
Formula \eqref{ratiopsimpsin} just comes from solving the linear system $(y\,{\rm Id}_{q\times q}-\mathcal D_n(x))\vec\psi_n=0$.
In the window $\mathcal W_n$, one can choose $\psi_n(z) = \det_{i\neq n; j\neq n} (y\,{\rm Id}_{q\times q}-\mathcal D_n(x)) f_n(z)$ where $f_n(z)$ is some Laurent polynomial of $z$. This gives
\beq
\forall\,m\in \mathcal W_n \, , \qquad
\psi_m(z) = (-1)^{n-m}\,f_n(z) \,\det_{i\neq n; j\neq m}(y\,{\rm Id}_{q\times q}-\mathcal D_n(x))
\eeq
which is a Laurent polynomial of $z$ for all $m\in \mathcal W_n$.
Then by the recursion $\psi_{m+1}(z)=X(z)\psi_m(z) - \sum_{k=m-q+1}^{m} Q_{m,k}\psi_k(z)$, we see that $\psi_m$ will remain a Laurent polynomial for all $m\geq n$.

We have
\beq
\frac{f_{n+1}(z)}{f_n(z)} = - \frac{\det_{i\neq n; j\neq n}(y\,{\rm Id}_{q\times q}-\mathcal D_n(x))}{\det_{i\neq n+1; j\neq n}(y\,{\rm Id}_{q\times q}-\mathcal D_{n+1}(x))}.
\eeq

}

\paragraph{Behavior at double points.}If some $z^{k}(x)$ approaches $w_a$, there exists another branch $z^{\bar k}(x)$ approaching $\bar w_a$:
\beq
z^k(x) \to w_a \quad \Rightarrow \exists \bar k\neq k\, , \quad z^{\bar k}(x) \to \bar w_a .
\eeq
Notice from formula \eqref{ratiopsimpsin} that, up to a multiplicative constant, the eigenvectors are rational functions of $x$ and $y$ and therefore, at the double points, the two eigenvectors $\vec \psi(z^k(x))$ and $\vec \psi(z^{\bar k}(x))$ tend to be proportional. This implies that when $(x,y)=(X(w_a),Y(w_a))=(X(\bar w_a),Y(\bar w_a))$ is a double point, we have
\beq
\dim \mathcal V_x(Q) \cap \mathcal V_y(\pi \circ P)=1.
\eeq
This implies
\bl
\beq
\psi_n(\bar w_a)/\psi_n(w_a) = \lambda_a
\eeq
is independent of $n$, for $n$ large enough.

Similarly
\beq
\phi_n(\bar w_a)/\phi_n(w_a) = \mu_a
\eeq
is independent of $n$, for $n$ large enough.
\el
\proof{
Using eq~\eqref{ratiopsimpsin} in the window $\mathcal W_n$, we see that for all $m\in \mathcal W_n$ we have
\beq
\frac{\psi_m(\bar w_a)}{\psi_n(\bar w_a)} = \frac{\psi_m(w_a)}{\psi_n(w_a)},
\eeq
and thus, and choosing $m=n-1$, we get
\beq
\frac{\psi_m(\bar w_a)}{\psi_m(w_a)} = \frac{\psi_n(\bar w_a)}{\psi_n(w_a)} = \frac{\psi_{n-1}(\bar w_a)}{\psi_{n-1}(w_a)} .
\eeq
It then holds in all windows larger than $n$, and therefore
by recursion on $n$, it implies that $\lambda_a$ is independent of $n$ after a certain rank.

The proof for the left eigenvectors $\phi_n$ is similar, using the matrix $\tilde{\mathcal D_n}(x)$.
}

\paragraph{Complete basis of eigenvectors.}

As we already saw, given $x$, there are generically $q$ distinct values of $z$, denoted $z^i(x), i=0,1,\dots,q-1$ such that $X(z^i(x))=x$.
Then, for all $i$, $Y(z^i(x))$ is an eigenvalue of $\mathcal D_n(x)$, which implies that $\psi(z^i(x))\in \mathcal V_x(Q)$.
In particular the $q\times q$ matrix
\beq
\Psi_n(x):=
\begin{pmatrix}
\psi_{n-q+1}(z^0(x)) & \psi_{n-q+1}(z^1(x)) & \dots & \psi_{n-q+1}(z^{q-1}(x)) \cr
\vdots & & & \vdots \cr
\psi_{n-1}(z^0(x)) & \psi_{n-1}(z^1(x)) & \dots & \psi_{n-1}(z^{q-1}(x)) \cr
\psi_{n}(z^0(x)) & \psi_{n}(z^1(x)) & \dots & \psi_{n}(z^{q-1}(x)) \cr
\end{pmatrix}
\eeq 
is a matrix whose columns are eigenvectors of $\mathcal D_n(x)$, and thus
\beq
\mathcal D_n(x) \Psi_n(x) = \Psi_n(x) .{\rm diag}(Y(z^0(x)),\dots,Y(z^{q-1}(x))).
\eeq
For generic $x$, the $z^i(x)$ are all distinct, the $Y(z^i(x))$ are all distinct and all those vectors are linearly independent. We thus have   $\det\Psi_n(x)\neq 0$, and 
\beq
\label{eq:DnPsi}
\mathcal D_n(x) = \Psi_n(x) .{\rm diag}(Y(z^0(x)),\dots,Y(z^{q-1}(x)))\,\Psi_n(x)^{-1}.
\eeq
By construction we have
\beq
\label{eq:PsintoPsin1}
\Psi_{n+1}(x) = \Lambda_n(x)  \Psi_n(x) ,
\eeq
and thus using \eqref{eq:QLambdan}
\bea
\frac{\det \Psi_{n+1}(x) }{\det \Psi_n(x)}
&=& \det \Lambda_n(x) \cr
&=& (-1)^{q-1}\,Q_{n,n-q+1} \cr
&=& (-1)^{q-1} \tilde g_{p} R_n R_{n-1}\dots R_{n-q+2}.
\eea

\bp[Large $n$]
From \eqref{eq:Lambdainfty}, we see that $\Lambda_\infty$ is the companion matrix of the Laurent polynomial $X(z)$, i.e. its eigenvalues are $ z^i(x)$ for $i=0,\ldots,q-1$, and its eigenvectors form the Vandermonde matrix of the $1/z^i(x)$, namely
\beq
\Lambda_\infty(x) = \mathcal V(x) .Z(x). \mathcal V(x)^{-1}
\eeq
where
\beq
Z(x) = {\rm diag}(z^0(x),\dots,z^{q-1}(x)), 
\eeq
and 
\beq
\mathcal V(x) = \begin{pmatrix}
z^0(x)^{1-q} & z^1(x)^{1-q} & & z^{q-1}(x)^{1-q} \cr
z^{0}(x)^{2-q} & z^1(x)^{2-q} & & z^{q-1}(x)^{2-q} \cr
\vdots & & & \vdots \cr
z^0(x)^{-1} & z^1(x)^{-1} & & z^{q-1}(x)^{-1} \cr
1 & 1 & \dots & 1 \cr
\end{pmatrix}.
\eeq
This implies that at large $n$ we have
\beq
\mathcal D_\infty(x) = \mathcal V(x) \left(R Z(x)^{-1} + \sum_{k=0}^{q-1} \beta_k Z(x)^{k}  \right) \mathcal V(x)^{-1}
\eeq
and
\beq
\label{eq:Psilargen}
\Psi_n(z) \underset{n\to\infty}{=} \mathcal V(x)  Z(x)^{n} C(x)  (1+o(1))
\eeq
where $C(x)=\operatorname{diag}(C(z^0(x)),\dots,C(z^{q-1}(x)))$ is a diagonal matrix, independent of $n$, i.e. $C(z)$ is a function of $z$, independent of $n$.
This is equivalent to
\beq
\psi_n(z) \underset{n\to\infty}{=} C(z) z^{n} (1+o(1)).
\eeq
Similarly we define
\beq
\tilde{\mathcal V}(x) = \begin{pmatrix}
1 & 1 & \dots & 1 \cr
z^0(x)^{-1} & z^1(x)^{-1} & & z^{q-1}(x)^{-1} \cr
\vdots & & & \vdots \cr
z^{0}(x)^{2-q} & z^1(x)^{2-q} & & z^{q-1}(x)^{2-q} \cr
z^0(x)^{1-q} & z^1(x)^{1-q} & & z^{q-1}(x)^{1-q} \cr
\end{pmatrix}.
\eeq
and we get
\beq
\Phi_n(z) \underset{n\to\infty}{=} \tilde{\mathcal V}(x)  Z(x)^{-n-N+1} \tilde C(x)  (1+o(1))
\eeq
where $\tilde C(x)=\operatorname{diag}(\tilde C(z^0(x)),\dots,\tilde C(z^{q-1}(x)))$ is a diagonal matrix, independent of $n$, i.e. $\tilde C(z)$ is a function of $z$, independent of $n$.
This is equivalent to
\beq
\phi_n(z) \underset{n\to\infty}{=} \tilde C(z) z^{-n-N+1} (1+o(1)).
\eeq

\ep
\proof{
Let's prove first eq \eqref{eq:Psilargen}, we define:
\begin{equation}
    \td\Psi_n(x)=Z(x)^{-n}\mathcal V(x)^{-1}\Psi_n(z).
\end{equation}
From eq \eqref{eq:DnPsi} we get
\begin{equation}
\label{eq:tdPsin}
    \td\Psi_n(x){\rm diag}(Y(z^0(x)),\dots,Y(z^{q-1}(x)))\td\Psi_n(x)^{-1}=Z^{-n}(x)\mathcal{V}^{-1}(x)\mathcal{D}_n(x) \mathcal{V}(x)Z^{n}.
\end{equation}
For large $n$, eq \eqref{eq:tdPsin} leads to the commutation relation
\begin{equation}
\label{eq:commuPsinY}
    \left[\lim_{n\rightarrow \infty} \td\Psi_n(x),{\rm diag}(Y(z^0(x)),\dots,Y(z^{q-1}(x)))\right]=0.
\end{equation}
In addition from eq \eqref{eq:PsintoPsin1} we get 
\begin{equation}
\label{eq:tdPsintotdPsin1}
    \td \Psi_{n+1}(x) \td \Psi_n(x)^{-1}=Z^{-n-1}(x)\mathcal{V}^{-1}(x)\Lambda_n(x)\mathcal{V}(x) Z^{n}(x).
\end{equation}
Taking the limit, eq \eqref{eq:tdPsintotdPsin1} implies that
\begin{equation}
\label{eq:Psinindpn}
   \lim_{n\rightarrow \infty} \td\Psi_{n+1}(x)\td\Psi_{n}(x)^{-1}={\rm Id}_{q\times q}. 
\end{equation}
Eq \eqref{eq:commuPsinY} implies that, for generic $x$, at large $n$ $\td \Psi_n(x)$ is a diagonal matrix. In addition, eq \eqref{eq:Psinindpn} implies that for large $n$,  $\td \Psi_n(x)$ is independent of $n$. Therefore, $\td\Psi_n (x)$ is equivalent to a diagonal matrix independant of $n$ for each generic $x$. Hence, we can write:
\begin{equation}
    \td\Psi_n=C(x)(1+o(1))
\end{equation}
where $C(x)=\operatorname{diag}(C(z^0(x)),\dots,C(z^{q-1}(x)))$ is a diagonal matrix, independent of $n$. Equivalently, 
\begin{equation}
    \psi_{m}(z^j(x))=(z^j(x))^m C(z^j(x))(1+o(1))
\end{equation}
for all $m=n-q+1,...,n$ and $j=0,...,q-1$, recall that the points $(z^j(x))_{0\leq j\leq q-1}$ are such that $X(z^j(x))=x$. Since this is true for every generic point $x$, we deduce that 
\begin{equation}
    \psi_{m}(z)=z^m C(z)(1+o(1))
\end{equation}
for all $z \in \mathbb C P^1$. 

The same works for matrix wave function of left eigenvectors $\Phi_n(x)$.

}
\begin{remark}
We must have 
\beq
C(\bar w_a) = \lim_{n\to \infty} \lambda_a C(w_a) X_a^n = 0,
\eeq
therefore $C(\bar w_a)=0$ and thus $C(z)$ must be proportional to $\bar \Delta(z)$.
\end{remark}

\subsection{Reconstruction formula and Baker-Akhiezer functions}
\label{sec:recformBAfct}
\bl
We have
\beq\label{eq:psinHnkSnk}
\frac{\psi_n(z)}{\bar\Delta(z)} = z^n\left(1+\sum_{a=1}^N \frac{H_{n,a}}{z-\bar w_a} + S_n(1/z) \right) 
\eeq
where $S_n(1/z)\in \mathbb C[1/z]$. Moreover
\beq
1+S_{n}(1/z) = \sum_{k=0}^{k_n} S_{n,n-k}z^{-k} \quad , \quad S_{n,n}=1,
\eeq
with $k_n$ some fixed integer less than or equal to $N$. 

Similarly
\beq\label{eq:phinHnkSnk}
\frac{\phi_n(z)}{\Delta(z)} = z^{-n}\left(1+\sum_{a=1}^N \frac{\td H_{n,a}}{z-w_a} + \td S_n(z) \right) 
\eeq
where $\td S_n(z)\in \mathbb C[z]$. Moreover
\beq
1+\td S_{n}(z) = \sum_{k=0}^{\td k_n} \td S_{n,n+k}z^{k} \quad , \quad \td S_{n,n}=1,
\eeq
with $\td k_n$ some fixed integer less than or equal to $N$.
\el

\proof{
We know that $\psi_n(z)$'s highest degree is $z^{n+c}$ where $c$ is independent of $n$. Up to multiplying by a power of $z$, we may assume that the highest degree is $z^{n+N}$ and thus $z^{-n}\psi_n(z)/\bar\Delta(z) = 1+O(1/z)$.
It follows that $z^{-n}\psi_n(z)/\bar\Delta(z)$ is a rational fraction of $z$ that can have simple poles at $z=\bar w_1,\dots,\bar w_N$, and possible poles at $z=0$.
}

\bd

Let us set $w_0=z$ and $\bar w_0=z'$, $\lambda_0=\mu_0=1$ and let the 
$(N+1)\times (N+1)$ matrices $S_n(z',z)$ and $\td S_n(z',z)$:
\beq
S_n(z,z')_{a,b}
:= 
\delta_{a,b}-\delta_{a,0}\delta_{b,0} - \lambda_a \frac{w_a^n \bar w^{-n}_b}{w_a-\bar w_b}   
\eeq
\beq
\td S_n(z,z')_{a,b}
:= 
\delta_{a,b}-\delta_{a,0}\delta_{b,0} - \mu_a \frac{w_a^n \bar w^{-n}_b}{w_a-\bar w_b},
\eeq
for $a,b=0,\ldots,N$.

Let us also define
\beq
\label{eq:hnsec3}
h_n := \det\left(\delta_{a,b}- \lambda_a \frac{w_a^N \bar w^{-n}_b}{w_a-\bar w_b}   
\right)_{a,b=1,\dots,N},
\eeq
\beq
\label{eq:tdhnsec3}
\td h_n := \det\left(\delta_{a,b}- \mu_a \frac{w_a^N \bar w^{-n}_b}{w_a-\bar w_b}   
\right)_{a,b=1,\dots,N}.
\eeq
\ed

\bl
Let us define the so-called Baker-Akhiezer functions:
\beq\label{def:hatpsindetlambdaa}
\hat\psi_n(z) := \lim_{z'\to\infty} \frac{\bar\Delta(z)}{h_n} \ z'^{n+1} \ \det S_n(z',z)
 = \frac{z^n \bar\Delta(z)}{h_n} \ \det
\begin{pmatrix}
1 & \frac{\bar w^{-n}_b}{z-\bar w_b} \cr
\lambda_a {w_a^n} & \delta_{a,b} - \lambda_a \frac{w_a^n\bar w^{-n}_b}{w_a-\bar w_b}
\end{pmatrix}
\eeq
\beq
\label{def:hatphindetmua}
\hat\phi_n(z') := \lim_{z\to\infty} \frac{\Delta(z)}{\td h_n} \ z^{-n+1} \ \det \td S_n(z',z)
 = \frac{z'^{-n} \Delta(z')}{\td h_n} \ \det
\begin{pmatrix}
-1 & \bar w^{-n}_b \cr
\mu_a \frac{w_a^n}{z'-w_a} & \delta_{a,b} - \mu_a \frac{w_a^n\bar w^{-n}_b}{w_a-\bar w_b}
\end{pmatrix}.
\eeq
They satisfy for all $a=1,\dots,N$:
\beq
\hat\psi_n(\bar w_a) = \lambda_a \hat\psi_n(w_a)
 , \qquad
\hat\phi_n(\bar w_a) = \mu_a \hat\phi_n(w_a).
\eeq

We have
\beq\label{eq:psinphin}
\psi_n(z) = \frac{\sum_{k=0}^{k_n} S_{n,n-k} h_{n-k} \hat \psi_{n-k}(z) }{\sum_{k=0}^{k_n} S_{n,n-k} h_{n-k}  }
 , \qquad
\phi_n(z) = \frac{\sum_{k=0}^{\td k_n} \td S_{n,n+k} \td h_{n+k} \hat \phi_{n+k}(z) }{\sum_{k=0}^{\td k_n} \td S_{n,n+k} \td h_{n+k}  }.
\eeq

\el
\proof{
Eq \eqref{eq:psinphin} is just the solution of the linear equations $\psi_n(\bar w_a)=\lambda_a\psi_n(w_a)$ for the residues $H_{n,a}$ in terms of the $S_{n,k}$ of \eqref{eq:psinHnkSnk}, written as a Cramer's determinant. The Cramer's determinant coefficient of  $S_{n,k}$ is $\hat \psi_{n-k}$.
Same thing for $\phi_n$.
}

\bt[Reconstruction formula]
We have $S_{n,n-k}=\delta_{k,0}$ and $\td S_{n,n+k}=\delta_{k,0}$. This implies that
\beq\label{eq:psinpsihatn}
\psi_n(z) =  \hat \psi_{n}(z),\quad 
\phi_n(z) = \hat \phi_n(z).
\eeq
\et

\proof{
This theorem is rather important in the theory of integrable systems.
Here we shall admit the proof (in fact all the steps of the proof will appear in  \autoref{sec:mainth} below).
Let us just give a sketch of the proof:

1) we shall find a scalar product for which $<\hat\phi_m,\hat\psi_n>=\delta_{n,m}$. This will be used to show that $X(z)\hat\psi_n(z)=\sum_m \hat Q_{n,m} \hat\psi_m(z)$ and that $\hat Q_{n,m} = <\hat\phi_m(z),X(z)\hat\psi_n(z)>$ is a band matrix of the size that we want for $Q$. Similarly $Y(z)\hat\psi_n(z)=\sum_m \hat P_{n,m} \hat\psi_m(z)$ and $\hat P_{n,m} = <\hat\phi_m(z),Y(z)\hat\psi_n(z)>$ is a band matrix of the size that we want for $P$.

2) The matrices $S_{n,m}$ (lower triangular) and $\td S_{n,m}$ (upper triangular) are a change of basis from $\hat\psi_n$ to $\psi_n$ and $\hat\phi_n$ to $\phi_n$. This means that the band matrices $Q$ and $P$ will be equal to conjugations of the band matrices $\hat Q$, $\hat P$ by these triangular matrices, and this will change their band widths unless the triangular matrices are in fact diagonal.
This implies that $S$ and $\td S$ must be diagonal, and thus identity.

All this scalar product technique is described below in \autoref{sec:mainth}.

}

\textbf{Conclusion of this section:}

By necessary conditions, we have found that the wave functions $\psi_n$ and $\phi_n$ must take the forms of determinants \eqref{def:hatpsindetlambdaa} and \eqref{def:hatphindetmua}.
This allows us to recover $Q$ and $P$ as we shall see below.

However, let us remark that these determinantal formulas depend on $2N$ parameters $\lambda_a$ and $\mu_a$ which we still need to find.
Keeping $\lambda_a$ and $\mu_a$ arbitrary gives the solution to the \textit{doubly infinite} problem with $Q_{n,m}$ and $P_{n,m}$ doubly infinite band matrices, with indices $n$ and $m$ ranging from $-\infty$ to $+\infty$.
Let us call it the ``general solution" of the eigenvector problem.

In  our case we need a ``special solution" that satisfies the following constraint: we must impose that $Q_{n,m}$ and $P_{n,m}$ have entries only for $n\geq 0$ and $m\geq 0$.
This is a very strong constraint that fixes the coefficients $\lambda_a$ and $\mu_a$, as we shall see in the next section.

\section{Proof of the main theorem}
\label{sec:mainth}

So far, we have seen that the combinatorics of mobiles falls into the framework of integrable systems, and thus a necessary condition for finding a solution to our combinatorics problem is to find a solution of the integrable system \eqref{eq:PQ} with appropriate boundary conditions at $n=0$. The purpose of this section is to exhibit explicitly this ``combinatorial'' solution.

Even though we will use some of the concepts introduced in \autoref{sec:Intsys}, we will redefine them when required so that the present section is self-contained (apart from use of \autoref{lemma:eqwbarw}).

\subsection{Spectral curve and Baker-Akhiezer functions}
\label{sec:spcurv2}

Our first ingredient is given by the spectral curve and its double points.
We recall that the spectral curve, and more precisely the functions $X(z)$ and $Y(z)$ can be found from the potentials $\tilde V(x)$ and $V(y)$ of \eqref{eq:potentials} as follows:

\bp
There exist unique coefficients $\alpha_j$, $\beta_j$ and $R$, formal series of the $g_k$'s and the $\tilde g_k$'s, such that the Laurent polynomials
\beq
X(z) = z+\sum_{j=0}^{q-1}\alpha_j z^{-j}
\label{eq:XX}
\eeq
\beq
Y(z) = \frac{R}{z}+\sum_{j=0}^{p-1}\beta_j z^{j}
\label{eq:YY}
\eeq
satisfy \autoref{th:alphabeta}, i.e. obey the equations:
\begin{equation}
\begin{split}
& \alpha_j=\sum_{k=1+j}^{q} g_k \times [z^{-j}] Y(z)^{k-1}\ , \quad j=0,\cdots,q-1\\
& \beta_j=\sum_{k=1+j}^{p} \tilde{g}_k \times [z^{j}] X(z)^{k-1}\ , \quad j=0,\cdots,p-1\ \\
& \sum_{j=1}^{{\rm min}(p,q)-1}\,\, j \alpha_j \beta_j = R-1.
\label{eq:alphabeta}
\end{split}
\end{equation} 
\ep
\proof{
It is well known that the equations \eqref{eq:XX}, \eqref{eq:YY} and  \eqref{eq:alphabeta} determine $R$, the $\alpha_j$'s and the $\beta_j$'s as algebraic functions of the $g_k$'s and $\tilde g_k$'s.
We note that there exist in general several solutions to these equations, but exactly one of them is such that $R$, the $\alpha_j$'s and the $\beta_j$'s are formal
power series of the  $g_k$'s and the $\tilde g_k$'s, as easily seen by setting $g_k=g \lambda_k$, $\tilde{g}_k=g \td\lambda_k$ and developing the equations recursively in powers of $g$, starting 
with $R=1+O(g)$, $\alpha_j=g \lambda_{j+1}+O(g^2)$ and $\beta_j=g \td\lambda_{j+1}+O(g^2)$.}

From now on, it will always be implicitly assumed when referring to the spectral curve that we take for $R$ and for the $\alpha_j$'s and the $\beta_j$'s this latter ``combinatorial'' solution.

\medskip
We recall that the spectral curve has $N=(p-1)(q-1)-1>0$ double points, i.e. pairs $(w_a,\bar w_a)$ such that
\beq
X(\bar w_a)=X(w_a) \quad \text{and} \quad Y(\bar w_a)=Y(w_a),
\eeq
and we choose for $w_a$ the one with smallest modulus $|w_a|<|\bar w_a|$ (the moduli are generically not equal). We define $X_a = w_a/\bar w_a$, and thus $|X_a|<1$.
We also define
\beq
\Delta(z) := \prod_{a=1}^N(z-w_a) \quad , \quad \bar\Delta(z) := \prod_{a=1}^N(z-\bar w_a).
\eeq
We recall that the double points satisfy \autoref{lemma:eqwbarw}:
\beq
\frac{w_a^{N-1}}{\Delta'(w_a)\bar\Delta(w_a)} = - \frac{\bar w^{N-1}_a}{\Delta(\bar w_a)\bar\Delta'(\bar w_a)}.
\label{eq:wbarw}
\eeq

\paragraph{The Baker-Akhiezer functions.}

\bd
We define the $N$-dimensional vectors
\beq
\xi_n := (\bar w_1^{n}-w_1^n,\bar w_2^n-w_2^n,\dots,\bar w_N^n-w_N^n)
\eeq
$($remark that $\xi_0=\vec 0$$\ )$.
We then define:
\beq
h_n := \xi_{n+1}\wedge \xi_{n+2} \wedge \dots \wedge \xi_{n+N} = \det(\xi_{n+1},\xi_{n+2} ,\dots ,\xi_{n+N}).
\eeq
and, for $n\geq 0$, the Baker-Akhiezer functions:
\beq
\begin{split}
\psi_n(z) & := \frac{z^{n+1}}{h_n} (z\xi_{n+1}  -\xi_{n+2})\wedge (z\xi_{n+2} - \xi_{n+3})\wedge \dots \wedge (z\xi_{n+N}  - \xi_{n+N+1}), \\
\phi_n(z) & := \psi_{-2-n-N}(z) =  \frac{z^{-n-N-1}}{h_{-n-N-2}} (z\xi_{-n-N-1}  -\xi_{-n-N})\wedge \dots \wedge (z\xi_{-n-2}  - \xi_{-n-1}) . \\
\end{split}
\eeq

\ed
As we will see later (eq \eqref{eq:hnsimp}) the above definitions of $h_n$, $\psi_n(z)$, $\phi_n(z)$ match (up to unimportant global factors) the definitions given in eqs \eqref{eq:hnsec3} to \eqref{def:hatphindetmua} for some appropriate choice of $\lambda_a$ and $\mu_a$.

Remark that $h_{-1}=...=h_{-N}=0$ and that $\psi_n$ is ill-defined for $n=-1,...,-N$ while $\phi_n$ is ill-defined for $n=-2,...,-(N+1)$.

\bp[Generically defined]
Remark that $w_a$ and $\bar w_a$ are algebraic functions of the $g_k$'s and $\tilde g_k$'s, therefore $h_0$ is an algebraic functions of the $g_k$'s and $\tilde g_k$'s.
It is not identically vanishing, this implies that for generic $g_k$'s and $\tilde g_k$'s, we can assume that
\beq
h_0\neq 0
\eeq
$($the set of $g_k$'s and $\tilde g_k$'s such that $h_0=0$ is at least of codimension 1\ $)$.
\ep
\proof{
We only have to prove that there is some $g_k$'s and $\tilde g_k$'s for which $h_0\neq 0$.

This is a consequence of \autoref{prop:formalhn} below in the particular case $n=0$
}

\bp[Asymptotic behaviors]\label{prop:asymppsi}
We have
\beq
\psi_n(z)  \underset{z\to\infty}{\sim }z^{n+N+1}
, \qquad
\phi_n(z) \underset{z\to\infty}{\sim }z^{-n-1},
\eeq
\beq
\psi_n(z) \underset{z\to 0}{\sim } \frac{(-1)^N h_{n+1}}{h_n} \ z^{n+1}
,\qquad 
\phi_n(z) \underset{z\to 0}{\sim } \frac{(-1)^N h_{-n-N-1}}{h_{-n-N-2}} \ z^{-n-1-N}.
\eeq
\ep

\bl\label{lemma:psiwbarw}
We have 
\beq
\psi_n(\bar w_a) = \psi_n(w_a)
\quad , \quad
\phi_n(\bar w_a) = \phi_n(w_a).
\eeq

\el
\proof{
 We will prove it for the case $a=1$. The other cases are clearly similar.
 We have:
 \begin{equation}
    \psi_n(w_1)=\frac{w_1^{n+1}}{h_n} \det(w_1 \xi_{n+1}- \xi_{n+2},...,w_1 \xi_{n+N}- \xi_{n+N+1})\ .
 \end{equation}
Let's denote by $D$ the determinant $\det(w_1 \xi_{n+1}- \xi_{n+2},...,w_1 \xi_{n+N}- \xi_{n+N+1})$.
We denote also $\xi_{n,a}=(\bar w_a^n- w_a^n)$.
We have
\begin{equation}
\begin{split}
   D&=\bar w_1 ^{n+1}(w_1-\bar w_1)\begin{vmatrix}
   1&\bar w_1&\hdots&\bar w_1^{N-1}\\
   w_1\xi_{n+1,2}-\xi_{n+2,2}&w_1\xi_{n+2,2}-\xi_{n+3,2}&\hdots&w_1\xi_{n+N,2}-\xi_{n+N+1,2}\\
   \vdots&&&\vdots\\
    w_1\xi_{n+1,N}-\xi_{n+2,N}&w_1\xi_{n+2,N}-\xi_{n+3,N}&\hdots&w_1\xi_{n+N,N}-\xi_{n+N+1,N}
   \end{vmatrix}\\
   &\\ 
   &=\bar w_1 ^{n+1}(w_1-\bar  w_1)\begin{vmatrix}
   1&\bar w_1+w_1&\hdots&\sum_{i=1}^{N}w_1^{N-i}\bar w_1^{i-1}\\
   w_1\xi_{n+1,2}-\xi_{n+2,2}&w_1^2\xi_{n+1,2}-\xi_{n+3,2}&\hdots&w_1^N\xi_{n+1,2}-\xi_{n+N+1,2}\\
   \vdots&&&\vdots\\
    w_1\xi_{n+1,N}-\xi_{n+2,N}&w_1^2\xi_{n+1,N}-\xi_{n+3,N}&\hdots&w_1^N\xi_{n+1,N}-\xi_{n+N+1,N}
   \end{vmatrix}\\
   &\\
   \end{split}
   \end{equation}
   where, to go from the first to the second line we performed recursively the operation $C_i\to  C_i +w_1C_{i-1}$ for each columm $C_i$.
   We then have :
   \begin{equation}
   \begin{split}
   D&=-\bar w_1 ^{n+1}\begin{vmatrix}
   \bar w_1-w_1&\bar w_1^2-w_1^2&\hdots&\bar w_1^{N}-w_1^{N}\\
   w_1\xi_{n+1,2}-\xi_{n+2,2}&w_1^2\xi_{n+1,2}-\xi_{n+3,2}&\hdots&w_1^N\xi_{n+1,2}-\xi_{n+N+1,2}\\
   \vdots&&&\vdots\\
    w_1\xi_{n+1,N}-\xi_{n+2,N}&w_1^2\xi_{n+1,N}-\xi_{n+3,N}&\hdots&w_1^N\xi_{n+1,N}-\xi_{n+N+1,N}
   \end{vmatrix}\\
   &\\
   &=-\bar w_1 ^{n+1}\begin{vmatrix}
   \bar w_1-w_1&\bar w_1^2-w_1^2&\hdots&\bar w_1^{N}-w_1^{N}\\
   \bar w_1\xi_{n+1,2}-\xi_{n+2,2}&\bar w_1^2\xi_{n+1,2}-\xi_{n+3,2}&\hdots&\bar w_1^N\xi_{n+1,2}-\xi_{n+N+1,2}\\
   \vdots&&&\vdots\\
    \bar w_1\xi_{n+1,N}-\xi_{n+2,N}&\bar w_1^2\xi_{n+1,N}-\xi_{n+3,N}&\hdots&\bar w_1^N\xi_{n+1,N}-\xi_{n+N+1,N}
   \end{vmatrix}.\\
\end{split}
\end{equation}
If we denote by $\tilde D$ the determinant $\det(\bar w_1 \xi_{n+1}- \xi_{n+2},...,\bar w_1 \xi_{n+N}- \xi_{n+N+1})$,
then we have : $\psi_n(\bar w_1)=\frac{\bar w_1^{n+1}}{h_n} \td D$, with

\begin{equation}
   \tilde D=-w_1 ^{n+1}\begin{vmatrix}
   \bar w_1-w_1&\bar w_1^2-w_1^2&\hdots&\bar w_1^{N}-w_1^{N}\\
   \bar w_1\xi_{n+1,2}-\xi_{n+2,2}&\bar w_1^2\xi_{n+1,2}-\xi_{n+3,2}&\hdots&\bar w_1^N\xi_{n+1,2}-\xi_{n+N+1,2}\\
   \vdots&&&\vdots\\
    \bar w_1\xi_{n+1,N}-\xi_{n+2,N}&\bar w_1^2\xi_{n+1,N}-\xi_{n+3,N}&\hdots&\bar w_1^N\xi_{n+1,N}-\xi_{n+N+1,N}
   \end{vmatrix}.
\end{equation}
Therefore we obtain $\psi_n(w_1)=\psi_n(\bar w_1)$ and similarly $\psi_n(w_a)=\psi_n(\bar w_a)$ for any $a=2,\ldots,N$.
Since $\phi_n = \psi_{-2-n-N}$ we deduce that $\phi_n(\bar w_a) = \phi_n(w_a)$ for $a=1,\ldots,N$.
}

\subsection{Scalar product}
\label{sec:scprod}

\bd
We define the scalar product
\beq
<f(z),g(z)> := -\Res_{z\to\infty} \frac{f(z)g(z) \ z^{N-1} dz}{\Delta(z)\bar\Delta(z)}.
\eeq
\ed

\bl
For any bivariate polynomial ${\mathcal P}(x,y)\in \mathbb C[x,y]$, and for any $n,m \geq 0$, we have
\beq
\begin{split}
<\phi_m(z), {\mathcal P}(X(z),Y(z)) \psi_n(z)> 
&:=  -\Res_{z\to\infty} \frac{\phi_m(z) {\mathcal P}(X(z),Y(z)) \psi_n(z) \ z^{N-1} dz}{\Delta(z)\bar\Delta(z)} \cr
&=  \ \Res_{z\to 0} \frac{\phi_m(z) {\mathcal P}(X(z),Y(z)) \psi_n(z) \ z^{N-1} dz}{\Delta(z)\bar\Delta(z)} \cr
\end{split}
\eeq
i.e.\ the scalar product can be calculated either from the residue at $\infty$ or by that at $0$.
\el
\proof{
Remark that $\phi_m(z) {\mathcal P}(X(z),Y(z)) \psi_n(z)$ is a Laurent polynomial of $z$, it has poles only at $z=0$ and at $z=\infty$.
Since the sum of residues at all poles must vanish, we only need to prove that the residues at all the poles $w_a$ and $\bar w_a$ cancel.
In other words it is enough to prove that
\beq
0 =  \Res_{z\to \bar w_a} \frac{\phi_m(z) {\mathcal P}(X(z),Y(z)) \psi_n(z) \ z^{N-1} dz}{\Delta(z)\bar\Delta(z)} 
 + \Res_{z\to w_a} \frac{\phi_m(z) {\mathcal P}(X(z),Y(z)) \psi_n(z) \ z^{N-1} dz}{\Delta(z)\bar\Delta(z)}. 
\eeq
This is true due to eq \eqref{eq:wbarw} (\autoref{lemma:eqwbarw}) and to \autoref{lemma:psiwbarw}.
}

\bp 
We have:
\beq
<\phi_m(z),\psi_n(z)> = \delta_{n,m}.
\eeq
We also have:
\beq\label{eq:hhDelta0}
\Delta(0)\bar\Delta(0) = \frac{h_{n+1}h_{-n-N-1}}{h_n h_{-n-N-2}} .
\eeq

\ep
\proof{
If $n<m$, from the asymptotic behaviors in \autoref{prop:asymppsi}, we see that at large $z$
\beq
\frac{\phi_m(z) \psi_n(z) \ z^{N-1} }{\Delta(z)\bar\Delta(z)} \sim O(z^{N-1+n+N+1-m-1-2N}) = O(z^{n-m-1}) = O(z^{-2})
\eeq
and thus the residue at $\infty$ vanishes.
On the contrary, if $n>m$, we have at $z\to 0$
\beq
\frac{\phi_m(z) \psi_n(z) \ z^{N-1} }{\Delta(z)\bar\Delta(z)} \sim O(z^{N-1+n+1-m-1-N}) = O(z^{n-m-1}) = O(z^0)
\eeq
and thus the residue at $0$ vanishes.
This implies that $<\phi_m(z),\psi_n(z)>=0$ if $n\neq m$.

If $n=m$, the residue at $\infty$ gives
\beq
<\phi_n(z),\psi_n(z)>  = 1.
\eeq
We can also compute the same residue at $z\to 0$: this gives equation \eqref{eq:hhDelta0}.
}
\bd
Let
\beq
\mathfrak W = \{ \pi \in \mathbb C[z] \ | \ \forall a=1,\dots N, \ \pi(\bar w_a)=\pi(w_a) \}
\eeq
\beq
\tilde {\mathfrak W} = \{ \pi\in \mathbb C[1/z] \ | \ \forall a=1,\dots N, \ \pi(\bar w_a)=\pi(w_a) \}
\eeq

\ed

\bl
We have the following properties:
\beq
\forall n\geq 0, \quad \psi_n\in \mathfrak W \quad 
\hbox{and} \quad \phi_n\in \tilde{\mathfrak W}.
\eeq
If $\pi\in \mathbb C[z]$ is a polynomial of degree $\leq N$ which is vanishing at $z=0$ and $\pi\in \mathfrak W$, then $\pi=0$, namely:
\beq
\pi\in \mathfrak W,\quad \pi(0)=0, \quad \deg \pi\leq N \quad \implies \ \pi=0.
\eeq
If $\pi\in \mathbb C[1/z]$ is a polynomial of degree $\leq N$ which is vanishing at $1/z=0$ and $\pi\in \tilde{\mathfrak W}$, then $\pi=0$, namely:
\beq
\pi\in \tilde{\mathfrak W},\quad \pi(\infty)=0, \quad \deg \pi\leq N \quad \implies \ \pi=0.
\eeq
Finally we have:
\beq
\mathfrak W = \mathbb C\ \oplus \mathrm{span}(\{\psi_n\}_{n\geq 0})
\quad , \quad
\tilde{\mathfrak W} = \mathbb C\ \oplus \mathrm{span}(\{\phi_n\}_{n\geq 0}) 
\eeq
\beq
\dim (\mathbb C[z]/\mathfrak W) = N 
\quad , \quad
\dim (\mathbb C[1/z]/\tilde{\mathfrak W}) = N .
\eeq

\el
\proof{
It is obvious that for $n\geq0$, $\psi_n\in \mathfrak W$.
It is obvious also that a polynomial of degree 0, i.e. a constant is in $\mathfrak W$.
Let now $\pi\in \mathfrak W$, such that $\pi(0)=0$ and $\deg \pi\leq N$.
Let us write:
\beq
\pi(z) = \sum_{k=1}^N p_k z^k.
\eeq
Since $\pi\in \mathfrak W$, we must have
\beq
\Xi \begin{pmatrix}
p_1 \cr p_2 \cr \vdots \cr p_N
\end{pmatrix} = 0
\qquad \text{where} \quad
\Xi = \begin{pmatrix}
\xi_1,\xi_2,\dots,\xi_N
\end{pmatrix}
\eeq
We have $\det\Xi=h_0\neq 0$, therefore $\Xi$ is invertible, and thus $\pi=0$.

Then, if $\pi\in \mathfrak W$, this implies that $\pi-\pi(0)$ is a polynomial that vanishes at $0$. If $\deg \pi>N$, we can recursively subtract from it a linear combination of $\psi_n$($n \geq 0$) that kills recursively the highest degree terms, and since $\deg \psi_n=N+n+1$ and $\psi_n(0)=0$, end up with a polynomial of degree $\leq N$ and no constant term. By the result above it is vanishing.
This proves that every $\pi\in \mathfrak W $ can be uniquely written as a linear combination of $\psi_n$'s and a constant.

The same works for $\tilde{\mathfrak W}$.
}

\subsection{Construction of the operators $Q$ and $P$}
\label{sec:constPQ}
Remark that $X(z)\psi_n(z)$ is a Laurent polynomial of $z$.
If $n\geq q-1$, then $X(z)\psi_n(z)$ has only positive powers of $z$, hence is a polynomial of $z$ and, since  $X(w_a)\psi_n(w_a)=X(\bar w_a)\psi_n(\bar w_a)$, it belongs to $\mathfrak W$.

\bp[Operator $Q$]
There exist some coefficients $Q_{n,m}$ and some Laurent polynomial $U_n(z)\in  \mathbb C[z,1/z]$ of degree $\leq N$, such that
\beq
X(z) \psi_n(z) = \sum_{m\geq 0} Q_{n,m} \psi_m(z) + U_n(z)
 \eeq
with $U_n(z)=0$ if $n\geq q-1$.
We have more precisely
\beq
Q_{n,m} = <\phi_m(z),X(z)\psi_n(z)>
\eeq
and in particular
\beq
Q_{n,n+1}=1
\eeq
\beq
Q_{n,m}=0 \ \ \text{if} \ m<n-q+1 \ \text{or} \ m>n+1.
\eeq
In other words
$$X(z) \psi_n(z) = \sum_{m=\max(0,n-q+1)}^{n+1} Q_{n,m} \psi_m(z) + U_n(z)$$
\ep

\proof{
If $n\geq q-1$, we see that $X(z)\psi_n(z)=O(z^{n+2-q})$ is a polynomial in $\mathfrak W$ that vanishes at $z=0$, therefore it is a unique linear combination of $\psi_m$'s, it can be written $\sum_m Q_{n,m}\psi_m$.
For $0\leq n<q-1$, we subtract from $X(z)\psi_n(z)$ the linear combination of $Q_{n,m}\psi_m$ that makes it of degree $\leq N$. The remainder $U_n(z)$ is then a Laurent polynomial of degree $\leq N$.

In all cases, let us compute the scalar product as the residue at $\infty$:
\bea
<\phi_m(z),X(z)\psi_n(z)> 
&=& \sum_{k} Q_{n,k} <\phi_m(z),\psi_k(z)> - \Res_{z\to\infty} \frac{z^{N-1} U_n(z) \phi_m(z)}{\Delta(z)\bar\Delta(z)} dz \cr
&=& \sum_{k} Q_{n,k} \delta_{k,m} - \Res_{z\to\infty} O(z^{N-1+N-m-1-2N}) dz \cr
&=& Q_{n,m} - \Res_{z\to\infty} O(z^{-m-2}) dz \cr
&=& Q_{n,m} .
\eea
Since $X(z)\sim z$ at large $z$, we have $X(z)\psi_n(z)\sim \psi_{n+1}(z)$, and thus $Q_{n,n+1}=1$ and $Q_{n,m}=0$ if $m>n+1$.
If $n\geq q-1$ we have at $z\to 0$ $X(z)\psi_n(z) \sim \alpha_{q-1}O(z^{1-q+n+1})$ i.e. it can be a linear combination of $\psi_m$ only with $m\geq n-q+1$, therefore $Q_{n,m}=0$ if $m<n-q+1$. 
}

\bp[Operator $P$]
There exist some coefficients $P_{n,m}$ such that
\beq
Y(z) \psi_n(z) = \sum_{m\geq 0} P_{n,m} \psi_m(z) + R\frac{(-1)^N h_{1}}{h_0} \ \delta_{n,0} .
\eeq
We have more precisely
\beq
P_{n,m} = <\phi_m(z),Y(z)\psi_n(z)>
\eeq
and
\beq
P_{n,m}=0 \ \ \text{if} \ m<n-1 \ \text{or} \ m>n+p-1
\eeq
i.e., we may write 
$$Y(z) \psi_n(z) = \sum_{m=\max(0,n-1)}^{n+p-1} P_{n,m} \psi_m(z) + R\frac{(-1)^N h_{1}}{h_0} \ \delta_{n,0} .$$
We also have
\beq
P_{n,n-1} = R_{n} = R \frac{h_{n-1}h_{n+1}}{h_n^2}.
\eeq
\ep
\proof{
Remark that $Y(z) \psi_n(z)\in \mathfrak W$. It can be uniquely decomposed as a linear combination of a constant and the $\psi_m$'s.
The constant can occur only if $n=0$, and is then worth $R\frac{(-1)^N h_{n+1}}{h_n} \delta_{n,0}$.

Remark that $<\phi_m(z),1>=0$, so that
\beq
P_{n,m} = <\phi_m(z),Y(z)\psi_n(z)>.
\eeq
Moreover, $Y(z)$ can at most raise the degree by $p-1$ and lower it by $1$, which implies that $P_{n,m}=0$ if $m>n+p-1$ or $m<n-1$.

By computing the residue at $z\to 0$, we get
\bea
R_{n}=P_{n,n-1}
&=& \Res_{z\to 0} \frac{h_{-n-N}h_{n+1}}{h_{-n-N-1}h_n} \frac{z^{N-1} z^{-N-n} z^{n+1} R z^{-1} (1+O(z))} {\Delta(z)\bar\Delta(z)} dz \cr
&=& R \frac{h_{-n-N}h_{n+1}}{h_{-n-N-1}h_n} \ \frac{1}{\Delta(0)\bar\Delta(0)} \cr
&=& R \frac{h_{n-1}h_{n+1}}{h_n^2}.
\eea

}
\bl
For $l\geq 1$
\beq
Y(z)^l \ \psi_n(z) = \sum_{m} (P^l)_{n,m} \psi_m(z) + R\frac{(-1)^N h_{1}}{h_0} \ \left( \frac{Y(z)^l-P^l}{Y(z)-P}\right)_{n,0} .
\eeq
\el
\proof{
Remark that $\frac{Y(z)^l-P^l}{Y(z)-P}$ is a polynomial of $Y(z)$, and thus a Laurent polynomial of $z$.
We prove the lemma by recursion on $l$.
For $l=1$, we have  $\left(\frac{Y(z)^l-P^l}{Y(z)-P}\right)_{n,0} = (\text{Id})_{n,0} = \delta_{n,0}$, so the statement is true for $l=1$.
Let us write:
\beq
c=R\frac{(-1)^N h_{1}}{h_0}.
\eeq

Assume that the statement holds for $l$.
Then compute
\bea
Y(z)^{l+1} \psi_n(z)
&=& Y(z)\left( \sum_{k} (P^l)_{n,k} \psi_k(z) + c \ \left( \frac{Y(z)^l-P^l}{Y(z)-P}\right)_{n,0}  \right) \cr
&=&  \sum_{k} (P^l)_{n,k} Y(z)\psi_k(z) + c \ \left( Y(z) \ \frac{Y(z)^l-P^l}{Y(z)-P}\right)_{n,0}   \cr
&=&  \sum_{k} (P^l)_{n,k} \left(\sum_m P_{k,m}\psi_m + c \delta_{k,0} \right) + c \ \left(  \frac{Y(z)^{l+1}-P^{l+1}}{Y(z)-P} - P^{l}\right)_{n,0}  \cr
&=&  \sum_{m} (P^{l+1})_{n,m} \psi_m + c P^l_{n,0} + c \ \left(  \frac{Y(z)^{l+1}-P^{l+1}}{Y(z)-P} - P^{l}\right)_{n,0}  \cr
&=&  \sum_{m} (P^{l+1})_{n,m} \psi_m + c \ \left(  \frac{Y(z)^{l+1}-P^{l+1}}{Y(z)-P} \right)_{n,0}  
\eea
which is the statement at rank $l+1$.
}

\bt
\label{thm:direct}
Let $T=Q-V'(P)$.
We have
\beq
(X(z)-V'(Y(z))) \psi_n(z) = \sum_{m=n+1}^{n+N} T_{n,m} \psi_m(z)
\eeq
where
\beq
T_{n,m} = <\phi_m(z),(X(z)-\tilde V'(Y(z))) \psi_n(z)>.
\eeq
In particular, the matrix $T$ is a strictly upper triangular matrix and
we have
\beq
T_{n,n+1}  = \frac{1}{R_{n+1}}.
\eeq
Otherwise stated, we have 
\beq
(Q-V'(P))_{-}=0\quad \hbox{and} \quad (Q-V'(P))_{n,n+1}=\frac{1}{R_{n+1}}
\eeq 
as wanted.
\et

\proof{
Remark that $(X(z)-V'(Y(z))) = \frac{z}{R}+O(z^2)$ is a polynomial of $z$, that vanishes at $z=0$. 
Therefore $(X(z)-V'(Y(z))) \psi_n(z) \in \mathfrak W$ and vanishes at $z=0$, this implies that it must be a linear combination of $\psi_m$'s.
We have
\bea
(X(z)-V'(Y(z))) \psi_n(z) 
&=& \sum_{m} Q_{n,m} \psi_m(z) - \sum_m (V'(P))_{n,m} \psi_m(z) \cr
&& + U_n(z) - c \ \left(  \frac{V'(Y(z))-V'(P)}{Y(z)-P} \right)_{n,0}  
\eea
Notice that $\frac{V'(Y(z))-V'(P)}{Y(z)-P}$ is a polynomial of $Y(z)$ of degree at most $q-2$, and thus a Laurent polynomial of $z$ of degree at most $(q-2)(p-1)=N+2-p$.
If we assume $p\geq 2$, this implies that it can have degree at most $N$.
We have
\beq
(X(z)-V'(Y(z))) \psi_n(z)  - \sum_m T_{n,m} \psi_m(z) = U_n(z) - c \ \left(  \frac{V'(Y(z))-V'(P)}{Y(z)-P} \right)_{n,0}  .
\eeq
Both terms in the right hand side are Laurent polynomials of $z$, with possibly negative powers of $z$, however, since the left hand side has no negative powers, all the negative powers must cancel.
Therefore the right hand side is a polynomial of $z$, of degree $\leq N$.
The left hand side vanishes at $z=0$ and is in $\mathfrak W$, therefore the right hand side must be a polynomial in $\mathfrak W$, of degree $\leq N$ and vanishing at $0$, therefore it must vanish.
This implies
\beq
(X(z)-V'(Y(z))) \psi_n(z)  = \sum_m T_{n,m} \psi_m(z).
\eeq
By taking the scalar product we have
\beq
T_{n,m} = <\phi_m(z),(X(z)-V'(Y(z))) \psi_n(z)>.
\eeq
It then follows that $T_{n,m}$ is strictly upper triangular.

Finally we have:
\bea
T_{n,n+1}
&=& <\phi_{n+1}(z),(X(z)-V'(Y(z))) \psi_n(z)> \cr
&=& \Res_0 \frac{h_{-n-N-2}h_{n+1}}{h_{-n-N-3}h_n} \frac{z^{N-1} z^{-N-n-2} z^{n+1} R^{-1} z^{1} (1+O(z))} {\Delta(z)\bar\Delta(z)} dz \cr
&=& \frac{1}{R} \ \frac{h_{-n-N-2}h_{n+1}}{h_{-n-N-3}h_n} \ \frac{1}{\Delta(0)\bar\Delta(0)} \cr
&=& \frac{1}{R} \  \frac{h_{n+1}^2}{h_{n+2}h_n} \cr
&=& \frac{1}{R_{n+1}}.
\eea
}
\subsection{The operators $Q$ and $P$ transposed}
\label{sec:constPQt}

We redo the same with the left eigenvectors $\phi_n$'s.

\bp[Operator $P$]
There exist some coefficients $\tilde P_{n,m}$ and some Laurent polynomial $\tilde U_n(z)\in  \mathbb C[z,1/z]$ of degree $\geq -N$, such that
\beq
Y(z) \phi_n(z) = \sum_{m=\max(0,n-p+1)}^{n+1}\tilde P_{m,n} \phi_m(z) + \tilde U_n(z)
\eeq
with $\tilde U_n(z)=0$ if $n\geq p-1$.
We have
\beq
\tilde P_{n,m} = <\phi_m(z)Y(z),\psi_n(z)> =  <\phi_m(z),Y(z)\psi_n(z)>  = P_{n,m}
\eeq
hence $\tilde P$ is the same matrix as that, $P$, computed in the previous section.
\ep

\proof{
If $n\geq p-1$, we see that $Y(z)\phi_n(z)=O(z^{-n+p-2})$ is a polynomial of $1/z$ in $\tilde{\mathfrak W}$ that vanishes at $z=\infty$, therefore it is a unique linear combination of $\phi_m$'s, it can be written $\sum_m \tilde P_{m,n}\phi_m$.
For $0\leq n<p-1$, we subtract to $Y(z)\phi_n(z)$ the unique linear combination $\tilde P_{m,n}\phi_m$ that makes it of degree $\leq N$ in $1/z$. The remainder $\tilde U_n(z)$ is then a Laurent polynomial of degree $\geq -N$.

In all cases, let us compute the scalar product as the residue at $0$. The remainder does not contribute and we get
\beq
<\phi_m(z)Y(z),\psi_n(z)> 
= \tilde P_{n,m} =  <\phi_m(z),Y(z)\psi_n(z)>  = P_{n,m}
\eeq
therefore $\tilde P_{n,m}=P_{n,m}$. 
}

\bp[Operator $Q$]
There exist some coefficients $\tilde Q_{n,m}$ such that
\beq
X(z) \phi_n(z) = \sum_{m=\max(0,n-q+1)}^{n+1} \tilde Q_{m,n} \phi_m(z) +   \delta_{n,0} .
\eeq
We have
\beq
\tilde Q_{n,m} = <\phi_m(z)X(z),\psi_n(z)>= <\phi_m(z),X(z)\psi_n(z)>=Q_{n,m}
\eeq
hence $\tilde Q$ coincides with the matrix $Q$ computed before.
\ep
\proof{
Remark that $X(z) \phi_n(z)\in \tilde{\mathfrak W}$. It can be uniquely decomposed as a linear combination of a constant and the $\phi_m$'s.
The constant can occur only if $n=0$, and is then worth $ \delta_{n,0}$.

Remark that $<1,\psi_m(z)>=0$, so that
\beq
Q_{n,m} = <\phi_m(z)X(z),\psi_n(z)> = <\phi_m(z),X(z)\psi_n(z)>
\eeq
which then coincides with the matrix found before.
}

\bl
For $l\geq 1$
\beq
X(z)^l \ \phi_n(z) = \sum_{m} (Q^l)_{m,n} \phi_m(z) +  \ \left( \frac{X(z)^l-Q^l}{X(z)-Q}\right)_{0,n} .
\eeq
\el
\proof{
Remark that $\frac{X(z)^l-Q^l}{X(z)-Q}$ is a polynomial of $X(z)$, and thus a Laurent polynomial of $z$.
We prove the lemma by recursion on $l$.
For $l=1$, we have  $\left(\frac{X(z)^l-Q^l}{X(z)-Q}\right)_{0,n} = (\text{Id})_{0,n} = \delta_{n,0}$, so the statement is true for $l=1$.

Assume that the statement holds for $l$.
Then compute
\bea
X(z)^{l+1} \phi_n(z)
&=& X(z)\left( \sum_{k} (Q^l)_{k,n} \phi_k(z) +  \left( \frac{X(z)^l-Q^l}{X(z)-Q}\right)_{0,n}  \right) \cr
&=&  \sum_{k} (Q^l)_{k,n} X(z)\phi_k(z) +  \left( X(z) \ \frac{X(z)^l-Q^l}{X(z)-Q}\right)_{0,n}   \cr
&=&  \sum_{k} (Q^l)_{k,n} \left(\sum_m Q_{m,k}\phi_m +  \delta_{k,0} \right) +  \left(  \frac{X(z)^{l+1}-Q^{l+1}}{X(z)-Q} - Q^{l}\right)_{0,n}  \cr
&=&  \sum_{m} (Q^{l+1})_{m,n} \phi_m +  Q^l_{0,n} +  \left(  \frac{X(z)^{l+1}-Q^{l+1}}{X(z)-Q} - Q^{l}\right)_{0,n}  \cr
&=&  \sum_{m} (Q^{l+1})_{m,n} \phi_m +  \left(  \frac{X(z)^{l+1}-Q^{l+1}}{X(z)-Q} \right)_{0,n}  
\eea
which is the statement at rank $l+1$.
}

\bt
\label{thm:transpose}
Let $\tilde T=P-\tilde V'(Q)$.
We have
\beq
(Y(z)-\tilde V'(X(z))) \phi_n(z) = \sum_{m=n+1}^{n+N} \tilde T_{m,n} \phi_m(z).
\eeq
\beq
\tilde T_{n,m} = <\phi_m(z),(Y(z)-\tilde V'(X(z))) \psi_n(z)> .
\eeq
In particular, $\tilde T$ is a strictly lower triangular matrix
and we have
\beq
\tilde T_{n,n-1}  = 1.
\eeq
Otherwise stated, we have
\beq
(P-\tilde V'(Q))_+=0\quad \hbox{and} \quad (P-\tilde V'(Q))_{n,n-1}=1
\eeq
as wanted.
\et

\proof{
Remark that $(Y(z)-\tilde V'(X(z))) = \frac{1}{z}+O(z^{-2})$ is a polynomial of $1/z$, that vanishes at $z=\infty$. 
Therefore $(Y(z)-\tilde V'(X(z))) \phi_n(z) \in \tilde{\mathfrak W}$ and vanishes at $z=\infty$, this implies that it must be a linear combination of $\phi_m$'s.
We have
\bea
(Y(z)-\tilde V'(X(z))) \phi_n(z) 
&=& \sum_{m} P_{m,n} \phi_m(z) - \sum_m (\tilde V'(Q))_{m,n} \phi_m(z) \cr
&& + \tilde U_n(z) -  \left(  \frac{\tilde V'(X(z))-\tilde V'(Q)}{X(z)-Q} \right)_{0,n}.
\eea
Notice that $\frac{\tilde V'(X(z))-\tilde V'(Q)}{X(z)-Q}$ is a polynomial of $X(z)$ of degree at most $p-2$, and thus a Laurent polynomial of $z$ of degree $\geq -(p-2)(q-1)=-(N+2-q)$.
If we assume $q\geq 2$, this implies that it can have degree $\geq -N$.
We have
\beq
(Y(z)-\tilde V'(X(z))) \phi_n(z)  - \sum_m \tilde T_{m,n} \phi_m(z) = \tilde U_n(z) -  \left(  \frac{\tilde V'(X(z))-\tilde V'(Q)}{X(z)-Q} \right)_{0,n}  .
\eeq
Both terms in the right hand side are Laurent polynomials of $z$, with possibly positive powers of $z$, however, since the left hand side has no positive powers, all the positive powers must cancel.
Therefore the right hand side is a polynomial of $1/z$, of degree $\geq -N$.
The left hand side vanishes at $z=\infty$ and is in $\tilde{\mathfrak W}$, therefore the right hand side must be a polynomial of $1/z$ in $\tilde{\mathfrak W}$, of degree $\geq -N$ and vanishing at $\infty$, therefore it must vanish.
This implies
\beq
(Y(z)-\tilde V'(X(z))) \phi_n(z)  = \sum_m \tilde T_{m,n} \phi_m(z).
\eeq
By taking the scalar product we have
\beq
\tilde T_{n,m} = <\phi_m(z),(Y(z)-\tilde V'(X(z))) \psi_n(z)>.
\eeq
It then follows that $\tilde T_{n,m}$ is strictly lower triangular.

Finally we have:
\bea
\tilde T_{n,n-1}
&=& <\phi_{n-1}(z),(Y(z)-\tilde V'(X(z))) \psi_{n}(z)> \cr
&=& -\Res_{z\to \infty}  \frac{z^{N-1} z^{-n} z^{n+1+N}  z^{-1} (1+O(1/z))} {\Delta(z)\bar\Delta(z)} dz \cr
&=& -\Res_{z\to \infty}  z^{-1} (1+O(1/z)) dz \cr
&=& 1.
\eea
}

\subsection{Conclusion of the proof}
\label{sec:concproof}

The combinatorial mobile enumeration problem has a unique solution for which the mobile or half-mobile generating functions are power series of the $g_k$'s and $\tilde g_k$'s,
with a well-defined limit when $n\to \infty$.

As in \autoref{sec:Mwg}, we consider mobiles weighted by a weight $g$ per labeled vertex by setting $g_1=\tilde{g}_1=0$ and 
\begin{equation}
\label{eq:gktdgk}
\begin{split} 
g_k&=g^{\frac{k-2}{2}}\lambda_k ,\quad k=2,...,q,\\ 
\quad \tilde{g}_k&=g^{\frac{k-2}{2}}\tilde{\lambda}_k,\quad k=2,...,p.
\end{split}
\end{equation}

From what precedes, we may now state the following theorem, which is our main result:

\bt[Main theorem]
\label{th:mainth} Assume $g_1=\td g_1=0$ and $g_k$, $\td g_k$ as in eq \eqref{eq:gktdgk}.  Then the semi-infinite matrices $Q$ and $P$ whose elements are given by the scalar products
\beq
Q_{n,m} = <\phi_m(z),X(z)\psi_n(z)>
\quad , \quad
P_{n,m} = <\phi_m(z),Y(z)\psi_n(z)>\ , \quad n,m\geq 0,
\eeq
are the solution to the combinatorial mobile problem.
In particular, we have the expression
\beq
R_n  = R \ \frac{h_{n-1}h_{n+1}}{h_{n}^2},\quad h_{n}=\det_{1\leq a,b\leq N}\left(\wb_a^{n+b}-\w_a^{n+b}\right).
\label{eq:Riexp}
\eeq
\et

\proof{
That the above scalar products indeed satisfy the equations \eqref{eq:PQ} is a direct consequence of \autoref{thm:direct} and \autoref{thm:transpose}.
In order to prove that they correspond to the wanted combinatorial solution, we only have to verify that they define formal power series in $\sqrt{g}$.
In practice, this boils down to prove that the $R_n$'s are indeed power series in $g$. 

Let us first discuss a number of alternative expressions for $R_n$. We first note that, as proved in \autoref{app:identity},
\beq
h_n=\prod_{a<b} (\bar w_a-\bar w_b)\, \prod_{a=1}^N \wb_a^{n+1}\, 
\det_{1\leq a,b\leq N}\left(\delta_{a,b}-\frac{\prod\limits_{c\neq b} (\w_a-\wb_c)}{\prod\limits_{c\neq a}(\wb_a-\wb_c)}\left(\frac{\w_a}{\wb_a}\right)^{n+1} \right)
\label{eq:hnsimp}
\eeq 
and we may therefore write
\beq
R_n  = R \ \frac{\bar h_{n-1}\bar h_{n+1}}{\bar h_{n}^2},\quad \bar h_{n}=\det_{1\leq a,b\leq N} \left(\delta_{a,b}-\frac{\rho_a}{\w_a-\wb_b} X_a^{n+1} \right)
\label{eq:barhrhoa}
\eeq
with
\beq
\rho_a=\frac{\prod\limits_{c\in \{1,\cdots,N\}}(\w_a-\wb_c)}{\prod\limits_{\substack{c\in \{1,\cdots,N\} \\ c\neq a}}(\wb_a-\wb_c)}=\frac{\bar\Delta(\w_a)}{\bar \Delta'(\wb_a)} \ .
\eeq
From \eqref{eq:wbarw}, we may also write
\beq
\rho_a=-X_a^{N-1}\frac{\prod\limits_{c\in \{1,\cdots,N\}}(\wb_a-\w_c)}{\prod\limits_{\substack{c\in \{1,\cdots,N\} \\ c\neq a}}(\w_a-\w_c)}=-X_a^{N-1}\frac{\Delta(\wb_a)}{\Delta'(\w_a)} .
\eeq\\
so that, using the identity 
\beq
\det_{1\leq a,b\leq N} \left(\delta_{a,b}-\frac{\rho_a}{\w_a-\wb_b} X_a^{n+1} \right)=\det_{1\leq a,b\leq N} \left(\delta_{a,b}-\frac{\rho_a}{\w_b-\wb_a} X_a^{n+1} \right)
\label{eq:id1}
\eeq
 we get the alternative expression
\beq
\bar h_n=\det_{1\leq a,b\leq N}\left(\delta_{a,b}-\frac{\prod\limits_{c\neq b} (\wb_a-\w_c)}{\prod\limits_{c\neq a}(\w_a-\w_c)}\left(\frac{\w_a}{\wb_a}\right)^{n+N}\right)\ .
\label{eq:barhrhoabis}
\eeq
In order to prove that $R_n$'s are power series in $g$ it is enough to prove that: 
\bp
\label{prop:formalhn}
For $n\geq 0$, the $\bar{h}_n$'s given by \eqref{eq:barhrhoa}, or equivalently by \eqref{eq:barhrhoabis} are formal power series in ${g}$.
\ep
This is done in \autoref{app:formalseries}. 
}

\section{Applications}
\label{sec:apps}
\subsection{The case of general planar map}
\label{sec:genplamap}
The case of {\it general}, i.e. non necessarily face bicolorable maps, is obtained by specializing
the black face weights of Eulerian maps to $\tilde{g_k}=\delta_{k,2}$ so that only black faces of degree $2$ are allowed
and receive weight $1$. These Eulerian maps are clearly in bijection with general maps upon squeezing the bivalent black faces 
into single edges, keeping only as faces the original white faces. The faces of the general maps are weighted 
by $g_{k}$ according to their degree $k$. Note that the canonical orientation on the Eulerian map is such that, after squeezing,
each edge of the general map is oriented {\it both ways}. In particular, the oriented geodesic distance on the (supposedly pointed)
Eulerian map is the true geodesic distance from the root vertex on the associated pointed general map (i.e.\ the graph distance using paths
on un-oriented edges).  
Let us now discuss the enumeration of the mobiles corresponding to this specialization, with
a special emphasis on the generating function $R_i$. Note that, in the map language, we may now 
interpret $R_i^{(0)}=R_i-R_{i-1}$ for $i\geq 2$ (or simply $R_1^{(0)}=R_1$ if $i=1$) as 
the generating function for pointed general planar maps with a marked edge $e$ connecting a vertex $v$ at distance 
$i-1$ from the root vertex to a vertex $v'$ at distance $i$. 

\subsubsection{Characteristic equation}
From \eqref{eq:BWeq}, we deduce that the only non-vanishing elements of $P$ are
\begin{equation}
P_{i,i-1}=R_i\ , \qquad P_{i,i}=B_{i,i}=W_{i,i}=: S_i\ ,\quad P_{i,i+1}=B_{i,i+1}=W_{i,i+1}=1
\end{equation} 
with $R_i$ and $S_i$ given in terms of $P$ only via \eqref{eq:Rexp} and
\begin{equation}
S_i=\sum_{k\geq 1} g_k (P^{k-1})_{i,i}.
\label{eq:Sexp}
\end{equation}
This in turn implies that
\begin{equation}
\begin{split}
&Y(z)=\frac{R}{z}+S+z\\
&X(z)=z+\sum_{j\geq 0} z^{-j} \sum_{k\geq 1+j}g_k \, \pi_{-j}(k-1;R,S)\\
\end{split}
\end{equation}
where 
\begin{equation}
\pi_{m}(n;R,S):= [z^{m}] Y(z)^{n}
\end{equation} 
denotes the generating function for three-step paths, i.e.\ lattice paths in the discrete Cartesian plane $\mathbb Z\times \mathbb Z$, starting at $(0, 0)$ and ending at $(n, m)$, 
and made of elementary up-steps $(1, 1)$, level-steps $(1, 0)$ and down-steps $(1, -1)$ with a weight $R$ attached 
to each down-step and a weight $S$ attached to each level-step.
Here $R$ and $S$ are obtained through
\begin{equation}
R=1/(1-\sum_{k\geq 1} g_k \pi_{1}(k-1;R,S))\ , \qquad 
S=\sum_{k\geq 1} g_k \pi_0(k-1;R,S)
\label{eq:RS}
\end{equation}
which are the large $i$ limit counterparts of eqs.~\eqref{eq:Rexp} and \eqref{eq:Sexp}.

Writing $Y(\w)=Y(\wb)$ with $\w\neq\wb$, we deduce
\begin{equation}
0=\frac{Y(\w)-Y(\wb)}{\w-\wb}=-\frac{R}{\w \wb}+1\ ,
\end{equation}
hence
\begin{equation}
\w\,\wb=R
\end{equation}
while, writing  $X(\w)=X(\wb)$, we deduce
\begin{equation}
\begin{split}
0&=\frac{X(\w)-X(\wb)}{\w-\wb}= 1+\sum_{j\geq 1} \frac{\w^{-j}-\wb^{-j}}{\w-\wb}
\sum_{k\geq 1+j}g_k \, \pi_{-j}(k-1;R,S)\\
&=1-\sum_{j\geq 1}R^{-\frac{j+1}{2}}\left(\sum_{\substack{n=-j+1 \\ n=-j+1\, [2]}}^{j-1}x^n\right)
 \sum_{k\geq 1+j} g_k \, \pi_{-j}(k-1;R,S) \\
\end{split}
\end{equation}
where we have set
\begin{equation}
x=\sqrt{X}= \sqrt{\frac{\w}{\wb}}\ .
\end{equation}
Multiplying the above equation by $R$ and writing $j=|n|+1+2m$ (with $m\geq 0$),
we deduce the characteristic equation
\begin{equation}
0=\sum_{n\in\mathbb Z} B_n x^n\qquad {\rm with}\ B_n=R\, \delta_{n,0}
-\sum_{k\geq 2+|n|} g_k \sum_{m=0}^{\lfloor\frac{k-2-|n|}{2}\rfloor}\, \pi_{-2m-|n|-1}(k-1;R,S) R^{-m-\frac{|n|}{2}}\ .
\label{eq:chareq}
\end{equation}
We may compare this equation with the characteristic obtained in \cite{PMCF} by a totally different approach using
continued fractions. There, the characteristic equation has exactly the same form as above, 
but with another expression for $B_n$, namely
\begin{equation}
B_n=\sum_{s\geq|n|}A_s \pi_{|n|}(s;R,S) R^{\frac{|n|}{2}}\ , 
\qquad A_s=R\left(\delta_{s,0}-\sum_{k\geq s+2}g_k \pi_0(k-s-2;R,S)\right)\ ,
\end{equation}
or equivalently
\begin{equation}
B_n=R\, \delta_{n,0}-\sum_{k\geq 2+|n|}g_k \sum_{s=|n|}^{k-2} \pi_0(k-s-2;R,S)\, \pi_{|n|}(s;R,S) R^{1+\frac{|n|}{2}}\ .
\end{equation}
This apparently different expression for $B_n$ turns out to be fully equivalent to our expression, as a consequence of the identity
\begin{equation}
\sum_{s=n}^{k-2} \pi_0(k-s-2;R,S)\, \pi_{|n|}(s;R,S)=
\sum_{m=0}^{\lfloor\frac{k-2-|n|}{2}\rfloor}\, \pi_{-2m-|n|-1}(k-1;R,S) R^{-m-|n|-1}\ ,
\label{eq:pathident}
\end{equation}
proved in \autoref{app:proofid}.

\subsubsection{Expression for $R_i$}
We shall consider here general maps whose face degrees are bounded by a fixed integer, say $q$, larger that or equal to $3$.
In other words, we set  $g_{k}=0$ for $k>q$. Then $X(z)$ has terms ranging from $z$ to $z^{-q+1}$ and 
the spectral curve has generically $N=q-2$ pairs of double-points $(\w_a,\wb_a)$, $a=1,\cdots q-2$
(with $|\w_a/\wb_a|\leq 1$ by convention).
These points satisfy $\w_a\wb_a=R$ and we may thus write
\begin{equation}
\w_a= \sqrt{R}\, x_a\ , \qquad \wb_a=\frac{\sqrt{R}}{x_a}\ , \qquad x_a=\sqrt{\frac{\w_a}{\wb_a}}\, 
\end{equation}
where the $x_a$, $a=1,\cdots q-2$ are the solutions with modulus less than $1$ of the characteristic equation
\eqref{eq:chareq}. Note that $B_n=0$ for $|n|>q-2$, so the characteristic equation has $2(q-2)$ solutions
(the $x_a$ and their inverses $1/x_a$), as expected.
Our general expression \eqref{eq:Riexp} for $R_i$ reads 
\begin{equation}
R_i=R \frac{h_{i-1} \, h_{i+1}}{h_{i}^2}
\end{equation}
with $R$ given by \eqref{eq:RS} and with
\begin{equation}
\begin{split}
h_i& =\det_{1\leq a,b\leq q-2}\left(\wb_a^{i+b}- w_a^{i+b}\right)\\
& =\sqrt{R}^{\, q-2}\det_{1\leq a,b\leq q-2}\left(\frac{1}{x_a^{i+b}}- x_a^{i+b}\right)\ .\\
\end{split}
\end{equation}
This allows to write eventually 
\begin{equation}
R_i=R \frac{\tilde{h}_{i-1} \, \tilde{h}_{i+1}}{\tilde{h}_{i}^2}\ ,\qquad
\tilde{h}_i=\det_{1\leq a,b\leq q-2}\left(\frac{1}{x_a^{i+b}}- x_a^{i+b}\right)\ .
\end{equation}
This form is precisely the general expression obtained in \cite{PMCF} in the framework of continued fractions.

\subsection{The case of $p$-constellations}
\label{sec:constel}
A $p$-constellation is an Eulerian map whose black faces are all of degree $p$ and whose white faces have a degree
multiple of $p$. The generating functions for the corresponding mobiles are obtained by setting $\tilde{g}_k=\delta_{k,p}$
(we don't give a non-trivial weight to black faces as their number can be obtained from the numbers of white faces of all allowed
degrees) and $g_k=\hat{g}_{m}$ if $k=p\, m$ for some $m$ and $g_k=0$ otherwise. 
\subsubsection{Characteristic equation}
From \eqref{eq:BWeq}, we now deduce that the only non-vanishing elements of $P$ are
\begin{equation}
P_{i,i-1}=R_i\ ,\quad P_{i,i+p-1}=B_{i,i+p-1}=(Q^{p-1})_{i,i+p-1}=1
\end{equation} 
with $R_i$ in terms of $P$ only via
\begin{equation}
 R_i=1/\left(1-\sum_{m\geq 1} \hat{g}_m (P^{p\, m-1})_{i-1,i}\right)
 \label{eq:Rexpconst}
\end{equation}
As for $Q$, its only non-vanishing elements are $Q_{i,i+1}=1$ and $Q_{i,i-p\, m+1}=W_{i,i-p\, m+1}$ for $m\geq 1$.
This leads to
\begin{equation}
\begin{split}
&Y(z)=\frac{R}{z}+z^{p-1}\\
&X(z)=z+\sum_{j\geq 1} z^{-p\, j+1} \sum_{m\geq j}\hat{g}_m \, \pi^{(p)}_{-p\, j+1}(p\, m-1;R)\\
\end{split}
\end{equation}
where 
\begin{equation}
\pi^{(p)}_{m}(n;R,S)\equiv [z^{m}] Y(z)^{n}
\end{equation} 
denotes the generating function for $p$-paths, i.e.\ lattice paths in the discrete Cartesian plane $\mathbb Z\times \mathbb Z$,
starting at $(0, 0)$ and ending at $(n, m)$ and made of elementary up-steps $(1, p-1)$ and elementary down-steps $(1, -1)$, with a weight $R$ attached 
to each down-step.
Here $R$ itself is obtained through
\begin{equation}
R=1/(1-\sum_{m\geq 1} \hat{g}_m \pi^{(p)}_{1}(p\, m-1;R))
\label{eq:Rconst}
\end{equation}
which is the large $i$ counterpart of eq.~\eqref{eq:Rexpconst}.

If we now consider $p$-constellations whose white face degrees are bounded, say by $p\, \ell$
(i.e.\ $\hat{g}_{m}=0$ for $m>\ell$ and therefore $q=p\, \ell$ with the notations of \autoref{sec:bounded}), 
then $X(z)$ has terms ranging from $z$ to $z^{-p\, \ell+1}$ and 
the spectral curve has generically $(p-1)(p\, \ell-1)-1=p (p\ell-\ell-1)$ pairs of double-points.
Now, writing $Y(\w)=Y(\wb)$ and $X(\w)=X(\wb)$ displays a clear $p$-fold symmetry, so the double-points
may be classified in $p$-uples of pairs, namely
\begin{equation}
 (\w_a \Omega^s,\wb_a \Omega^s), \quad 
 \left\{\begin{matrix}
 & a=1,\cdots, p\ell-\ell-1\\  & s=0, \cdots,p-1\\
 \end{matrix}
 \right. , \quad \Omega=e^{2i\pi/p}
 \end{equation}
with again $|\w_a/\wb_a|\leq 1$ by convention. Note that we have made implicitly a choice of pair in each $p$-uple (the
pair corresponding to $s=0$)  but this choice is irrelevant in what follows as all quantities involved below are invariant under
the $p$-fold symmetry. For instance, we define
\begin{equation}
X_a=\frac{\w_a}{\wb_a}
\end{equation}
which is also ${\w_a \Omega^s}/{\wb_a \Omega^s}$ for any $s$. Writing $Y(\w_a)=Y(\wb_a)$ with $\w_a\neq\wb_a$, 
we get 
\begin{equation}
R=\wb_a^p(X_a+X_a^2+\cdots+X_a^{p-1})=\w_a^p(X_a^{-1}+X_a^{-2}+\cdots+X_a^{-p+1})
\end{equation}
and we may take the choice
\begin{equation}
\w_a=\left(\frac{R}{X_a^{-1}+X_a^{-2}+\cdots+X_a^{-p+1}} \right)^{1/p}\, \qquad
\wb_a=\left(\frac{R}{X_a+X_a^{2}+\cdots+X_a^{p-1}} \right)^{1/p}\ .
\label{eq:wwbp}
\end{equation}
Writing  $X(\w_a)=X(\wb_a)$, we deduce
\begin{equation}
\begin{split}
0&=\frac{X(\w_a)-X(\wb_a)}{\w_a-\wb_a}= 1+\sum_{j= 1}^\ell \frac{\w_a^{-p\, j+1}-\wb_a^{-p\, j+1}}{\w_a-\wb_a}
\sum_{m= j}^{\ell}\hat{g}_m \, \pi^{(p)}_{-p\, j+1}(p\, m-1;R)\\
&=1-\sum_{j=1}^\ell R^{-j}
(X_a+\cdots+X_a^{p-1})^{\frac{j}{2}}(X_a^{-1}+\cdots+X_a^{-p+1})^{\frac{j}{2}}
\left(\sum\limits_{\substack{n=-p\, j+2 \\ n=-p\, j+2\, [2]}}^{p\, j-2}X_a^{\frac{n}{2}}\right)\\
& \qquad \qquad \qquad \qquad\qquad \qquad \qquad \qquad \qquad \qquad\qquad \qquad\times
 \sum_{m=j}^{\ell}\hat{g}_m \, \pi^{(p)}_{-p\, j+1}(p\, m-1;R) \\
&=1-\sum_{j=1}^\ell R^{-j}(X_a+\cdots+X_a^{p-1})^j
\left(\sum\limits_{n=1}^{p\, j-1}X_a^{-n}\right)
 \sum_{m=j}^{\ell}\hat{g}_m \, \pi^{(p)}_{-p\, j+1}(p\, m-1;R) \\ 
\end{split}
\end{equation}
hence the characteristic equation
\begin{equation}
X_a^{p\ell-\ell-1}=\sum_{j=1}^\ell R^{-j}
(1+\cdots+X_a^{p-2})^{j}
\left(\sum\limits_{n=1}^{p\, j-1}X_a^{p\ell-\ell-1+j-n}\right)
 \sum_{m=j}^{\ell}\hat{g}_m \, \pi^{(p)}_{-p\, j+1}(p\, m-1;R) 
\end{equation}
which is a polynomial equation of degree $2(p\ell-\ell-1)$ in $X_a$ (the coefficients in the sum over $n$
are non-negative), hence gives $2(p\ell-\ell-1)$ solutions,
as wanted (the $X_a$ and $1/X_a$ for $a=1,\cdots,p\ell-\ell-1$).

As a simple example, consider Eulerian $p$-angulations, made of black and white
faces of degree $p$ only, with a weight $g$ per white face (the number of black and white faces are necessarily
the same). The corresponding mobiles are obtained by setting $\hat{g}_m=g\delta_{m,1}$ (so that $\ell=1$) and
the characteristic equation reads
 \begin{equation}
X_a^{p-2}= R^{-1}
(1+\cdots+X_a^{p-2})\left(\sum\limits_{n=1}^{p\,-1}X_a^{p-1-n}\right)
g\, R^{p-1}\ ,
\end{equation}
or equivalently
 \begin{equation}
(1+X_a+\cdots+X_a^{p-2})(1+X_a^{-1}+\cdots+X_a^{-p+2})= 
\frac{1}{
g\, R^{p-2}}\
\end{equation}
with $R=1/(1-g (p-1) R^{p-2})$ from \eqref{eq:Rconst}.
Using this last equation, we may write the following alternative form for the characteristic equation:
\begin{equation}
\sum_{n=1}^{p-2} (p-1-n) \left(X_a^n+X_a^{-n}\right)= 
\frac{1}{
g\, R^{p-1}}\ .
\end{equation}

\subsubsection{Expression for $R_i$}
Let us denote by $N_0=p\ell-\ell-1$ the number of $p$-uples of pairs of double-points of the spectral curve.
From the general expressions \eqref{eq:barhrhoabis} and \eqref{eq:barhrhoa} respectively, we have, $R_i=R\bar h_{i-1} \bar h_{i+1}/(\bar h_{i}^2)$ with the following two equivalent expressions for $\bar h_i$:
\begin{equation}
\bar h_i=\det\limits_{\substack{1\leq a,b\leq N_0\\0\leq s,t\leq p-1}}\left( \delta_{a,b}\delta_{s,t}-
\frac{\prod\limits_{(c,u)\neq (b,t)}(\wb_a \Omega^s-\w_c \Omega^u)}{\prod\limits_{(c,u)\neq (a,s)}(\w_a \Omega^s-\w_c \Omega^u)} 
X_a^{i+N_0p}\right)
\label{eq:hiconst}
\end{equation}
and
\begin{equation}
\bar h_i=\det\limits_{\substack{1\leq a,b\leq N_0\\0\leq s,t\leq p-1}}\left( \delta_{a,b}\delta_{s,t}-
\frac{\prod\limits_{(c,u)\neq (b,t)}(\w_a \Omega^s-\wb_c \Omega^u)}{\prod\limits_{(c,u)\neq (a,s)}(\wb_a \Omega^s-\wb_c \Omega^u)} 
X_a^{i+1}\right)
\label{eq:hiconstbis}
\end{equation}
Using the first expression \eqref{eq:hiconst} and performing the products, we deduce
\begin{equation}
\bar h_i=\det\limits_{\substack{1\leq a,b\leq N_0\\0\leq s,t\leq p-1}}\left( \delta_{a,b}\delta_{s,t}-
U^{(i)}_{a,b} \times\frac{1}{p}\sum_{r=0}^{p-1}\left( \Omega^{t-s} V_{a,b} \right)^r\right)
\label{eq:circblok}
\end{equation}
where
\begin{equation}
U^{(i)}_{a,b}=\frac{\prod\limits_{c\neq b}(\wb_a^p -\w_c^p)}{\prod\limits_{c\neq a}(\w_a^p-\w_c^p)} 
X_a^{i+1+p(N_0-1)}\ , \qquad V_{a,b}=\frac{\w_b}{\wb_a}\ .
\end{equation}
Here we used
\begin{equation}
\begin{split}
\frac{\prod\limits_{u\neq t}(\wb_a \Omega^s-\w_b \Omega^u)}{\prod\limits_{u\neq s}(\w_a \Omega^s-\w_a \Omega^u)} 
= X_a^{1-p}\frac{\prod\limits_{u\neq t}( \Omega^s-\frac{\w_b}{\wb_a} \Omega^u)}{\prod\limits_{u\neq s}(\Omega^s-\Omega^u)} 
&= \frac{X_a^{1-p}}{p} \prod\limits_{u\neq t}( 1-\frac{\w_b}{\wb_a} \Omega^{u-s})
\\&= \frac{X_a^{1-p}}{p}\sum_{r=0}^{p-1}\left( \Omega^{t-s} \frac{\w_b}{\wb_a} \right)^r\ .\\
\end{split}
\end{equation}
Now the matrix in eq.~\eqref{eq:circblok} is a block-circulating matrix, so we may write its determinant as a product of 
determinants (see \autoref{app:prooffac} for details):
\begin{equation}
\begin{split}
\bar h_i&=\prod_{s=0}^{p-1}\det_{1\leq a,b\leq N_0} \left(\delta_{a,b}-
U^{(i)}_{a,b} V_{a,b}^s \right)=\prod_{s=0}^{p-1}\det_{1\leq a,b\leq N_0} \left(\delta_{a,b}-
U^{(i)}_{a,b} \left(\frac{\w_b}{\wb_a}\right)^s \right)\\
&=\prod_{s=0}^{p-1}\ \frac{\prod\limits_{1\leq b\leq N_0} \w_b^s}{\prod\limits_{1\leq a \leq N_0} \wb_a^s} \det_{1\leq a,b\leq N_0}\left(\delta_{a,b}
 \left(\frac{\w_a}{\wb_a}\right)^{-s}- U^{(i)}_{a,b}\right)=\prod_{s=0}^{p-1}\det_{1\leq a,b\leq N_0} \left(\delta_{a,b}-
U^{(i)}_{a,b} X_a^s \right)\\
&=\prod_{s=1}^{p} u_{i+s}\\
\label{eq:facto}
\end{split} 
\end{equation}
with
\begin{equation}
u_i=\det_{1\leq a,b\leq N_0} \left(\delta_{a,b}-\frac{\prod\limits_{c\neq b}(\wb_a^p -\w_c^p)}{\prod\limits_{c\neq a}(\w_a^p-\w_c^p)} 
X_a^{i+p(N_0-1)}\right)\ .
\end{equation}
Using this factorization of $\bar h_i$, we deduce
\begin{equation}
R_i=R \frac{u_i\, u_{i+p+1}}{u_{i+1}\, u_{i+p}}
\end{equation}
with $u_{i}$ as above.

Using now the second expression \eqref{eq:hiconstbis} for $\bar h_i$, we get instead the alternative expression
\begin{equation}
\bar h_i= \prod_{s=1}^{p} v_{i+s}\ , \qquad R_i=R \frac{v_i\, v_{i+p+1}}{v_{i+1}\, v_{i+p}}\ , 
\quad  {\rm with}\ v_i=\det_{1\leq a,b\leq N_0} \left(\delta_{a,b}-\frac{\prod\limits_{c\neq b}(\w_a^p -\wb_c^p)}{\prod\limits_{c\neq a}(\wb_a^p-\wb_c^p)} 
X_a^{i}\right)\ .
\end{equation}
Using the explicit form \eqref{eq:wwbp} of $\w_a$ and $\wb_a$ in terms of $X_a$, we may express $u_i$ and $v_i$ in terms
of $X_a$ only, namely
\begin{equation}
\begin{split}
& u_i=\det_{1\leq a,b\leq N_0} \left(\delta_{a,b}-\frac{\prod\limits_{c\neq b}(\xi_a-\chi_c)}{\prod\limits_{c\neq a}(\chi_a-\chi_c)} 
X_a^{i}\right)\\
& v_i= \det_{1\leq a,b\leq N_0} \left(\delta_{a,b}-\frac{\prod\limits_{c\neq b}(\chi_a -\xi_c)}{\prod\limits_{c\neq a}(\xi_a-\xi_c)} 
X_a^{i+p(N_0-1)}\right) \\
& \xi_a \equiv X_a+\cdots+X_a^{p-1} \ , \qquad \chi_a\equiv  X_a^{-1}+\cdots+X_a^{-p+1}\ .\\
\end{split}
\end{equation} 
In particular, $u_i$ reads
\begin{equation}
\begin{split}
& u_i=\det_{1\leq a,b\leq N_0} \left(\delta_{a,b}-\frac{\sigma_a}{(\xi_a-\chi_b)} 
X_a^{i}\right) \\
& = \sum_{K\subset \{1,\cdots,N_0\}} \prod_{a\in K} (-\sigma_a X_a^{i} ) \times \det_{a,b\in K} \left(\frac{1}{\xi_a-\chi_b}\right)  \\
& = \sum_{K\subset \{1,\cdots,N_0\}} \prod_{a\in K} (-\tau_a X_a^{i} ) \prod\limits_{\substack{a,b\in K \\ a< b}} 
\frac{(\xi_a-\xi_b)(\chi_a-\chi_b)}{(\xi_a-\chi_b)(\chi_a-\xi_b)} \\
\end{split}
\end{equation}
with
\begin{equation}
\sigma_a\equiv \frac{\prod\limits_{c}(\xi_a-\chi_c)}{\prod\limits_{c\neq a}(\chi_a-\chi_c)}\ , \qquad
\tau_a\equiv \frac{\prod\limits_{c\neq a}(\xi_a-\chi_c)}{\prod\limits_{c\neq a}(\chi_a-\chi_c)}
\end{equation}
and similarly
\begin{equation}
\begin{split}
& v_i=\det_{1\leq a,b\leq N_0} \left(\delta_{a,b}-\frac{\tilde{\sigma}_a}{(\chi_a-\xi_b)} 
X_a^{i}\right) \\
& = \sum_{K\subset \{1,\cdots,N_0\}} \prod_{a\in K} (-\tilde{\sigma}_a X_a^{i} ) \times \det_{a,b\in K} 
\left(\frac{1}{\chi_a-\xi_b}\right)  \\
& = \sum_{K\subset \{1,\cdots,N_0\}} \prod_{a\in K} (-\tilde{\tau}_a X_a^{i} ) \prod\limits_{\substack{a,b\in K \\ a< b}} 
\frac{(\xi_a-\xi_b)(\chi_a-\chi_b)}{(\xi_a-\chi_b)(\chi_a-\xi_b)} \\
\end{split}
\end{equation}
with
\begin{equation}
\tilde{\sigma}_a\equiv X_a^{p(N_0-1)}\frac{\prod\limits_{c}(\chi_a-\xi_c)}{\prod\limits_{c\neq a}(\xi_a-\xi_c)}\ , \qquad
\tilde{\tau}_a\equiv X_a^{p(N_0-1)} \frac{\prod\limits_{c\neq a}(\chi_a-\xi_c)}{\prod\limits_{c\neq a}(\xi_a-\xi_c)}
\end{equation}
which is precisely the form that was conjectured in \cite{GDPG} and \cite{GDPGIA}.

\section{Conclusion}
\label{sec:conc}
In this paper we showed that the system of equations satisfied by the mobile generating functions falls into the category of integrable systems. In particular we exhibited an explicit formula for the generating function $R_i$ of mobiles rooted at a labeled vertex $i$ as the ratio of appropriate ($N\times N$) determinants involving the double points of the spectral curve. Note that in our derivation, the fact that the operator $P$ and $Q$ take the form of band matrices was crucial in our analysis. In our case this property was guaranteed  by the fact that our mobiles have unlabeled vertices of bounded degrees. We know however from the continued fraction analysis of \cite{PMCF} that for $p=2$ this result can be extended to white unlabeled vertices of unbounded degree: in that case the expression for $R_i$ involves Hankel determinants which are typically of size $i\times i$ hence growing with $i$. It is therefore likely that our results may be extended to the case of mobiles with undounded black and white vertex degrees, but the explicit expression for $R_i$ is yet to be determined. A first step in this direction would be to study mobile enumeration problems for which the derivatives of the potentiels ($\td V(x)$ and $V(y)$) are rational fractions of $x$ and $y$. Indeed we expect in this case that the $P$ and $Q$ operators still remain band matrices, allowing us to extend our construction in a quite straightforward way.

Another direction of study would be to consider the case of (face-bicolored) maps drawn on higher genus and/or nonorientable surfaces, in correspondence with unicellular mobiles \cite{bettinelli2022bijection}.

\section*{Acknowledgements}
This work is supported by the ERC synergy grant ERC-2018-SyG  810573, ``ReNewQuantum''. We thank Rapha\"el Belliard, J\'er\'emie Bouttier and Taro Kimura for fruitfull discussions on this topic.

\appendix

\section{Proof of \autoref{prop:psRBW}}
\label{app:prop2.5}
Let us rewrite the equations \eqref{eq:BWeq} and \eqref{eq:Rexpbis} in a way which shows that $R_i$, $W_{i,j}$ and $B_{i,j}$ have an expansion as power series in $\sqrt{g}$.

We first use \eqref{eq:Rexpbis} to write: 
\beq
R_i=1+\td{\lambda}_2 W_{i,i-1}+\sum_{k\geq 3}\td{\lambda}_k\sqrt{g}^{k-2}\left(Q^{k-1}\right)_{i,i-1}
\eeq
while from \eqref{eq:BWeq} we get:
\beq
W_{i,i-1}={\lambda}_2 R_{i}+\sum_{k\geq 3}{\lambda}_k\sqrt{g}^{k-2}\left(P^{k-1}\right)_{i,i-1}.
\eeq 
From these two equations we deduce that:
\beq
\label{eq:2.5eq1}
\begin{split}
R_i&=\frac{1}{1-\td{\lambda}_2{\lambda}_2}+\frac{1}{1-\td{\lambda}_2{\lambda}_2}\sum_{k\geq3} \sqrt{g}^{k-2}\bigg\{\td{\lambda}_k\left(Q^{k-1}\right)_{i,i-1}+\td{\lambda}_2{\lambda}_k\left(P^{k-1}\right)_{i,i-1}\bigg\}\\
W_{i,i-1}&=\frac{{\lambda}_2}{1-\td{\lambda}_2{\lambda}_2}+\frac{1}{1-\td{\lambda}_2{\lambda}_2}\sum_{k\geq3} \sqrt{g}^{k-2}\bigg\{\td{\lambda}_k{\lambda}_2\left(Q^{k-1}\right)_{i,i-1}+{\lambda}_k\left(P^{k-1}\right)_{i,i-1}\bigg\}.
\end{split}
\eeq
As for $B_{i,i+1}$ we immediatly obtain from \eqref{eq:BWeq} that:
\beq
\label{eq:2.5eq2}
B_{i,i+1}=\td{\lambda}_2+\sum_{k\geq 3}\td{\lambda}_k\sqrt{g}^{k-2}\left(Q^{k-1}\right)_{i,i+1}.
\eeq
We now use:
\beq
\begin{split}
W_{i,i}&={\lambda}_2B_{i,i}+\sum_{k\geq 3}{\lambda}_k\sqrt{g}^{k-2}\left(P^{k-1}\right)_{i,i}\\
B_{i,i}&=\td{\lambda}_2W_{i,i}+\sum_{k\geq 3}{\td\lambda}_k\sqrt{g}^{k-2}\left(Q^{k-1}\right)_{i,i}
\end{split}
\eeq
to deduce that:
\beq
\label{eq:2.5eq3}
\begin{split}
W_{i,i}&=\frac{1}{1-\td{\lambda}_2{\lambda}_2}\sum_{k\geq3} \sqrt{g}^{k-2}\bigg\{\td{\lambda}_k{\lambda}_2\left(Q^{k-1}\right)_{i,i}+{\lambda}_k\left(P^{k-1}\right)_{i,i}\bigg\}\\
B_{i,i}&=\frac{1}{1-\td{\lambda}_2{\lambda}_2}\sum_{k\geq3} \sqrt{g}^{k-2}\bigg\{\td{\lambda}_k\left(Q^{k-1}\right)_{i,i}+\td{\lambda}_2{\lambda}_k\left(P^{k-1}\right)_{i,i}\bigg\}.
\end{split}
\eeq
Finaly for $j<i-1$ we may use the property that $(P^{k-1})_{i,j}=0\quad \text{for $k-1<i-j$}$ to write:
\beq
\label{eq:2.5eq4}
W_{i,j}=\sum_{k\geq i-j+1} \sqrt{g}^{k-2} \lambda_k (P^{k-1})_{i,j}
\eeq
which involves only strictly positive power of $\sqrt{g}$ since $i-j+1>2$. 

Similarly for $j>i+1$ we may use the property that $(Q^{k-1})_{i,j}=0\quad \text{for $k-1<j-i$}$ to write:
\beq
\label{eq:2.5eq5}
B_{i,j}=\sum_{k\geq j-i+1} \sqrt{g}^{k-2} \td{\lambda}_k (Q^{k-1})_{i,j}
\eeq
which involves only strictly positive power of $\sqrt{g}$ since $j-i+1>2$. 

From \eqref{eq:2.5eq1},\eqref{eq:2.5eq2},\eqref{eq:2.5eq3},\eqref{eq:2.5eq4} and \eqref{eq:2.5eq5} it is clear that the generating functions $R_i$, $B_{i,j}$ and $W_{i,j}$ have formal series expansions in powers of $\sqrt{g}$ and their coefficients $\sqrt{g}^n$  are uniquely determined recursively from the knowledge of the coefficients $\sqrt{g}^m$ of these generating functions for all $m<n$. In other words these equations determine uniquely the expansions of $B_{i,j}$, $W_{i,j}$ and $R_i$ to all orders in $\sqrt{g}$ from the initial values given by:
\beq
\begin{split}
R_i&=\frac{1}{1-\td{\lambda}_2{\lambda}_2}+O(\sqrt{g}),\quad
W_{i,i-1}=\frac{\lambda_2}{1-\td{\lambda}_2{\lambda}_2}+O(\sqrt{g})\quad
B_{i,i+1}=\td{\lambda}_2+O(\sqrt{g})\\
B_{i,j}&=O(\sqrt{g})\quad \text{for}\quad j=i \quad\text{or}\quad j>i+1,\\
W_{i,j}&=O(\sqrt{g})\quad \text{for}\quad j=i \quad\text{or}\quad j<i-1.
\end{split}
\eeq
It is clear from the form of the recursive equations \eqref{eq:2.5eq1},\eqref{eq:2.5eq2},\eqref{eq:2.5eq3},\eqref{eq:2.5eq4} and \eqref{eq:2.5eq5} that the coefficients in these expansions are polynomial in $\lambda_2,...,\lambda_q,\td{\lambda}_2,...,\td{\lambda}_q$ and $\frac{1}{1-\td{\lambda}_2\lambda_2}$ with nonnegative integer coefficients. 
\section{Proof of the equality (\ref{eq:hnsimp})}
\label{app:identity}
We adapt and extend here a similar proof given in Appendix B of \cite{GDPG}. Consider the $N\times N$  matrices $M^{(n)}$ and $H$
with entries
\begin{equation}
\begin{split}
& M^{(n)}_{a,b}=\delta_{a,b}-\frac{\prod\limits_{c\neq b} (\w_a -\wb_c)}{\prod\limits_{c\neq a}(\wb_a-\wb_c)}\, \left(\frac{\w_a}{\wb_a}\right)^{n+1}\\
&H_{a,b}=\frac{\wb_a^{b-1}}{\prod\limits_{c\neq a}(\wb_a-\wb_c)} \ ,\\
\end{split}
\end{equation}
$a,b=1,\cdots,N$. We have 
\begin{equation}
(M^{(n)}H)_{a,b}=\frac{\wb_a^{b-1}}{\prod\limits_{c\neq a}(\wb_a-\wb_c)}\left(
1-\left(\frac{\w_a}{\wb_a}\right)^{n+1} \frac{1}{\wb_a^{b-1}}\sum_{e=1}^{N} \wb_e^{b-1}\prod\limits_{c\neq e}\frac{(\w_a- \wb_c)}{(\wb_e-\wb_c)}\right)\ .
\end{equation}
Now the function
\begin{equation}
f_b(x)=\sum_{e=1}^{N} \wb_e^{b-1}\prod\limits_{c\neq e}\frac{(w- \wb_c)}{(\wb_e-\wb_c)}
\end{equation}
is a polynomial in $w$ of degree at most $N-1$ which satisfies $f_b(\wb_a)=\wb_a^{b-1}$, so that the polynomial $f_b(w)-w^{b-1}$ 
(of degree at most
$N-1$ since $1\leq b\leq N$) vanishes at the $N$ distinct points $\wb_a$, $a=1,\cdots,N$. We deduce that $f_b(w)=w^{b-1}$
for $1\leq b\leq N$ and
\begin{equation}
(M^{(n)}H)_{a,b}=\frac{\wb_a^{-n-1}}{\prod\limits_{c\neq a}(\wb_a-\wb_c)}\left(
\wb_a^{n+b}-\w_a^{n+b}\right)\ .
\end{equation}
The identity~\eqref{eq:hnsimp} follows immediately since 
\begin{equation}
\prod_a\prod_{c\neq a} (\wb_a-\wb_c)\times \det_{1\leq a,b\leq N} H_{a,b}= \prod_{1\leq a< b\leq N} (\wb_a-\wb_b)\ .
\end{equation}
 
\section{Proof of \autoref{prop:formalhn}}
\label{app:formalseries}
Let us set $g_1=\td{g}_1=0$ and 
\beq
\label{eq:B2}
\begin{split}
g_k&=g^{\frac{k-2}{2}} \lambda_k\ , \qquad  k=2,\ldots, q\ ,\\
\tilde{g}_k&=g^{\frac{k-2}{2}}  \td{\lambda}_k\ , \qquad  k=2,\ldots, p\ ,\\
\end{split}
\eeq
where it is implicitly assumed that $\lambda_q\neq 0$ and $\td{\lambda}_p\neq 0$ and $|\lambda_2\td{\lambda}_2|<1$.

From eq \eqref{eq:alphabeta} we first deduce that 
\beq
\label{eq:B3}
\begin{split}
\alpha_0&=\lambda_2 \beta_0+O(\sqrt{g})\\
\beta_0&=\td{\lambda}_2\alpha_0+O(\sqrt{g})
\end{split}
\eeq
hence $\alpha_0=O(\sqrt{g})$ and $\beta_0=O(\sqrt{g})$. We also deduce that for $j>0$, 
\beq
\label{eq:B4}
\begin{split}
\alpha_j&=\lambda_{j+1}\sqrt{g}^{j-1}R^{j}+O(\sqrt{g}^{j}),\\
\beta_j&=\td\lambda_{j+1}\sqrt{g}^{j-1}+O(\sqrt{g}^{j})
\end{split}
\eeq
From the third equation in \eqref{eq:alphabeta} we deduce:  
\beq
R-1=\lambda_2\td{\lambda}_2 R+O(\sqrt{g})
\eeq
and 
\beq
\label{eq:B5}
R=\frac{1}{1-\lambda_2\td{\lambda}_2}+O(\sqrt{g}).
\eeq
In practice, \eqref{eq:B2},\eqref{eq:B3},\eqref{eq:B4} and \eqref{eq:B5} are the first terms of a systematic expansion of $\alpha_j$, $\beta_j$ and $R$ in power of $\sqrt{g}$. We now look for double points $w_a$, $\bar w_a$ solutions of $X(w_a)=X(\bar{w}_a)$ and $Y(w_a)=Y(\bar{w}_a)$, i.e:
\beq
\label{eq:w_abarw_a}
\begin{split}
w_a+\sum_{j=0}^{q-1} \alpha_j w_a^{-j}&=\bar w_a+\sum_{j=0}^{q-1} \alpha_j \bar {w}_a^{-j},\\
\frac{R}{w_a}+\sum_{j=0}^{p-1} \beta_j w_a^j&=\frac{R}{\bar w_a}+\sum_{j=0}^{p-1} \beta_j \bar {w}_a^j.\\
\end{split}
\eeq
where we impose by convention that $|w_a|<|\bar{w}_a|$. From these equations we deduce: 
\beq
w_a \sim \frac{\sqrt{g}}{\eta_a}R ,\quad \bar {w}_a \sim \frac{\xi_a}{\sqrt{g}}
\eeq\\
where $\eta_a$ and $\xi_a$ are solutions of the system of equations:
\beq
\label{eq:vV'}
\begin{split}
\sum_{j=1}^{q-1} \lambda_{j+1} \eta_a^j &=\xi_a\\
\sum_{j=1}^{p-1} \td{\lambda}_{j+1} \xi_a^j &=\eta_a
\end{split}
\eeq
as obtained by equating the leading order of the lhs and rhs of the equations in \eqref{eq:w_abarw_a} (all of order $\frac{1}{\sqrt{g}}$). Finding the solutions of the system is equivalent to finding the roots of polynomial of degree $(p-1)(q-1)=N+1$. Removing the unwanted trivial solution $(\xi_a,\eta_a)=(0,0)$, we are left with exactly $N$ solutions (generically) which yield the desired double points $w_a$,$\bar{w}_a$.
We have in particular
\beq
X_a=\frac{w_a}{\bar{w}_a}=\frac{gR}{\eta_a\xi_a}(1+O(\sqrt{g}))
\eeq

We finally deduce that:
\beq
\begin{split}
\bar{h}_n&=\det_{1\leq a,b\leq N}\left(\delta_{a,b}-\frac{\prod\limits_{c\neq b} (\w_a-\wb_c)}{\prod\limits_{c\neq a}(\wb_a-\wb_c)}X_a^{n+1} \right)\\
&=1-\left(\frac{g}{1-\lambda_2\td{\lambda}_2}\right)^{n+1}\sum_{a=1}^{N}\frac{1}{(\xi_a\eta_a)^{n+1}}\prod_{c\neq a}\frac{1}{1-\frac{\xi_a}{\xi_c}}\left(1+O(\sqrt{g})\right)
\end{split}
\eeq
which is a first term of a systematic expansion in powers of $\sqrt{g}$. Let us now see in details how this expansion works:

From eq \eqref{eq:alphabeta} we have
\begin{align}
    \alpha_j&=\lambda_{j+1}g^{\frac{j-1}{2}}R^j+\lambda_{j+2}g^{\frac{j}{2}}(j+1)\beta_0R^j+o(g^{\frac{j+1}{2}})\\
    \beta_j&=\td\lambda_{j+1}g^{\frac{j-1}{2}}+\td\lambda_{j+2}g^{\frac{j}{2}}(j+1)\alpha_0+o(g^{\frac{j+1}{2}})
\end{align}
with
\begin{align}
 \alpha_0&=\sqrt{g}\frac{2\lambda_2\lambda_3}{1-\lambda_2\tilde{\lambda}_2}+o(\sqrt{g})\\
 \beta_0&=\sqrt{g}\frac{2\tilde{\lambda}_2\tilde{\lambda}_3}{1-\lambda_2\tilde{\lambda}_2}+o(\sqrt{g}).
\end{align}
  then we may thus write
  \begin{align}
    \alpha_j&=\lambda_{j+1}g^{\frac{j-1}{2}}R^j+\gamma_{j}g^{\frac{j+1}{2}}R^j+o(g^{\frac{j+1}{2}})\\
    \beta_j&=\td\lambda_{j+1}g^{\frac{j-1}{2}}+\td\gamma_{j}g^{\frac{j+1}{2}}+o(g^{\frac{j+1}{2}}).
\end{align}
Let us set $w_a=g^{\frac{1}{2}}z_a$ and $\bar w_a=g^{\frac{-1}{2}}\bar z_a$: the system of equations \eqref{eq:w_abarw_a} becomes
\beq
\label{eq:zbarz}
\begin{split}
gz_a+g\sum_{j=1}^{q-1}\gamma_j R^j z_a^{-j}+\sum_{j=1}^{q-1}\lambda_{j+1}R^jz_a^{-j} &=\bar z_a+\sum_{j=1}^{q-1}g^j \lambda_{j+1}R^j \bar {z}_a^{-j}+\sum_{j=1}^{q-1} \gamma_j g^{j+1}R^j\bar z_a^{-j}+O(g^2),\\
\frac{R}{z_a}+\sum_{j=1}^{p-1} g^j\tilde{\lambda}_ {j+1}z_a^j+\sum_{j=1}^{p-1}g^{j+1}\tilde{\gamma}_j z_a^j &=g\frac{R}{\bar z_a}+\sum_{j=1}^{p-1} \tilde{\lambda}_{j+1}\bar z_a^{j}+g\sum_{j=1}^{p-1} \tilde{\gamma}_j\bar z_a^{j}+O(g^2).\\
\end{split}
\eeq
This can be written as:
\beq
\label{eq:zbarz}
\begin{split}
\sum_{j=1}^{q-1}\lambda_{j+1}R^jz_a^{-j}-\bar z_a&=-gz_a-g\sum_{j=1}^{q-1} \gamma_j R^j z_a^{-j}+g \lambda_{2}R \bar {z_a}^{-1} +O(g^2),\\
\sum_{j=1}^{p-1} \tilde{\lambda}_{j+1}\bar z_a^{j}-\frac{R}{z_a}&= g\tilde{\lambda}_ {2}z_a -g\frac{R}{\bar z_a}-g\sum_{j=1}^{p-1} \tilde{\gamma}_j\bar z_a^{j}+O(g^2).\\
\end{split}
\eeq
Define the functions $f$ and $h$ as
\begin{equation}
    f(z,\bar z)=\sum_{j=1}^{q-1}\lambda_{j+1}R^jz^{-j}-\bar z \qquad,\qquad h(z,\bar z)=\sum_{j=1}^{p-1} \tilde{\lambda}_{j+1}\bar z^{j}-\frac{R}{z}
\end{equation}
As we have seen above $f(z,\bar z)=h(z,\bar z)=0$ fixes the leading order to $(z_a^*,\bar z_a^*)=(\frac{R}{\eta_a},\xi_a)$
Writing $z=z_a^*+\delta z_a$, $\bar z_a=\bar z_a^*+\delta \bar z_a$ and linearizing \eqref{eq:zbarz} to first order in $\delta z_a$ and $\delta \bar z_a$ we get a linear system involving the matrix 
\begin{equation}
    D=\begin{pmatrix}
\frac{\partial f}{\partial z}|_{z = z_a^*\atop \bar z=\bar z_a^*}&\frac{\partial f}{\partial \bar z}|_{z = z_a^*\atop \bar z=\bar z_a^*}\\
\frac{\partial h}{\partial z}|_{z = z_a^*\atop \bar z=\bar z_a^*}&\frac{\partial h}{\partial \bar z}|_{z = z_a^*\atop \bar z=\bar z_a^*}
    \end{pmatrix}.
\end{equation}
Its determinant is 
\begin{equation}
\det (D)=-\frac{1}{R}\sum_j j\lambda_{j+1}\eta_a^{j+1}\sum_j j \tilde{\lambda}_{j+1}\xi_a^{j-1}+\frac{\eta_a^2}{R}.
\end{equation}
For $\lambda_a=\lambda_q \delta_{q,a}$ and $\tilde \lambda_a=\tilde \lambda_p \delta_{p,a}$ we get 
\begin{equation}
    \det (D)=-\frac{N}{R}\eta_a^2\neq 0.
\end{equation}
We may deduce from this special case that the determinant does not vanish generically.
Therefore, the matrix D is invertible, which means that there is a second order expansion of $w_a$ and $\bar w_a$ and recursively there is an expansion to all orders. Furthermore, since the rhs of the equation \eqref{eq:zbarz} is an integer power of $g$, the expansion of $z_a$ and $\bar z_a$ is also an integer power of $g$.

Let us calculate the second order expansion. Defining $f_1$ and $h_1$ as
\begin{align*}
    f_1(z,\bar z)&=-z-\sum_{j=1}^{q-1} \gamma_j R^j z^{-j}+\lambda_{2}R \bar {z}^{-1}\\
    h_1(z,\bar z)&=\tilde{\lambda}_ {2}z -\frac{R}{\bar z}-\sum_{j=1}^{p-1} \tilde{\gamma}_j\bar z^{j}
\end{align*}
we have
\begin{equation}
    \begin{pmatrix}
        \delta z_a\\
        \delta \bar z_a
    \end{pmatrix}=gD^{-1} \begin{pmatrix}
         f_1(z_a^*,\bar z_a^*)\\
          h_1(z_a^*,\bar z_a^*)
    \end{pmatrix}.
\end{equation}

We can see now that the expansion of $\bar h_n$ is given by integer powers of $g$.

\section{Proof of the identity (\ref{eq:pathident})}
\label{app:proofid}
To prove \eqref{eq:pathident}, we first use $\pi_m(n;R,S)=R^{-m}\pi_{-m}(n;R,S)$ (as obtained by reversing
the height of the enumerated paths) to write the r.h.s. of this equation as
\begin{equation}
\begin{split}
\sum_{m=0}^{\lfloor\frac{k-2-|n|}{2}\rfloor}\, \pi_{-2m-|n|-1}(k-1;R,S) & R^{-m-|n|-1} =
\sum_{m=0}^{\lfloor\frac{k-2-|n|}{2}\rfloor}\, \pi_{2m+|n|+1}(k-1;R,S) R^{m}\\
&=\sum_{s=n}^{k-2} \pi_{|n|}(s;R,S)\, \sum_{m=0}^{\lfloor\frac{k-s-2}{2}\rfloor}\, \pi^{+}_{2m}(k-s-2;R,S) R^{m}\ ,\\
\end{split}
\end{equation}
where $\pi^{+}_{m}(n;R,S)$ denote tree-step paths from $(0,0)$ to $(n,m)$ {\it which remain above height $0$}. 
Indeed the r.h.s. in the first line above enumerates three-step paths from $(0,0)$ to $(k-1,2m+|n|+1)$ for all possible 
positive $m$, with an extra weight $R^m$. By marking the {\it last passage} at height $|n|$ (which is a step $(s,|n|)\to(s+1,|n|)$
 for some $s$ between $0$ and $k-2$), the path is decomposed into a first part enumerated by $\pi_{|n|}(s;R,S)$
and a last part enumerated by $\pi^{+}_{2m}(k-s-2;R,S)$ (paths of length $k-s-2$ from height $|n|+1$ to height $|n|+1+2m$
which remain above height $|n|+1$). Proving eq.~\eqref{eq:pathident} therefore reduces to proving
\begin{figure}[h!]
\begin{center}
\includegraphics[width=10.cm]{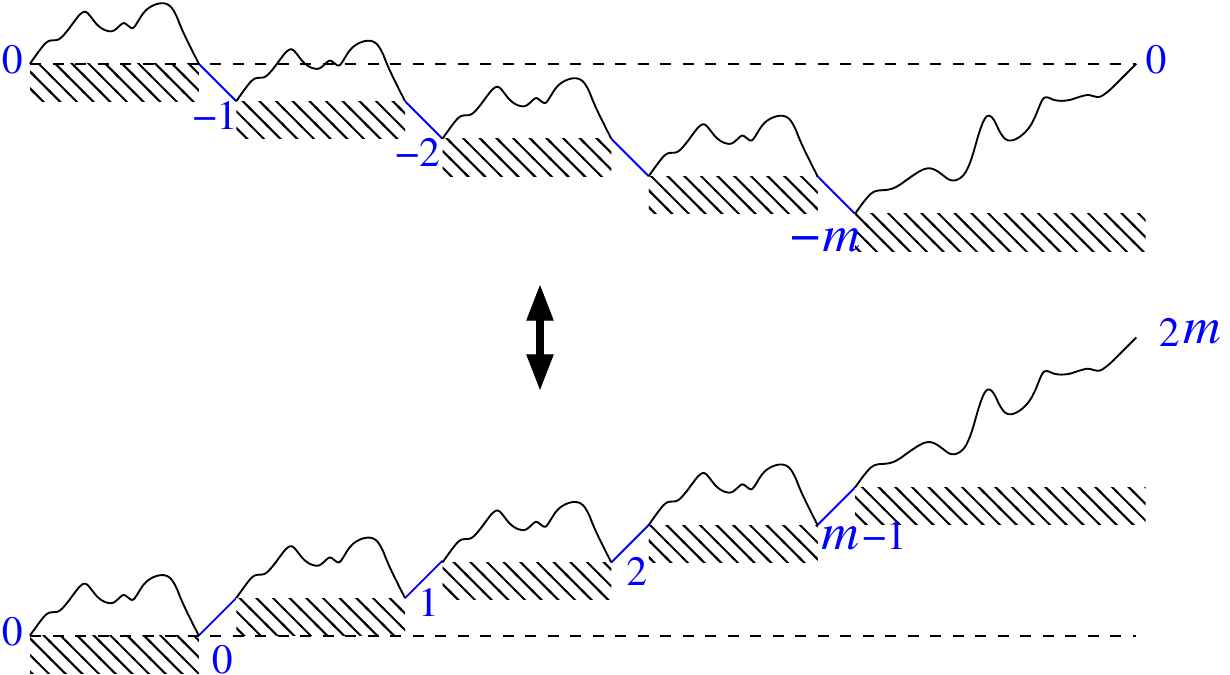}
\end{center}
\caption{\small{Sketch of the proof of Eq~\eqref{eq:pathidreduced} for a given fixed $r$: a path enumerated by $\pi_0(r;R,S)$ and minimum height $-m$ (top) 
is bijectively mapped onto a path enumerated by $\pi^{+}_{2m}(r;R,S)$ (bottom). The weight of the former path is $R^m$ times that of the latter.}} 
\label{fig:appa}
\end{figure}
\begin{equation}
\pi_0(r;R,S)=\sum_{m=0}^{\lfloor\frac{r}{2}\rfloor}\, \pi^{+}_{2m}(r;R,S) R^{m}
\label{eq:pathidreduced}
\end{equation}
for some arbitrary $r$, which is done as follows (see  \autoref{fig:appa} for an illustration): 
consider a path enumerated by $\pi_0(r;R,S)$ and mark the step before the {\it first}
passage at height $-1$ (which is a step $(s_1,0)\to (s_1+1,-1)$ for some $s_1\geq 0$), then the step before the first passage at height $-2$
(which is a step $(s_1+1+s_2,-1)\to (s_1+1+s_2+1,-2)$ for some $s_2\geq 0$) and so on up to the step before the first passage at height 
$-m$ where $-m$ is the minimum height reached by the path (this last marked step is of type 
$(s_1+1+s_2+1+\cdots +s_m,-m+1)\to (s_1+1+s_2+1+\cdots +s_m+1 ,-m)$ for some $s_m\geq 0$). The last part of the path is a path
from height $-m$ to height $0$ which remains above height $-m$. The $m$ marked steps are down steps, hence contribute
$R^m$ to $\pi_0(r;R,S)$. Replacing these steps by up steps creates paths of length $r$ from height $0$ to height $2m$
which remain above height $0$, as enumerated by  $\pi^{+}_{2m}(r;R,S)$. This mapping is bijective 
since the steps which have been reversed are easily identified as the steps just after the {\it last} passage at the heights 
$0,1,\cdots m-1$ in the image path. Eq~\eqref{eq:pathidreduced} follows immediately.

\section{Proof of the factorization (\ref{eq:facto})}
\label{app:prooffac}
Consider the matrix
\begin{equation}
M:= \left( \delta_{a,b}\delta_{s,t}-
U^{(i)}_{a,b} \times\frac{1}{p}\sum_{r=0}^{p-1}\left( \Omega^{t-s} V_{a,b} \right)^r\right)_{\substack{1\leq a,b\leq N_0\\0\leq s,t\leq p-1}}
\end{equation}
whose determinant gives the desired $\bar h_i$. This matrix is block-circulating (recall that $\Omega^p=1$), i.e.\ may be written as
\begin{equation}
M\!=\!\left(
\begin{matrix}
\left(m_{a,b}(0)\right)_{\scriptscriptstyle{1\leq a,b\leq N_0}} & \!\! \left(m_{a,b}(1)\right)_{\scriptscriptstyle{1\leq a,b\leq N_0}} &\!\!\left(m_{a,b}(2)\right)_{\scriptscriptstyle{1\leq a,b\leq N_0}}&
\cdots&\!\!\left(m_{a,b}(p\!-\!1)\right)_{\scriptscriptstyle{1\leq a,b\leq N_0}} \\
\left(m_{a,b}(p\!-\!1)\right)_{\scriptscriptstyle{1\leq a,b\leq N_0}} & \!\! \left(m_{a,b}(0)\right)_{\scriptscriptstyle{1\leq a,b\leq N_0}} &\!\!\left(m_{a,b}(1)\right)_{\scriptscriptstyle{1\leq a,b\leq N_0}}&
\cdots&\!\!\left(m_{a,b}(p\!-\!2)\right)_{\scriptscriptstyle{1\leq a,b\leq N_0}} \\
\left(m_{a,b}(p\!-\!2)\right)_{\scriptscriptstyle{1\leq a,b\leq N_0}} & \!\! \left(m_{a,b}(p\!-\!1)\right)_{\scriptscriptstyle{1\leq a,b\leq N_0}} &\!\!\left(m_{a,b}(0)\right)_{\scriptscriptstyle{1\leq a,b\leq N_0}}&
\cdots&\!\!\left(m_{a,b}(p\!-\!3)\right)_{\scriptscriptstyle{1\leq a,b\leq N_0}} \\
\vdots &  \vdots &\vdots &
\ddots&\vdots \\
\left(m_{a,b}(1)\right)_{\scriptscriptstyle{1\leq a,b\leq N_0}} & \!\! \left(m_{a,b}(2)\right)_{\scriptscriptstyle{1\leq a,b\leq N_0}} &\!\!\left(m_{a,b}(3)\right)_{\scriptscriptstyle{1\leq a,b\leq N_0}}&
\cdots&\!\!\left(m_{a,b}(0)\right)_{\scriptscriptstyle{1\leq a,b\leq N_0}} \\
\end{matrix}
\right)
\end{equation}
with blocks formed of $N_0\times N_0$ matrices $m(t)$ with elements $(m(t))_{a,b}:= m_{a,b}(t)$ given by
\begin{equation}
m_{a,b}(t)=\delta_{a,b} \delta_{t,0}-U^{(i)}_{a,b} \times \frac{1}{p}\sum_{r=0}^{p-1} \left(\Omega^t V_{a,b} \right)^r\ .
\end{equation}
Introducing the $p\times p$ matrix 
\begin{equation}
A=\left(
\begin{matrix}
0 & 1 & 0 & \cdots & 0 \\
0 & 0 & 1 & \cdots & 0 \\
\vdots & \ddots & \ddots & \ddots &  0 \\
0 & 0 & \ddots & \ddots & 1 \\
1 & 0 & 0 & \cdots & 0 \\
\end{matrix}
\right)\ ,
\end{equation}
we may write
\begin{equation}
M=\sum_{t=0}^{p-1} m(t) \otimes A^t\ .
\end{equation}
Now $A$ is easily diagonalized into $A=P\cdot D\cdot  P^{-1}$ with $D$ the $p\times p$ diagonal matrix with diagonal elements $\Omega^s$ for $s=0,\cdots, p-1$, and some unimportant invertible matrix $P$.
We deduce $M=(1\otimes P)\cdot (\sum_{t=0}^{p-1} m(t)\otimes D^t)\cdot (1\otimes P)^{-1}$ and $\det(M)=
\det(\sum_{t=0}^{p-1} m(t)\otimes D^t)$, which is the determinant of a block-diagonal matrix, hence
\begin{equation}
\det(M)=\prod_{s=0}^{p-1}\det\left(\sum_{t=0}^{p-1} m(t) (\Omega^s)^t\right)\ .
\end{equation}
Now 
\begin{equation}
\begin{split}
\sum_{t=0}^{p-1} m_{a,b}(t) (\Omega^s)^t &=\delta_{a,b}-\sum_{t=0}^{p-1}U^{(i)}_{a,b} \times 
\frac{1}{p}\sum_{r=0}^{p-1} \left(\Omega^ t V_{a,b}\right)^r (\Omega^s)^t \\
& =\delta_{a,b}-U^{(i)}_{a,b} \sum_{r=0}^{p-1}  V_{a,b}^r  \times \frac{1}{p} \sum_{t=0}^{p-1}  
\Omega^ {(r+s)t} \\
&= \left\{ 
\begin{matrix}
\delta_{a,b}- U^{(i)}_{a,b}  & {\rm if}\ s=0 \\
\delta_{a,b}- U^{(i)}_{a,b} V_{a,b}^{p-s} & {\rm if}\ s=1,\cdots p-1\\
\end{matrix} 
\right. \\
\end{split}
\end{equation}
and, upon reorganizing the terms, we end up with
\begin{equation}
\det(M)=\prod_{s=0}^{p-1}\det_{1\leq a,b\leq N_0} \left(\delta_{a,b}-
U^{(i)}_{a,b} V_{a,b}^s \right)
\end{equation}
as wanted.

\printbibliography
\end{document}